%% file: apssamp.tex
\documentclass[%
twocolumn,
 amsmath,amssymb,
 aps, physrev,
]{revtex4-2}

\usepackage{graphicx}
\usepackage{dcolumn}
\usepackage{bm}

\usepackage{quantikz}
\usepackage{subcaption}
\usepackage{booktabs}
\usepackage{comment}
\usepackage{multirow}
\usepackage{hyperref}
\usepackage{placeins, soul}
\newcommand{\ketbra}[2]{\left| #1 \middle\rangle \middle\langle #2 \right|}

\usepackage{xspace}
\newcommand{\idest}{i.e.,\xspace}
\newcommand{\eg}{e.g.,\xspace}

\newcommand{\reportnumber}{FERMILAB-PUB-26-0082-SQMS}
\usepackage[angle=0,scale=1,color=black,firstpage=true,opacity=1]{background}

\SetBgContents{\small\texttt{\reportnumber}}
\SetBgPosition{current page.north east}
\SetBgHshift{-3.5cm}
\SetBgVshift{-1.5cm}
\begin{document}

\preprint{APS/123-QED}

\title{\textbf{Sparse Phase Ansatzes for Resource-Efficient Qudit State Preparation via the SNAP-Displacement Protocol} 
}%

\author{Maurizio {Ferrari Dacrema}}
\email{maurizio.ferrari@polimi.it}
\affiliation{%
Politecnico di Milano, Milano, Italy
}%

\author{Do\u{g}a Murat K\"urk\c{c}\"uo\u{g}lu}
\affiliation{
 Superconducting Quantum Materials and Systems Center, Fermi National Accelerator Laboratory, Batavia, Illinois, USA
}%

\author{Andy C. Y. Li}
\affiliation{
 Superconducting Quantum Materials and Systems Center, Fermi National Accelerator Laboratory, Batavia, Illinois, USA
}%

\author{Tanay Roy}
\affiliation{
 Superconducting Quantum Materials and Systems Center, Fermi National Accelerator Laboratory, Batavia, Illinois, USA
}%

\author{Silvia Zorzetti}
\affiliation{
 Superconducting Quantum Materials and Systems Center, Fermi National Accelerator Laboratory, Batavia, Illinois, USA
}%



\begin{abstract}
Efficient preparation of nonclassical bosonic states is a central requirement for quantum computing, simulation, and precision metrology. We study resource-efficient quantum state preparation in bosonic qudit systems using the SNAP-displacement (SD) protocol. Existing SD-based approaches typically require a large number of gates and SNAP phases, resulting in complex control pulses with longer ansatz durations and amplified impact of photon-loss and control errors. In this work, we focus on the near- to medium-term regime, in which noisy quantum devices impose trade-offs on the fidelity that can be achieved, which must be taken into account. Specifically, we propose to optimize only a subset of the SNAP phases and introduce three progressively more general sparse ansatzes. To provide fine-grained control and identify the most suitable ansatz for a given target fidelity, we further employ a scalarized multi-objective optimization that trades off fidelity against either the number of phases or the duration of the ansatz. Numerical results for several target states and qudit dimensions up to $d=64$, evaluated through the hypervolume of the Pareto frontiers, show that these sparse ansatzes achieve favorable trade-offs over the fully parameterized SD protocol in both ideal and noisy settings. The advantage is strongest and most consistent when minimizing the number of phases, while improvements in ansatz duration are smaller and more dependent on the target-state family and noise level, suggesting a practical route to more efficient near- and medium-term bosonic state preparation.
\end{abstract}

\maketitle


\section{Introduction}
\label{sec:introduction}
Bosonic quantum computing encodes quantum information in harmonic oscillator modes, providing access to large Hilbert spaces. This setting allows for universal quantum computation in continuous variables when equipped with a suitable set of Gaussian and non-Gaussian operations~\cite{PhysRevLett.82.1784} or in discrete variables using ancilla qubits \cite{SNAP2015PRL,PhysRevA.92.040303,chakram2022multimode}. In practice, superconducting microwave and radio-frequency cavities coupled to nonlinear ancillae provide high-coherence bosonic modes with lifetimes that can substantially exceed those of typical superconducting qubits, making them attractive for use as long-lived quantum memories \cite{krayzman2024superconducting, Oriani2025coax} and quantum processors ~\cite{SNAP2015PRL, roy2024qudit,bornman2025benchmarking,kim2025ultracoherent}. Within a truncated Fock space of dimension $d$, a single cavity mode can be used to implement a $d$-level system also known as qudit, potentially reducing hardware overhead and circuit complexity \cite{Prakash2025lowoverheadqutrit,Wills2025,RevModPhys.97.021003}, simplifying the encodings required by some quantum algorithms~\cite{10.3389/fphy.2020.589504} including under noise \cite{Jankoviuc2024}, mitigating barren plateaus \cite{Ogunkoya_2025}, and enabling more efficient error correction~\cite{PRXQuantum.3.010335,PRXQuantum.4.020342,Brock2025,kurkccuoglu20252t}. Fock states with large photon numbers also play a crucial role in quantum metrology~\cite{deng2024quantum}, e.g., in dark matter detection~\cite{zhao2025DM, Agrawal2024DM}.

Several universal gate sets are available for bosonic qudits~\cite{Heeres2017, SNAP2015PRL, eickbusch2021ECD, smcc-t465}. One widely used option combines displacement operations with Selective Number-dependent Arbitrary Phase (SNAP) gates~\cite{SNAP2015PRL, Heeres2017}. Together, these gates have been used to implement high-fidelity state preparation and unitary synthesis in cavity systems~\cite{PRXQuantum.3.030301, SNAP2015PRL,fosel2020efficient,roy2024qudit,job2023efficient,vbh4-lysv,kurkccuoglu2024qudit}. Constructive methods exist to decompose arbitrary unitaries into SD gate sequences, for example, based on a sequence of Givens rotations~\cite{fosel2020efficient,job2023efficient}. While these constructions guarantee universality and can achieve very low infidelities in principle, they typically require a substantial number of gates as both target fidelity and $d$ increase. Each additional gate increases the total duration of the ansatz as well as the number of qudit control operations, both of which worsen the qudit performance in the presence of noise. 
Addressing multiple Fock levels simultaneously within a single multi-parameter SNAP gate requires application of multiple microwave pulses with close carrier frequencies which can cause cross-talk between nearby transitions, leading to coherent errors as well as heating issues~\cite{SNAP2015PRL,PRXQuantum.3.030301,roy2024qudit}. As a mitigation strategy, the gate may be split into multiple ones where only a portion of the phases are controlled but at the expense of increased gate duration. The development of optimal pulse control to suppress cross-talk errors is also an active area of research \cite{PhysRevLett.133.260802,smcc-t465,PRXQuantum.3.030301}. 
As a result, there is a nontrivial trade-off between achievable fidelity and resource requirements \cite{PRXQuantum.5.040307}, such as the number of SNAP phases and the effective duration of the ansatz, which becomes particularly important in the near- and medium-term regime. 

In this work, we investigate how to navigate this trade-off in two complementary ways. First, drawing inspiration from empirical observations, we propose several sparse and progressively more general ansatzes in which only a subset of the SNAP phases is optimized to provide explicit control over the number of degrees of freedom. We test those ansatzes on qudit sizes up to $d=64$, significantly larger than what is typically used in the existing literature, and a regime of interest for low-overhead fault-tolerant qudit schemes \cite{cervia2025magic}. Then, to balance state-preparation fidelity against resource requirements, we formulate a multi-objective optimization problem that combines fidelity with a tunable penalty on the number of non-zero phase angles or on the duration of the ansatz. We study the resulting Pareto frontiers across different target states and qudit dimensions, and quantify their quality through the corresponding hypervolume (HV). The results show that in both ideal conditions and under photon-loss noise, our sparse ansatzes can realize better trade-offs than the fully parameterized SD protocol, especially when the objective is to minimize the number of phases, whereas improvements obtained when minimizing the ansatz duration are smaller and less uniform. We provide an implementation of our ansatzes, optimization pipeline as well as the result tables in a GitHub repository \footnote{\url{https://github.com/qcpolimi/SparseSDAnzatz.git}}.

Overall, our results indicate that the internal arrangement of the SNAP phases across gates and Fock levels is an important design parameter, and suggest that exploiting structure and sparsity can be a promising route toward more resource-efficient qudit state preparation. Lastly, multi-objective optimization provides a practical knob for selecting operating points along this trade-off.

\section{The SNAP-Displacement Protocol}
\label{sec:SD_protocol}
In bosonic quantum computing, quantum information is encoded in harmonic oscillator modes, where the state space is described by an infinite-dimensional Fock space. The \emph{displacement gate} and the \emph{Selective Number-dependent Arbitrary Phase} (SNAP) gate form a universal gate set \cite{PhysRevLett.82.1784}. 

\subsection{SNAP and Displacement Gates}

The \textbf{displacement gate}, denoted as $D(\alpha)$, shifts the quantum state by a complex amplitude $\alpha$:
\begin{align}
D(\alpha) = \exp\left( \alpha \hat{a}^\dagger - \alpha^* \hat{a} \right),
\end{align}
where $ \hat{a} $ and $ \hat{a}^\dagger $ are the annihilation and creation operators, respectively, of a boson mode being used as a qudit.

The \textbf{SNAP gate} \cite{SNAP2015PRL} applies independent phase shifts to individual Fock states. It is defined as:
\begin{align}
S(\vec{\theta}) = \sum_{f=0}^{\infty} e^{i \theta_f} \ketbra{f}{f},
\end{align}
where $ \vec{\theta} = (\theta_0, \theta_1, \dots) $ is the set of phase angles applied to each qudit Fock state $\ket{f} $.

While the cavity has an infinite Hilbert space, in practical implementations in practical implementations, controlled SNAP phases are imposed only on the first $d$ Fock states for bandwidth and calibration reasons, at higher Fock levels SNAP phases are nominally zero.
However, the physical action of the displacement gate cannot be constrained to the controllable levels and will exhibit \emph{leakage} to the levels beyond $d$, which we will take into account by conducting our numerical simulations with an additional number of \emph{bumper states} (also referred to as \emph{buffer states}) beyond $d$. 
\subsection{The Protocol}
The SNAP-Displacement (SD) protocol is a widely used method for quantum state preparation in bosonic systems, exploiting sequences of displacement and SNAP gates to approximate target states within a truncated Fock space. In its full form, the protocol is structured as a sequence of $ B $ repeated blocks, each consisting of a displacement gate $ D(\alpha_b) $, a SNAP gate $ S(\vec{\theta}_b) $, and an inverse displacement $ D^\dagger(\alpha_b) $. 
The full sequence can be expressed as:
\begin{equation}
U(\vec{\alpha'}, \Theta) = \prod_{b=1}^{B} D^\dagger(\alpha'_b) S(\vec{\theta}_b) D(\alpha'_b),
\end{equation}
which can be contracted to:
\begin{equation}
U(\vec{\alpha}, \Theta) = D(\alpha_B) \left( \prod_{b=0}^{B-1} S(\vec{\theta}_b) D(\alpha_b)  \right) ,
\end{equation}
where $ \alpha_b \in \mathbb{C} $ are the displacement parameters and $ \vec{\theta}_b  \in \mathbb{R}^{d}$ are the phase angles applied by the SNAP gates at each block $ b $, note that the block index in the two forms differ. The contracted form, where the sequence alternates between SNAP and displacement, can be obtained by using the identities $ D^\dagger(\alpha) D(\beta) = e^{\frac{1}{2}\left(\alpha^* \beta - \alpha \beta^*  \right)}D(\beta - \alpha) \approx D(\beta - \alpha)$ and $D^\dagger(\alpha)=D(-\alpha)$. For the purpose of this study, the global phase that would accumulate through the blocks can be safely discarded.

The SD protocol defines a parameterized unitary operator $ U(\vec{\alpha}, \Theta) $, where:
\begin{itemize}
    \item $ \vec{\alpha} = (\alpha_1, \alpha_2, \dots, \alpha_B) $ is the vector of complex displacement amplitudes for each of the $ B $ blocks
    \item $ \Theta \in \mathbb{R}^{B \times d} $ is the matrix of SNAP gate parameters, where each row $ \vec{\theta}_b $ contains the phase angle applied to the individual $ d $ Fock states in block $ b $.
\end{itemize}
Given a fixed \emph{initial state} $ \ket{\psi_{\text{init}}} $, the prepared state is thus obtained as $\ket{\psi_{\text{prep}}} = U(\vec{\alpha}, \Theta) \ket{\psi_{\text{init}}}$. 
The number of blocks $B$ that are required is not known a priori and needs to be empirically determined based on the target fidelity. 

\subsection{Parameter Optimization}
Previous research has explored optimization methods for the gate parameters $\vec{\alpha}, \Theta$ targeting very high fidelity. \citet{fosel2020efficient} proposed a hierarchical initialization combined with gradient-based optimization of all displacement amplitudes and all SNAP phase angles in each block. For unitaries on a $d=10$ qudit, they find that the required number of blocks $B$ scales linearly with $d$, exhibiting a better scaling compared to earlier methods that required $\mathcal{O}(d^2)$. For state preparation tasks such as binomial (kitten) codes, they achieve fidelities $\mathcal{F} \geq 0.9999$ with only $B = 3$--$4$ blocks. However, each SNAP gate is parameterized to apply an independent phase to all $d$ Fock states, so the total number of phases still scales as $\mathcal{O}(dB) = \mathcal{O}(d^2)$. This method allows control of the trade-off between fidelity and number of blocks when $d \le 10$, but leaves open how to extend such dense SD parameterizations to larger qudits (\eg $d$ up to $64$) without incurring prohibitive overhead, in particular due to the large number of SNAP phases.

A complementary line of work focuses on efficient constructive compilation of arbitrary unitaries into SD sequences. \citet{PhysRevA.92.040303} established universal control of an oscillator dispersively coupled to a qubit and gave an explicit scheme to decompose any target unitary on a $d$-dimensional Fock subspace into a sequence of SNAP gates interleaved with $\mathcal{O}(d^2)$ Givens rotations on adjacent Fock levels, with an angle equal to that of the complex number in the corresponding component of the unitary. They also propose to prepare a unitary equivalent to a Givens rotation with the SD ansatz $D(\alpha)S_k(\pi)D(-2\alpha)S_k(\pi)D(\alpha)$, where $S_k(\pi) \equiv S(\theta_n = \pi, \mathbf{1}_{n \le k})$, and $\alpha$ is identified via gradient descent to maximize the fidelity between the SD ansatz and the target Givens rotation matrix. Building on this construction, \citet{job2023efficient} derived an analytic relation between the Givens rotation angle on levels $(k,k{+}1)$ and the displacement amplitude, enabling direct compilation of any $\mathrm{SU}(d)$ operation with the SD protocol without numerical optimization. The approximation error can be controlled by slicing large Givens rotations into multiple smaller ones, at the cost of a substantially higher number of blocks and therefore a longer ansatz. These results represent a major advance, but the resulting ansatz still contains $\mathcal{O}(d^2)$ Givens rotations, each of them implemented to high fidelity possibly requiring dozens of gates, and correspondingly very large numbers of SNAP phases, which remain challenging to implement in practice. For these reasons we focus on the short- to medium-term regime, where the physically attainable fidelity is constrained by the infidelity of the qubits and by control noise, as such obtaining an SD ansatz able to prepare a target state with a controllable target fidelity at a reduced resource cost (\idest number of blocks, number of phases) is more useful.

\section{Sparse Phase Ansatzes}
\label{sec:sparse_ansatzes}
\begin{figure}[ht!]
    \centering
    \caption{Absolute values of the learned SNAP phase angles and of the Fock state population after each displacement gate, for each block of a \texttt{Full} ansatz, preparing a Haar random state with $d=16$.     }
    \includegraphics[width=0.95\linewidth]{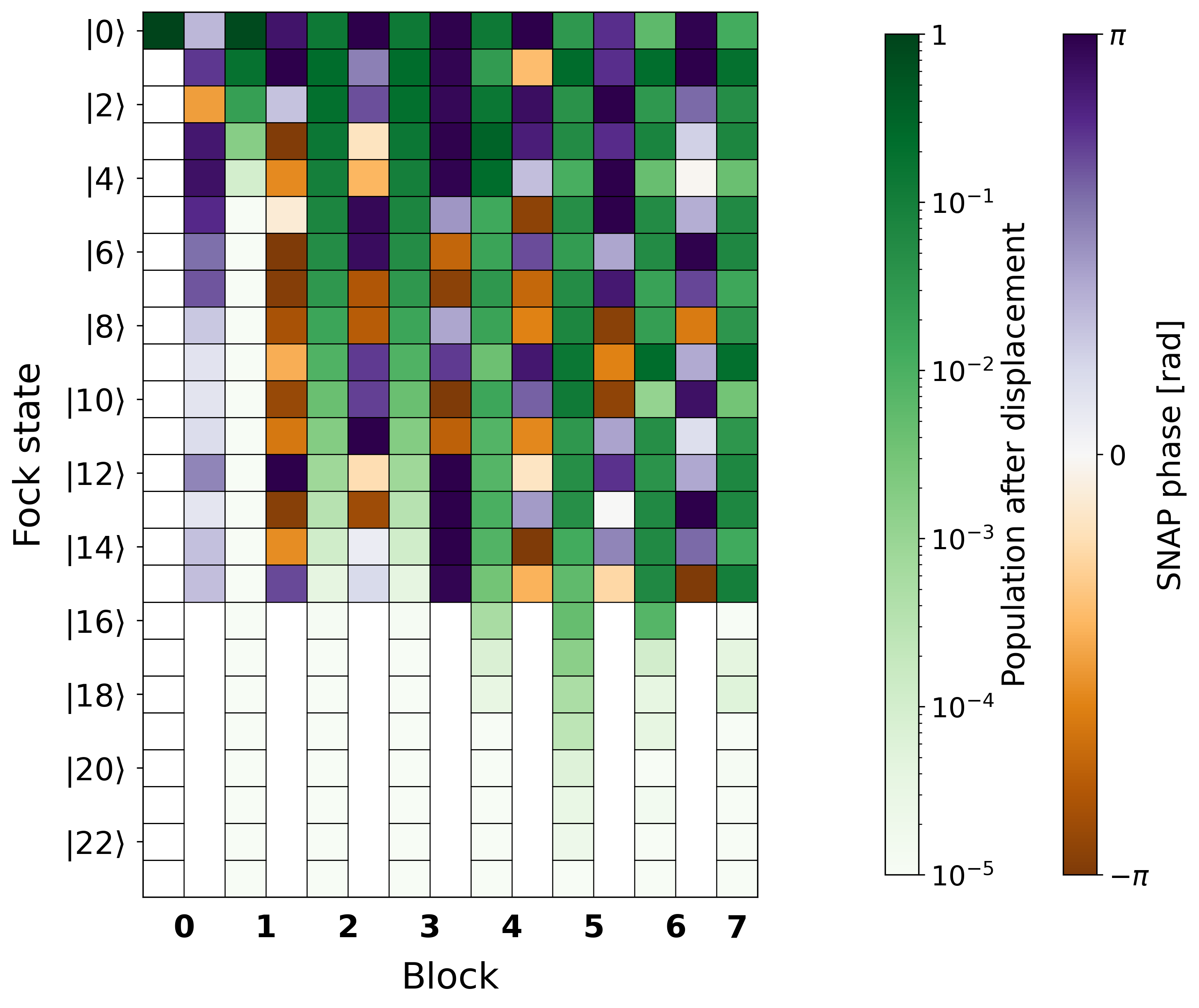}
    \label{fig:populated_levels}
\end{figure}

The SD protocol, in its general form, requires training all gate parameters $\vec{\alpha}, \Theta$. The number of parameters is dominated by the phase angles $\Theta$, though it can be argued that not all of them are necessary. Successive displacement gates tend to spread the population gradually across neighboring Fock states as the block index $b$ increases. As a result, particularly in the early blocks of the ansatz starting from the initial vacuum state, many phases are applied to Fock states that have negligible population, since the displacement gates have not yet populated them, see Figure \ref{fig:populated_levels}. Given that populating higher Fock levels already requires several displacement gates, a parameter-counting argument suggests that one does not need a fully parameterized SNAP gate with an independent phase on every Fock level at each block, since this would introduce far more gate parameters than the target state has degrees of freedom. Indeed, Figure \ref{fig:populated_levels} shows how, as the block index grows, repeated phases are applied to the same Fock states, creating redundancy. Finally, as illustrated in Appendix~\ref{app:single_parameter_ansatzes}, several useful Fock and equal-superposition states can be prepared to high fidelity using sequences with single-parameter SNAP gates that apply a fixed phase ($\pi$) to a certain Fock level~\cite{Roy:2024BW} at each iteration. 

Motivated by these observations, we introduce \emph{sparse phase ansatzes} with the goal of restricting the optimization to a subset of phases that align with the empirically observed population distribution induced by the displacement gates. The objective is to reduce the number of phase parameters in regions where they are unnecessary, while retaining them where they are required. We focus on the phase parameters, as they dominate the parameter count, while leaving the displacement parameters unaltered.

\begin{figure*}[ht!]
    \centering
    \caption{Location of the learnable SNAP phases on matrix $\Theta$ ($B\times d$), arranged on the block-Fock grid $(b,k)$, for a qudit of $d=8$ and $B=8$ blocks. Learnable phases are highlighted in black.}
    \begin{subfigure}[c]{0.32\linewidth}
        \centering
        \includegraphics[width=0.9\linewidth]{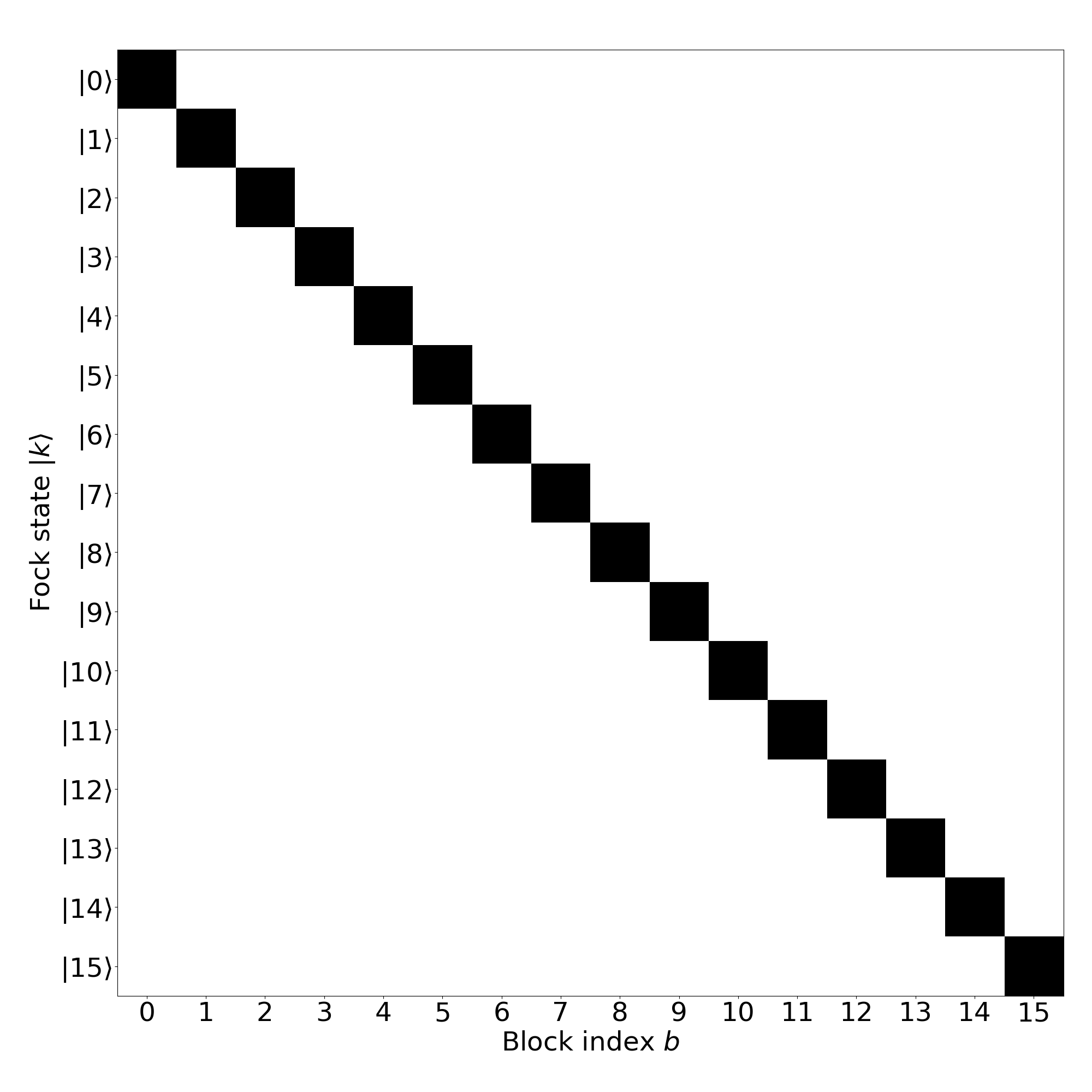}
        \caption{\texttt{Diagonal (adaptive)}}
        \label{fig:phase_locations/SAdaptiveDiagonal}
    \end{subfigure}\hfill
    \begin{subfigure}[c]{0.32\linewidth}
        \centering
        \includegraphics[width=0.9\linewidth]{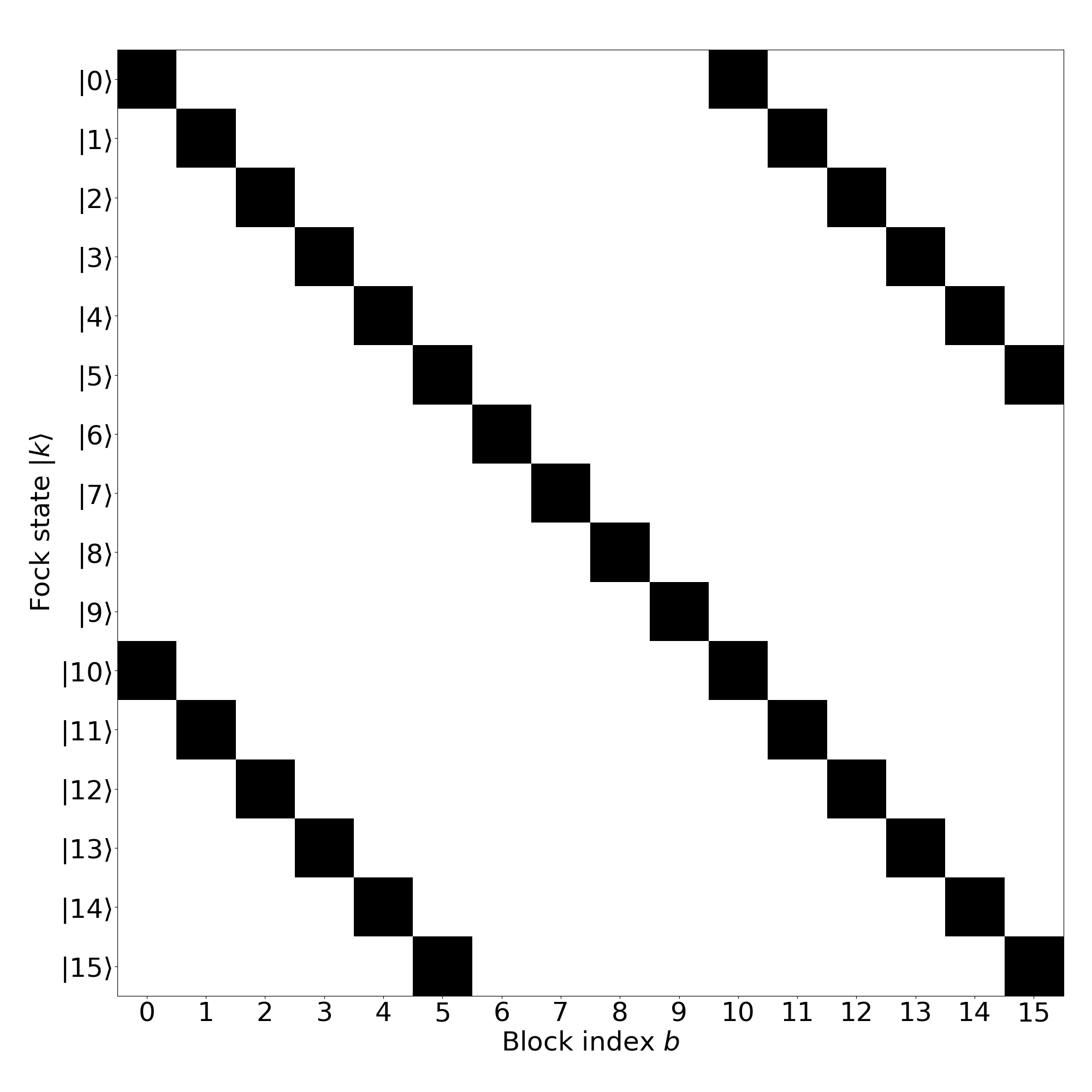}
        \caption{\texttt{Diagonal (multiple)}}
        \label{fig:phase_locations/SAdaptiveMultiDiagonal}
    \end{subfigure}\hfill
    \begin{subfigure}[c]{0.32\linewidth}
        \centering
        \includegraphics[width=0.9\linewidth]{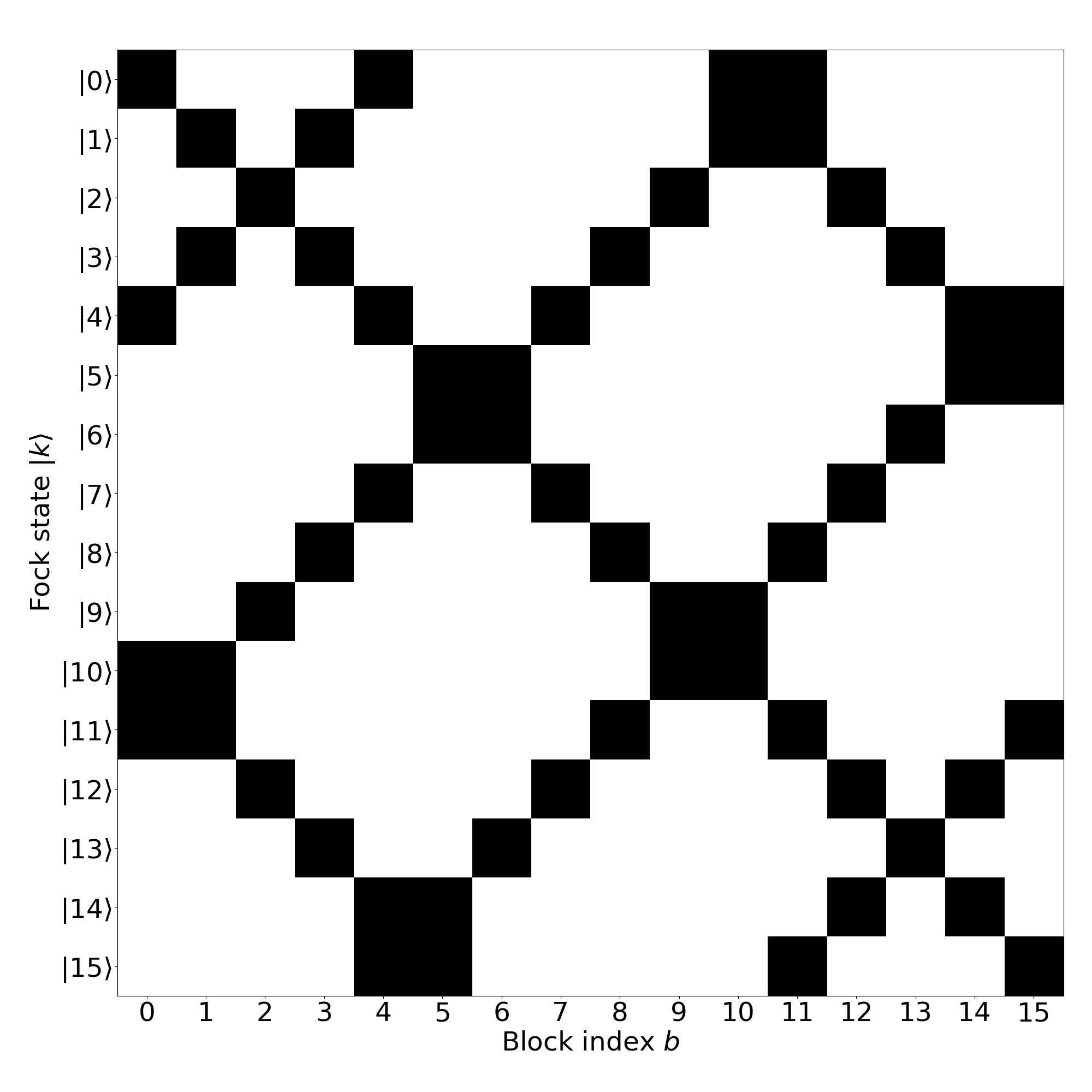}
        \caption{\texttt{Grid}}
        \label{fig:phase_locations/SAdaptiveGrid}
    \end{subfigure}
    \label{fig:phase_locations}
\end{figure*}

We define the \emph{sparse phase ansatzes} by treating the phase parameters as a $B\times d$ matrix $\Theta$ arranged on the block-Fock grid $(b,k)$, where only some of them are \emph{learnable}, while the others are set to zero. Only non-zero phase angles will need to be physically applied to the qudit, so the lower this number is, the smaller the impact of control operations and consequent errors will be. We define these sparse ansatzes as a family of progressively more general ones, based on a set of hyperparameters controlling their structure, so as to give as much flexibility as possible. Note that, while the values of the phase angles are learned during optimization, the locations of the learnable ones are fixed \emph{a priori}. Below, we provide a high-level overview of their construction, with the full description available in Appendix \ref{app:sparse_ansatzes}.

\begin{itemize}
    \item \texttt{Full}: The traditional fully parameterized SD ansatz, where each SNAP gate at block $b$ applies an independent phase to all $d$ Fock states (\idest the full phase matrix $\Theta$ is optimized). This structure maximizes flexibility but requires applying $Bd$ phases.

    \item \texttt{Diagonal (adaptive)}: A simple structure where the phases to optimize lie on the diagonal of the phase matrix, running from $(0,0)$ to $(B,d)$, see Figure \ref{fig:phase_locations/SAdaptiveDiagonal}. More precisely, the ansatz requires defining a diagonal $k = \big\lfloor \frac{b,d}{B} \big\rfloor$ and a tunable hyperparameter \emph{width} $w$, such that all phases within a perpendicular distance $\leq \frac{w}{2}$ from the diagonal (see Appendix \ref{app:sparse_ansatzes}) are optimized, defining a band. This simple ansatz can adapt to an arbitrary number of blocks, and the width hyperparameter $w$ allows control over how far the learnable phases should spread, thus determining the number of learnable phases. The number of phases has an upper bound of $wB$, though the actual number will be slightly lower due to clipping at the grid boundaries.

    \item \texttt{Diagonal (multiple)}: A generalization of the \texttt{Diagonal (adaptive)} ansatz, where the learnable phases lie on multiple parallel bands, see Figure \ref{fig:phase_locations/SAdaptiveMultiDiagonal}. The bands are spaced such that they intersect the anti-diagonal at evenly spaced points. This ansatz has two hyperparameters, the \emph{width} $w$ of the bands, and their number $m$. The number of phases has an upper bound of $mwB$, though the actual number will be much lower, as bands farther from the diagonal become shorter.

    \item \texttt{Grid}: A generalization of the \texttt{Diagonal (multiple)} ansatz, where the learnable phases lie on parallel bands along both the diagonal and anti-diagonal, forming a sparse grid structure over $\Theta$, see Figure \ref{fig:phase_locations/SAdaptiveGrid}. This ansatz has three hyperparameters, the \emph{width} of the bands, the number of diagonals $m_{\text{diag}}$, and of anti-diagonals $m_{\text{anti}}$. The number of phases has an upper bound of $w(m_{\text{diag}} + m_{\text{anti}})B$. However, the actual number will be much lower due to the shorter length of bands farther from the diagonal and overlaps between bands.
\end{itemize}

Figure \ref{fig:phase_locations} illustrates each structure by highlighting the locations of learnable phases on the $\Theta$ grid. All phase angles that are not learnable are set to zero.

As we previously mentioned, each ansatz has a set of tunable hyperparameters that control its structure, such as the number of blocks $B$, the width of each diagonal band $w$, the number of diagonals in the \texttt{Diagonal (multiple)} ansatz, and the number of anti-diagonals in the \texttt{Grid} ansatz. Because of this, an exhaustive search of all possible values is impractical, and it is necessary to identify an efficient procedure for optimizing them accounting for the two conflicting goals of optimizing fidelity while also minimizing the number of phases required.

\section{Methodology}
The goal of this study is to optimize the gate parameters of the SD protocol to maximize the fidelity $\mathcal{F}$ between a desired \emph{target state} $\ket{\psi_{\text{target}}} \in \mathbb{C}^{d} $ and the \emph{prepared state} $\ket{\psi_{\text{prep}}} \in \mathbb{C}^{d} $, which is obtained by applying the unitary $U(\vec{\alpha}, \Theta)$ implemented by the SD protocol to $\ket{0}$. All states reside in a truncated Hilbert space of dimension $d$ which represents the controllable Fock levels of the cavity. In addition to maximizing fidelity, we also aim to minimize the number of non-zero SNAP phase angles. To achieve this, we propose a two-step process: an inner \textbf{gate parameter} optimization step, and an outer \textbf{ansatz hyperparameter} optimization step. The ansatz hyperparameter optimization step will optimize both the hyperparameters that control the structure of the ansatz and those that affect the gradient ascent learning process. Meanwhile, the gate parameter optimization step will focus on finding the optimal gate parameters based on a set of hyperparameters.

The quality of the state preparation is measured by the infidelity $\mathcal{I}$:
\begin{equation}
\mathcal{I} = 1 - \mathcal{F} \in [0,1], \quad \mathcal{F} = |\braket{ \psi_{\text{target}}}{\psi_{\text{prep}}}|^2,
\end{equation}
where $ \mathcal{F} \in [0,1] $ represents the \emph{fidelity}, with $ \mathcal{F} = 1 $ and the corresponding $ \mathcal{I} = 0 $ indicating perfect overlap between the prepared and target states.

While in bosonic quantum computing the cavity possesses an infinitely large Hilbert space, only a finite number of Fock states, denoted by $d$, can be controlled. However, the application of displacement gates can cause population leakage beyond these $d$ levels. In order to account for this effect, all simulations are performed with an extended Hilbert space of dimension $\tilde{d} = d + \text{max}(64, 2d)$, where the additional levels are referred to as \emph{bumper states} or \emph{buffer states}.
The target state $ \ket{\psi_{\text{target}}} $ is embedded in this extended space by assigning zero amplitude to all bumper states:
\begin{equation}
\ket{\tilde{\psi}_{\text{target}}} = \begin{bmatrix} \psi_{\text{target}} \\ \mathbf{0}_{\text{max}(64, 2d)} \end{bmatrix},
\end{equation}
where $ \ket{\psi_{\text{target}}}$ is the original target state. Ensuring that the number of bumper states is at least $2d$ reflects empirical observations that fewer bumper states lead to underestimating leakage, which consequently results in an overestimation of the fidelity of the prepared states. This effect can become relevant if one is targeting very high fidelity where even small population leakage can alter the result to a significant extent.

\subsection{Gate Parameter Optimization}
\label{sec:gate_params}
The optimization of the gate parameters, given the set of learnable SNAP and displacement gate parameters, aims to optimize the fidelity of the prepared state. The loss function optimized is a combination of fidelity and both $\ell_1$ and $\ell_2$ regularization, which serve to penalize large parameter values (particularly to avoid large displacements) and encourage the selection of as few non-zero phase angles as possible. The loss function is defined as:
\begin{align}
\label{eq:gate_param_loss}
(\vec{\alpha}^*, \Theta^*)
= \arg\max_{\vec{\alpha}, \Theta}\;& \mathcal{F}(\vec{\alpha}, \Theta)
 - \lambda_1 \big(\lVert \vec{\alpha} \rVert_1 + \lVert \Theta \rVert_1 \big) \nonumber \\ 
& - \lambda_2 \big(\lVert \vec{\alpha} \rVert_2^2 + \lVert \Theta \rVert_2^2 \big),
\end{align}
where $\lambda_1$ and $\lambda_2$ are the regularization weights, which are hyperparameters, and $\vec{\alpha}^*, \Theta^*$ are the optimal learned gate parameters. Note that the regularization terms often counteract the optimization of the fidelity by introducing restrictions on the parameter values, so the corresponding weights should be kept small.
An important aspect to consider is how to represent the gate parameters so that they belong to the appropriate domain: displacement gates are parameterized by complex amplitudes $\alpha_b \in \mathbb{C}$, while SNAP gates require phase angles bounded in the range $[-\pi, \pi]$. 
To achieve this, we represent the gate parameters as functions of a second set of real-valued parameters, which are the actual learnable parameters.
The complex-valued displacements $\alpha$ are represented using two real-valued parameters in a polar-phase encoding. Note that any complex displacement can be rewritten as a real one surrounded by two SNAPs, \(D(r e^{i\phi}) = S(\{\theta_k=k\phi\})\, D(r)\, S(\{\theta_k=-k\phi\})\) with \(r\in\mathbb{R}\) \cite{PhysRev.131.2766}. In this study we keep the displacement parameter complex because our goal is to minimize the number of phases required in \(\Theta\), and restricting displacements to real parameters may push the protocol to compensate by requiring an increased number of phases, which is something we aim to avoid.
For block $b$, we have: 
\begin{align}
    \alpha_b &= \beta_b e^{i\varphi_b}, \quad
    \beta_b = \operatorname{ReLU}(r_b), \quad
    \varphi_b = \pi \tanh(s_b),
\end{align}
with the learnable unconstrained parameters $r_b, s_b \in \mathbb{R}$, and ReLU representing the Rectified Linear Unit function. This mapping ensures the radius $\beta_b \geq 0$ and the phase angle $\varphi_b \in [-\pi, \pi]$. 
Similarly, for each \emph{learnable} SNAP phase, we map a learnable real-valued parameter $t_{k,b}$ to a bounded phase angle via:
\begin{align}
    \theta_{k,b} = \pi \tanh(t_{k,b}).
\end{align}
Non-learnable phase angles are inactive and fixed to zero.
Note that there are alternative ways to ensure the gate parameters are correctly bounded. For example, a real-valued parameter can be wrapped modulo $2\pi$ and shifted to $[-\pi, \pi]$ to become a valid phase angle. Introducing a discontinuity at $\pm\pi$ in the domain of the learnable parameter could however make the training somewhat unstable for values near the boundary. In contrast, the $\tanh$ mapping we adopt is smooth but asymptotically saturates at the bounds, leading to vanishing gradients near $\pm\pi$. Therefore, there is a trade-off and, if the goal is to achieve the maximum fidelity, it would be beneficial to explore different mappings to identify the one that works best for the scenario at hand.

Gate parameter optimization is performed via gradient ascent using the Adam optimizer, with a maximum of $10^4$ steps. Early stopping interrupts the training if the optimal fidelity does not improve for $10^3$ consecutive steps. The minimum improvement in fidelity considered meaningful is $10^{-6}$. To account for the variance introduced by different random initializations of the gate parameters, which is sometimes high, we run 100 independent optimizations, which we call \emph{repetitions}, each starting from a different random initialization. Note that for the largest qudit $d=64$, due to the large computational cost of training 100 repetitions, which would be approximately 15 hours, depending on the number of blocks $B$, we limit the repetitions to 10. 
Because a small fraction of repetitions occasionally fails to converge within the allotted number of steps or becomes trapped in a poor local optimum, we report the mean and standard deviation computed over the top $75\%$ of repetitions ranked by achieved fidelity. 
While in practice one would simply select the gate parameters of the single repetition that yields the highest fidelity, here we aim to quantify the standard deviation across random initializations, without being overly affected by a limited number of poor outcomes that would be easy to discard. Even with this selection, the distribution of the infidelities across repetitions can easily span more than an order of magnitude, hence the standard deviation can be comparable to or larger than the mean. The results remain consistent when averaging over all repetitions, although the mean infidelity increases, and the single best repetition can be significantly better than the mean. 
Lastly, since the smallest number that can be exactly represented in single precision is around $10^{-7}$, which is too close to the infidelities we aim to study, all simulations are conducted in double precision. The simulations are implemented with PyTorch and run on NVIDIA A100 Tensor Cores. Further technical details related to the efficient use of GPU parallelism are reported in Appendix \ref{app:efficient_parallel_simulation}.

\subsection{Ansatz Hyperparameter Optimization}
\label{sec:ansatz_hyperparams}
The gate parameter optimization process depends on two key sets of hyperparameters: the first set controls the gate parameter learning process (learning rate, regularization weights $\lambda_1$ and $\lambda_2$), while the second set controls the structure of the ansatz (number of blocks $B$, diagonal width $w$, number of diagonals $m_{\text{diag}}$, number of anti-diagonals $m_{\text{anti}}$) and varies depending on the type of ansatz. These hyperparameters, especially those related to the structure of the ansatz, affect both the fidelity of the prepared state and the resource requirements, such as the number of blocks and phases. Given our goal of maintaining high fidelity with limited resources, we need to combine both objectives, which we do by formulating a multi-objective optimization problem where we aim to maximize fidelity while minimizing the number of SNAP phases. We combine both objectives into a single scalarized loss that balances the two, which is a popular strategy \cite{BazganRuzikaThielenVanderpooten2022}, allowing us to compute a single scalar value to evaluate a hyperparameter configuration $h$, defined as:
\begin{equation}
\label{eq:scalarized}
\mathcal{L}_{\text{h}}(\vec{\alpha}^*_h, \Theta^*_h, \beta) = \mathcal{F}(\vec{\alpha}^*_h, \Theta^*_h) - \beta \frac{p(\Theta^*_h)}{2d - 2},
\end{equation}
where $\vec{\alpha}^*_h, \Theta^*_h$ denote the optimal gate parameters learned according to the process described in Section \ref{sec:gate_params}, under the hyperparameter configuration $h$, $p(\Theta^*)$ is the number of non-zero phase angles, and $\beta \geq 0$ is the phase penalty term used to control the fidelity-number of phases trade-off. The normalization factor $2d - 2$ corresponds to the maximum number of degrees of freedom a pure quantum state of dimension $d$ can have and mainly allows for easier comparability of the values for $\beta$ when changing the number of qudit levels $d$, as well as improved interpretability in its selection. In simple terms, $\beta$ controls how much fidelity must increase to justify an increase in the number of phases. For example, if $\beta = 0.1$ and a candidate hyperparameter configuration $h$ leads to a $10\%$ increase in the number of phases, the minimum fidelity gain required for this increase to be acceptable would be $10^{-2}$. Note that the coefficient $\beta$ cannot be optimized automatically like the hyperparameters, as it is an integral part of our definition of the experimental conditions. Therefore, it is usually necessary to run multiple hyperparameter optimization rounds, each using a different value for $\beta$.

As a second study, we also explore a different penalty that is based on the duration of the ansatz and evaluate a hyperparameter configuration $h$ with: 
\begin{equation}
\label{eq:scalarized_time}
\mathcal{L}_{\text{h}}(\vec{\alpha}^*_h, \Theta^*_h, \beta) = \mathcal{F}(\vec{\alpha}^*_h, \Theta^*_h) - \beta \frac{t(B, \Theta^*_h)}{T_{\text{worst}}},
\end{equation}
Here we use $T_{\text{worst}}$ as a normalization factor, corresponding to the duration of an ansatz with the longest duration within a reasonable number of blocks, that is, one where $B = d - 1$ and each SNAP gate optimizes only 2 phases, for a total of $2d - 2$, corresponding again to the maximum number of degrees of freedom a pure quantum state of dimension $d$ can have. For current implementations, the SNAP duration scales approximately as
$T_{S_b} \propto \sqrt{p(\theta_b)}$, with a recent study reporting $T_{S_b} \approx 10\sqrt{p(\theta_b)} \ \mu\text{s}$~\cite{bornman2025benchmarking}, which is the value we will use, while the duration of a displacement gate depends on its parameter $\alpha$ but is very small and can be safely approximated as $0.1 \mu \text{s}$. The total duration of the ansatz, $t(\vec{\alpha}^*_h, \Theta^*_h)$, can be computed as the sum of the durations of all its gates, based on these quantities.  Hence, $T_{\text{worst}} \approx 10\sqrt{2} (d - 1) \mu\text{s}$. This normalization again serves to make it easier to choose and interpret the penalty coefficient $\beta$ for different $d$.

Note that the two strategies, while optimizing different goals, are strongly related. Optimizing the duration of the ansatz will also indirectly minimize the number of learnable phases because this reduces the duration of the SNAP gates, however, the actual distribution of the phases among the gates will also play a role, something that penalizing the number of phases alone will not take into account. As a result, we can expect, and indeed we will see, that optimizing the duration of the ansatz will also minimize the number of phases, while minimizing the number of phases will not consistently reduce the duration of the ansatz.

In our numerical simulations we perform hyperparameter optimization using the Tree-structured Parzen Estimator (TPE) algorithm \cite{Bergstra2011TPE}, implemented in the \texttt{optuna} library. We explore a maximum of 50 configurations and choose $\beta \in [0.0, 10^{-6}, 10^{-5}, 10^{-4}, 10^{-3}, 10^{-2}, 10^{-1}]$. The hyperparameter ranges are reported in Appendix \ref{app:ansatz_hyperparams}.
TPE is a sequential Bayesian optimizer that learns from past trials by constructing two smooth probability models over the hyperparameter space: one for the best results so far and one for the remaining results. It then samples new hyperparameter configurations $h$ from regions that are expected to yield better results, \idest those corresponding to high expected improvement. TPE naturally handles conditional (tree-structured) choices as well as mixed discrete and continuous variables, making it particularly well-suited for our scenario. It should be noted that exploring a fixed number of hyperparameter configurations does not affect all ansatzes equally. Sparse ansatzes with more hyperparameters, especially \texttt{Grid}, define a larger search space than \texttt{Full} and may be harder to optimize within the same number of trials. The protocol we adopt ensures a comparable computational effort across ansatzes, and therefore the gains reported for sparse ansatzes may be conservative.

\paragraph{Hypervolume indicator.}
\label{sec:ansatz_hyperparams_hypervolume}
In order to summarize the trade-off between the resources required by the ansatz and the infidelity of the prepared state, we report the
\emph{hypervolume} (HV) indicator, a widely used scalar measure for comparing sets of
solutions in multi-objective optimization \cite{DBLP:journals/csur/GuerreiroFP21,DBLP:journals/tcs/AugerBBZ12}.
In our setting each hyperparameter optimization trial of a given ansatz is associated with a point $(x,y)$, where $x$ is a resource cost metric (either the number of non-zero phase angles or the ansatz duration, as previously defined) and $y$ is the  mean of the prepared state infidelity computed over the top 75\% of repetitions, ranked by fidelity. Both objectives should be minimized. 
Given an ansatz, we collect all points explored by the hyperparameter optimization trials for all cost penalty coefficients $\beta$ in a set $\mathcal{S}=\{(x_i,y_i)\}$. Intuitively, given a reference point $r$ that should represent the \emph{worst} trade-off, HV measures the area defined between the points in set $\mathcal{S}$ and $r$. When comparing two ansatzes, the one with a larger HV corresponds to a set of points that are further away from $r$, and therefore exhibits a better trade-off. More formally, HV is
defined as the area of the region in the $(x,y)$ plane that is \emph{dominated} by
$\mathcal{S}$ and bounded by the reference point $r=(x_{\mathrm{ref}},y_{\mathrm{ref}})$, chosen so that its individual components are worse than the corresponding component of any point in the set, hence $r$ is outside $\mathcal{S}$. All points are then normalized so that $r=(1,1)$. Note that we compute the HV on the $log_{10}$ of both infidelity and cost metric, and for readability we report $100 \times HV$ in the tables. 
When the goal is to minimize, a point $(x_i,y_i)$ dominates $(x_j,y_j)$ if $x_i\le x_j$ and
$y_i\le y_j$, with at least one strict inequality. These non-dominated points are also called \emph{Pareto-optimal} \cite{Miettinen1999} and will be discussed in the results section. Then, each point $(x_i,y_i)$ defines a rectangle $[x_i,x_{\mathrm{ref}}]\times[y_i,y_{\mathrm{ref}}]$, and the HV is the area of the union of these rectangles. To quantify the uncertainty of the hypervolume, we compute the reference point using all repetitions and then use bootstrap resampling over the independent repetitions. For each hyperparameter configuration, we resample the repetitions with replacement, reapply the same aggregation rule used throughout the paper by computing the mean of the $75\%$ repetitions with the best fidelity, and then recompute the HV. We repeat this procedure 1000 times to obtain a distribution of HV values, from which we report the mean and a $95\%$ confidence interval.

\section{Results and Discussion }
The goal of this study is to explore how introducing sparsity within the phases of the SD protocol can offer new ways to better control the trade-off between fidelity and resource requirements, compared to simply choosing the number of blocks $B$ in the \texttt{Full} protocol.
To assess this, we select eight quantum states grouped into two types with different degrees of structure: complex-valued pure states obtained from a Haar-random unitary \cite{mezzadri2006generate}, and superposition states referred to as \emph{Fourier-5}, obtained by applying the Quantum Fourier Transform to the Fock state $\ket{5}$. Both types of states are defined for four qudit sizes $d \in \{8,16,32,64\}$. Note that previous studies such as \cite{fosel2020efficient} considered qudit sizes only up to $d=10$.
This section summarizes the key results, comprising more than twenty-two thousand numerical simulations (two scalarized losses, eight quantum states, seven phase penalty values, four ansatzes, and fifty hyperparameter-tuning trials). Accounting for repetitions, this corresponds to training more than one million and seven hundred thousand individual sets of gate parameters.

\begin{table*}[!ht]
    \centering
    \footnotesize
    \caption{Results of the best hyperparameter configuration $h$ obtained when  \textbf{penalizing the duration of the ansatz}, for each penalty coefficient $\beta$. The table reports the infidelity, number of non-zero phase angles and duration for the different ansatzes we compare. The infidelity reports the mean and standard deviation computed over the top 75\% of repetitions, ranked by fidelity.}
    \begin{subfigure}[c]{\linewidth}
    \centering
    \begin{tabular}{ll|ccc|ccc|ccc|ccc}
    \toprule
    &  & \multicolumn{3}{c}{\texttt{Full}} & \multicolumn{3}{c}{\texttt{Diagonal (adaptive)}} & \multicolumn{3}{c}{\texttt{Diagonal (multiple)}} & \multicolumn{3}{c}{\texttt{Grid}} \\
    \multicolumn{1}{c}{d} & \multicolumn{1}{c}{$\beta$} \vline & Infidelity & Phases & T [$\mu$s] & Infidelity & Phases & T [$\mu$s] & Infidelity & Phases & T [$\mu$s] & Infidelity & Phases & T [$\mu$s] \\
    \midrule
\multirow[c]{7}{*}{32} & $0.0$ & $ 3.8 \pm 2.6 \times 10^{-5} $ & 967 & 1756 & $ 6.6 \pm 5.0 \times 10^{-5} $ & 508 & 1358 & $ 1.1 \pm 1.0 \times 10^{-4} $ & 431 & 1137 & $ 4.1 \pm 3.6 \times 10^{-5} $ & 514 & 1337 \\
 & $ 10^{-6} $ & $ 1.3 \pm 0.9 \times 10^{-5} $ & 980 & 1756 & $ 3.6 \pm 4.6 \times 10^{-5} $ & 674 & 1349 & $ 4.2 \pm 8.4 \times 10^{-4} $ & 339 & 616 & $ 1.2 \pm 1.0 \times 10^{-4} $ & 560 & 1334 \\
 & $ 10^{-5} $ & $ 4.4 \pm 4.7 \times 10^{-5} $ & 662 & 1246 & $ 4.9 \pm 4.9 \times 10^{-5} $ & 673 & 1423 & $ 4.1 \pm 6.9 \times 10^{-4} $ & 344 & 620 & $ 1.6 \pm 1.6 \times 10^{-4} $ & 371 & 1072 \\
 & $ 10^{-4} $ & $ 4.1 \pm 8.1 \times 10^{-4} $ & 411 & 736 & $ 5.9 \pm 5.3 \times 10^{-5} $ & 771 & 1582 & $ 4.6 \pm 8.6 \times 10^{-4} $ & 262 & 579 & $ 4.0 \pm 3.0 \times 10^{-5} $ & 303 & 555 \\
 & $ 10^{-3} $ & $ 1.0 \pm 1.1 \times 10^{-3} $ & 243 & 453 & $ 0.8 \pm 1.1 \times 10^{-3} $ & 212 & 496 & $ 1.2 \pm 1.1 \times 10^{-3} $ & 223 & 508 & $ 8.8 \pm 9.3 \times 10^{-5} $ & 256 & 509 \\
 & $ 10^{-2} $ & $ 5.1 \pm 0.6 \times 10^{-3} $ & 95 & 170 & $ 3.8 \pm 1.1 \times 10^{-3} $ & 79 & 251 & $ 4.1 \pm 0.9 \times 10^{-3} $ & 118 & 218 & $ 4.9 \pm 0.4 \times 10^{-3} $ & 93 & 168 \\
 & $ 10^{-1} $ & $ 1.1 \pm 0.1 \times 10^{-2} $ & 61 & 113 & $ 8.4 \pm 1.1 \times 10^{-3} $ & 42 & 110 & $ 1.3 \pm 0.1 \times 10^{-2} $ & 39 & 85 & $ 6.8 \pm 0.4 \times 10^{-3} $ & 53 & 116 \\
    \bottomrule
    \end{tabular}
    \caption{Target state \emph{Fourier-5}.}
        \label{tab:hyperopt-results-32-fourier5}
    \end{subfigure}\hfill
    \begin{subfigure}[c]{\linewidth}
        \centering
    \begin{tabular}{ll|ccc|ccc|ccc|ccc}
    \toprule
    &  & \multicolumn{3}{c}{\texttt{Full}} & \multicolumn{3}{c}{\texttt{Diagonal (adaptive)}} & \multicolumn{3}{c}{\texttt{Diagonal (multiple)}} & \multicolumn{3}{c}{\texttt{Grid}} \\
    \multicolumn{1}{c}{d} & \multicolumn{1}{c}{$\beta$} \vline & Infidelity & Phases & T [$\mu$s] & Infidelity & Phases & T [$\mu$s] & Infidelity & Phases & T [$\mu$s] & Infidelity & Phases & T [$\mu$s] \\
    \midrule
\multirow[c]{7}{*}{32} & $0.0$ & $ 1.0 \pm 1.2 \times 10^{-5} $ & 757 & 1360 & $ 9.6 \pm 8.8 \times 10^{-6} $ & 674 & 1448 & $ 1.7 \pm 2.4 \times 10^{-5} $ & 289 & 544 & $ 7.6 \pm 8.2 \times 10^{-6} $ & 318 & 593 \\
 & $ 10^{-6} $ & $ 9.2 \pm 6.8 \times 10^{-6} $ & 658 & 1190 & $ 1.8 \pm 1.4 \times 10^{-5} $ & 496 & 1081 & $ 2.1 \pm 1.8 \times 10^{-5} $ & 261 & 549 & $ 7.7 \pm 4.8 \times 10^{-5} $ & 299 & 573 \\
 & $ 10^{-5} $ & $ 1.1 \pm 1.2 \times 10^{-5} $ & 472 & 850 & $ 4.4 \pm 4.0 \times 10^{-5} $ & 206 & 475 & $ 3.9 \pm 3.8 \times 10^{-5} $ & 405 & 1076 & $ 1.5 \pm 0.8 \times 10^{-5} $ & 496 & 933 \\
 & $ 10^{-4} $ & $ 2.9 \pm 2.6 \times 10^{-5} $ & 223 & 396 & $ 3.9 \pm 2.7 \times 10^{-5} $ & 196 & 522 & $ 6.4 \pm 7.4 \times 10^{-5} $ & 171 & 321 & $ 2.2 \pm 2.1 \times 10^{-5} $ & 193 & 367 \\
 & $ 10^{-3} $ & $ 1.0 \pm 1.0 \times 10^{-4} $ & 191 & 340 & $ 1.5 \pm 2.0 \times 10^{-4} $ & 136 & 330 & $ 1.5 \pm 1.2 \times 10^{-4} $ & 194 & 368 & $ 1.1 \pm 1.2 \times 10^{-4} $ & 149 & 279 \\
 & $ 10^{-2} $ & $ 1.9 \pm 1.7 \times 10^{-4} $ & 159 & 283 & $ 0.9 \pm 1.8 \times 10^{-3} $ & 111 & 235 & $ 3.5 \pm 4.1 \times 10^{-4} $ & 125 & 260 & $ 1.1 \pm 1.2 \times 10^{-3} $ & 132 & 257 \\
 & $ 10^{-1} $ & $ 1.0 \pm 1.3 \times 10^{-2} $ & 95 & 170 & $ 5.5 \pm 6.0 \times 10^{-3} $ & 75 & 193 & $ 3.4 \pm 4.1 \times 10^{-3} $ & 102 & 200 & $ 1.3 \pm 2.1 \times 10^{-3} $ & 111 & 210 \\
    \bottomrule
    \end{tabular}
    \caption{Target state \emph{Haar random}.}
    \label{tab:hyperopt-results-32-random}
    \end{subfigure}\hfill  
    \label{tab:hyperopt-results-32}
\end{table*}

\subsection{Cost Penalty Optimization}
\label{sec:results:phase_penalty}

\begin{figure*}[ht!]
    \centering        
    \caption{Infidelity and number of non-zero phase angles for the best hyperparameter configuration $h$ obtained when \textbf{penalizing the number of phases}, for each penalty coefficient $\beta$. 
    }
    \begin{subfigure}[c]{0.49\linewidth}
        \centering
        \includegraphics[width=0.9\linewidth]{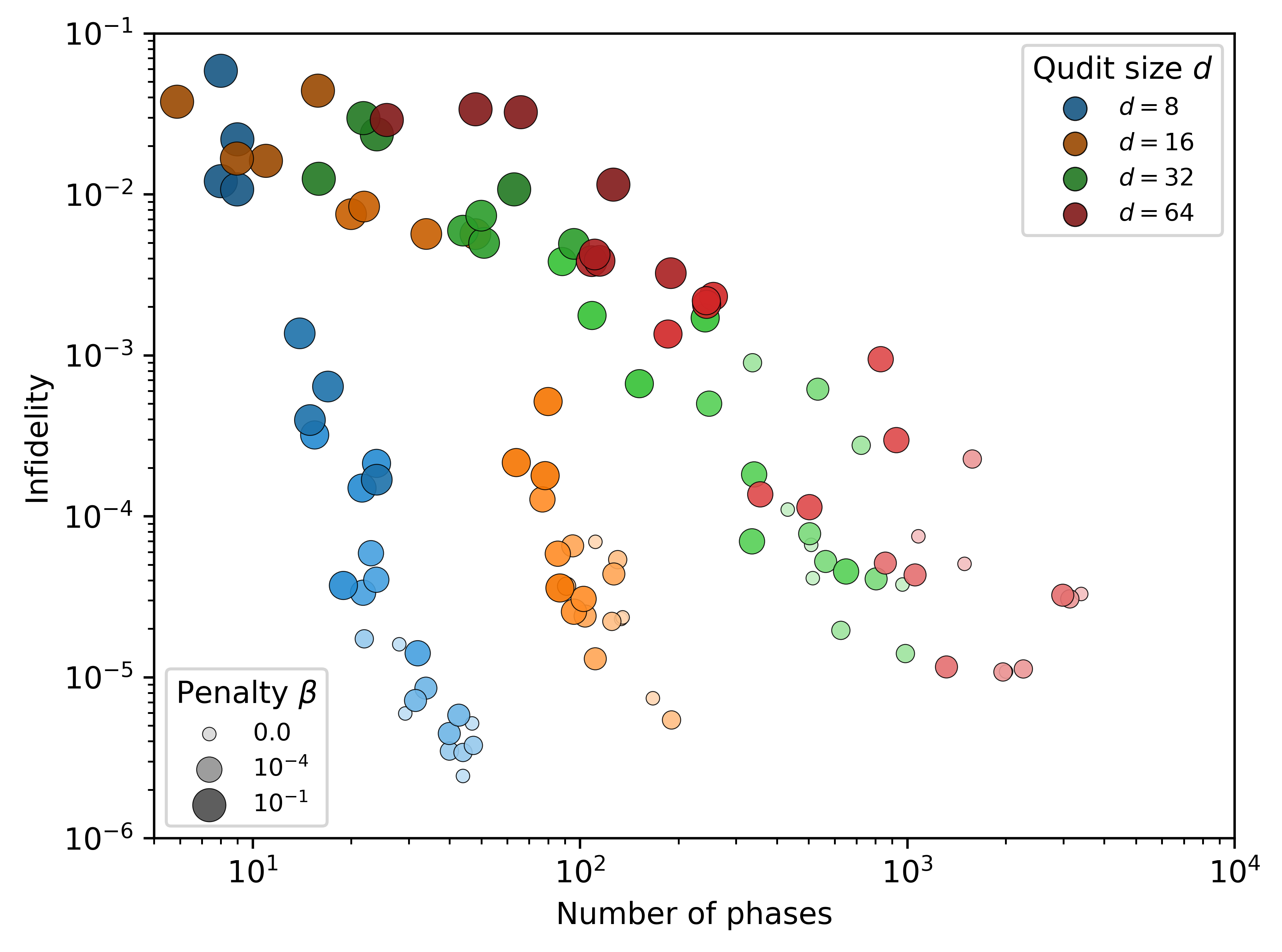}
        \caption{Target state \emph{Fourier-5}.}
        \label{fig:tables_phases/mean_75/fourier_5/initial_0/pareto_d=8_16_32_64_no_arch}
    \end{subfigure}\hfill
    \begin{subfigure}[c]{0.49\linewidth}
        \centering
        \includegraphics[width=0.9\linewidth]{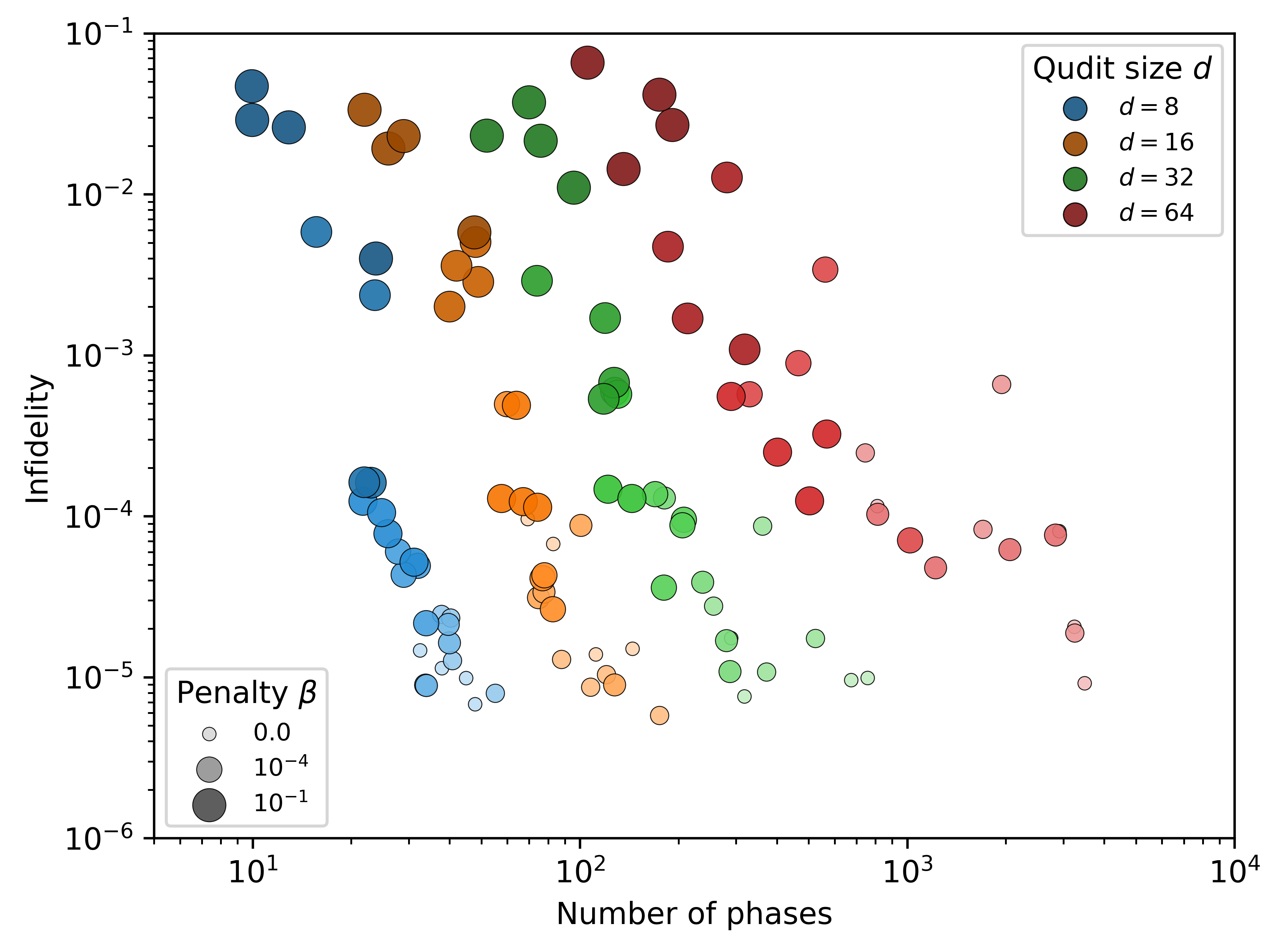}
        \caption{Target state \emph{Haar random}.}
        \label{fig:tables_phases/mean_75/random_gaussian/initial_0/pareto_d=8_16_32_64_no_arch}
    \end{subfigure}
    \label{fig:tables_phases/pareto_d=8_16_32_64_no_arch_phases}
\end{figure*}

\begin{figure*}[ht!]
    \centering
    \caption{Infidelity and duration of the ansatz for the best hyperparameter configuration $h$ obtained when \textbf{penalizing the duration of the ansatz}, for each penalty coefficient $\beta$. 
    }
    \begin{subfigure}[c]{0.49\linewidth}
        \centering
        \includegraphics[width=0.9\linewidth]{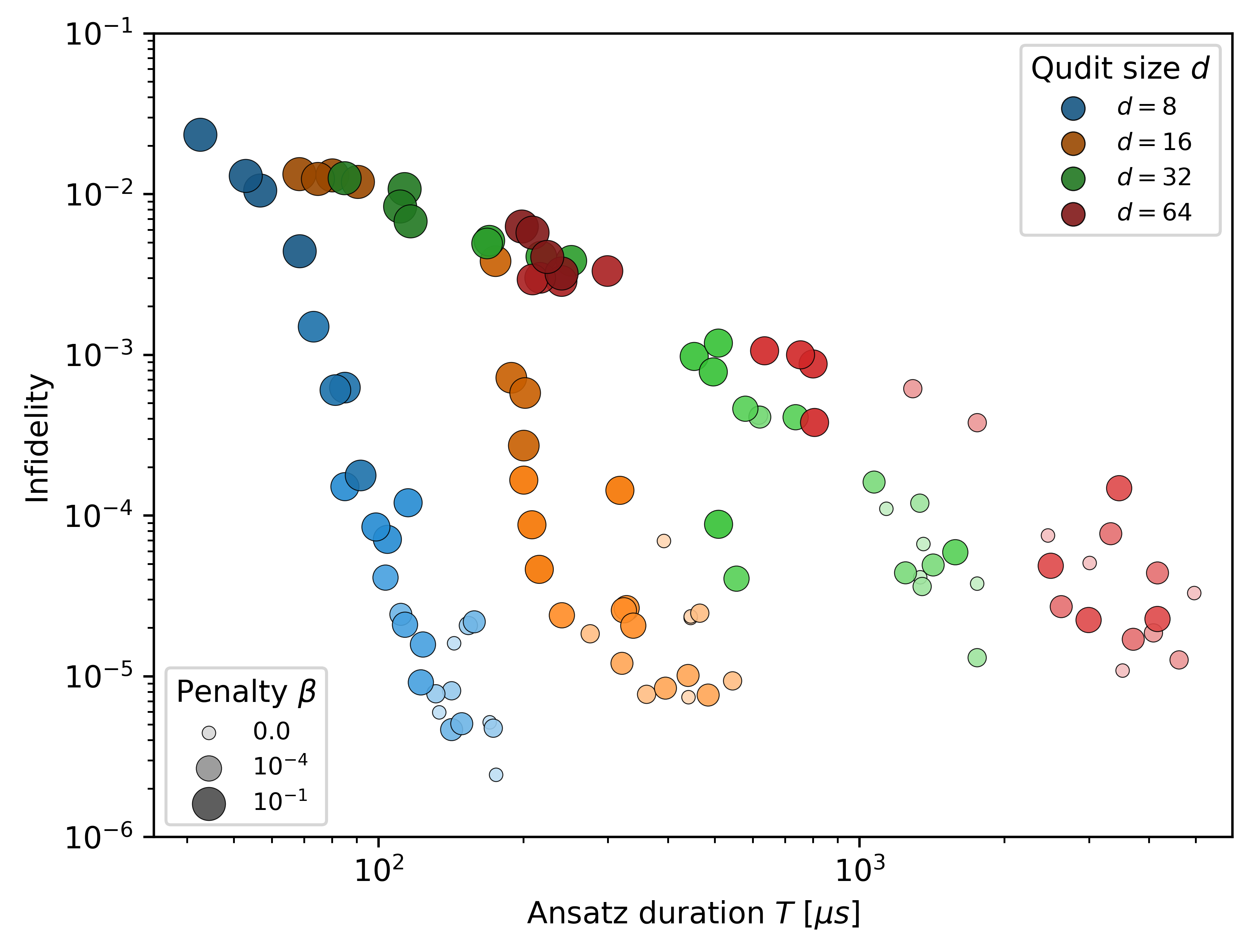}
        \caption{Target state \emph{Fourier-5}.}
        \label{fig:tables_time/mean_75/fourier_5/initial_0/pareto_d=8_16_32_64_no_arch_time}
    \end{subfigure}\hfill
    \begin{subfigure}[c]{0.49\linewidth}
        \centering
        \includegraphics[width=0.9\linewidth]{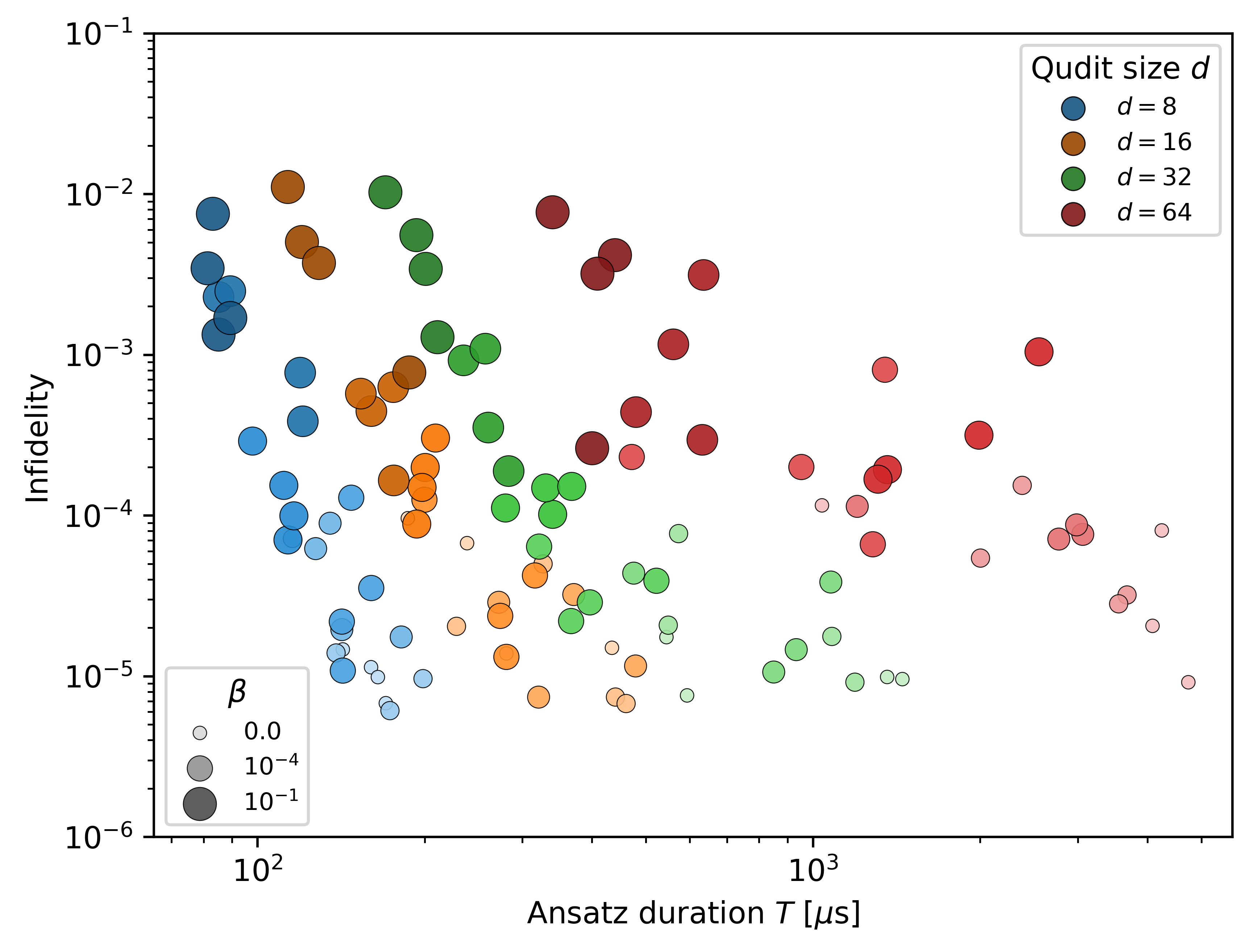}
        \caption{Target state \emph{Haar random}.}
        \label{fig:tables_time/mean_75/random_gaussian/initial_0/pareto_d=8_16_32_64_no_arch_time}
    \end{subfigure}
    \label{fig:tables_time/pareto_d=8_16_32_64_no_arch_time}
\end{figure*}

In order to control the trade-off between fidelity and resource requirements, we have introduced a scalarized loss which combines the two conflicting objectives. This section aims to compare the results of the hyperparameter optimization described in Section \ref{sec:ansatz_hyperparams} under different values for the cost penalty $\beta$ when applied to minimize the number of phases $p(\Theta)$, or the duration of the ansatz $t(B, \Theta)$. 
Note that the focus of this section is to discuss the effectiveness of the cost penalty term, while the ability of the sparse ansatzes to minimize more efficiently the number of phases or the duration of the ansatz compared to the \texttt{Full} ansatz will be discussed in the following Section \ref{sec:results:trade-off}.

While the full results of the hyperparameter optimization are reported in the Appendix, in this section we will provide a summary of their overall behavior and look at specific ones. The results of the hyperparameter optimization penalizing the number of phases are reported in Table \ref{tab:tables_phases/mean_75/fourier_5/initial_0-best_scalarized} (Fourier-5) and Table \ref{tab:tables_phases/mean_75/random_gaussian/initial_0-best_scalarized} (Haar random). The results of the hyperparameter optimization penalizing the duration of the ansatz are instead reported in Table \ref{tab:tables_time/mean_75/fourier_5/initial_0-best_scalarized} (Fourier-5) and Table \ref{tab:tables_time/mean_75/random_gaussian/initial_0-best_scalarized} (Haar random). 
First, we summarize the results of this study by reporting the infidelity and either the number of SNAP phases or the duration of the ansatz for the best hyperparameter configuration, selected by applying different penalty coefficients $\beta$. Figure \ref{fig:tables_phases/pareto_d=8_16_32_64_no_arch_phases} shows the results of the hyperparameter optimization when the scalarized loss penalizes the number of phases, while Figure \ref{fig:tables_time/pareto_d=8_16_32_64_no_arch_time} when the scalarized loss penalizes the duration of the ansatz. 
Both figures show a clear pattern with high values of the penalty coefficient $\beta$ (larger and darker dots) being associated to both a higher infidelity and a lower number of phases or ansatz duration, while as $\beta$ decreases, the results move smoothly across both dimensions improving the infidelity while increasing either the number of phases or the ansatz duration. Note that when the penalty coefficient becomes large, the jumps in infidelity may become quite large as can be expected in a highly constrained problem. The behavior differs slightly according to the target states, with Fourier-5 showing that with high penalty $\beta$ the cost (phases or ansatz duration) decreases more rapidly for large qudit sizes. In particular, it is possible to prepare at fidelity $0.99$ with 10 phases the Fourier-5 state for both $d=8$ and $d=16$. On the other hand, when $\beta$ decreases, the different qudit sizes tend to separate, with the smaller ones, as expected, requiring far fewer phases to reach infidelities below $10^{-5}$ compared to $d=32$ or $64$. A more consistent pattern can be observed for the Haar random state, where again increasing $\beta$ allows us to transition smoothly from a high fidelity-high cost region, to a low fidelity-low cost one. Here, however, the number of phases or the ansatz duration required by each qudit size remains separated, with $d=64$ always having a higher cost compared to smaller $d$. The different behavior of Fourier-5 and Haar random states can be easily explained by how it is easier to compress in fewer phases states that exhibit a strongly regular pattern. 

In the Appendix \ref{app:phase_penalty} we also report a summary of the results on one cost metric when the hyperparameter optimization penalizes the other, \idest the duration of the ansatz when the optimization penalizes the number of phases, and vice versa. Those results indicate that penalizing the duration of the ansatz enables us to smoothly control the reduction in the number of phases, but their total number remains higher than what can be obtained penalizing the number of phases directly. Furthermore, penalizing the number of phases is not an effective way to ensure the duration of the ansatz is consistently optimized, as for high values of the penalty coefficient $\beta$ the number of phases reduces but the duration of the ansatz oscillates, likely due to how the phases become spread across more blocks. The penalization can be chosen depending on which is the most significant source of noise, penalizing the number of phases is more appropriate if the dominant source of noise are control errors while penalizing the duration of the ansatz would be preferable if the cavity lifetime is the limiting factor.


For space reasons here we focus on a specific portion of the results, those optimizing the duration of the ansatz for $d=32$, see Table~\ref{tab:hyperopt-results-32}. Note that across many settings, different ansatzes can achieve similar infidelities. Our primary objective is not to obtain lower infidelities per se, but rather to improve \emph{resource efficiency}, \idest to meet a given target infidelity with fewer learnable phases or shorter ansatz duration. Accordingly, the key comparisons in the tables are the number of phases and the ansatz duration required, and we use infidelity mainly to verify that targets are met.
For the \texttt{Full} ansatz the trend is monotonic with the number of phases required to prepare the Fourier-5 state dropping from $967$ at $\beta=0$ to $61$ at $\beta=10^{-1}$, and a reduction in the duration of the ansatz from $1756 \mu s$ to $113 \mu s$, accompanied by a degradation of the infidelity from $10^{-5}$ to $10^{-2}$. This behavior is expected because at larger $\beta$ the hyperparameter search accepts higher infidelities in exchange for fewer phases or shorter ansatz duration, pushing the solution toward a highly constrained ansatz. Beyond a certain point, the resource penalty controlled by $\beta$ dominates the objective, limiting the fidelity that can be achieved. The same holds for preparing the Haar random state and for the sparse ansatzes, as well as for the other qudit sizes. Note that while the trend is largely similar for very small and very large $\beta$, there is occasional instability, mostly between small values of $\beta$, likely due to how introducing the cost penalty term creates a more difficult optimization problem and that such a small value is not affecting the optimization to a sufficiently significant extent, coupled with the stochastic nature of both the gate parameter optimization and the hyperparameter optimization process.

At \emph{no} penalty ($\beta=0$), the sparse ansatzes already achieve high fidelity with far fewer phases than the \texttt{Full} ansatz, \eg on the Haar random state most ansatzes reach an infidelity in the order of $10^{-5}$, however while the \texttt{Full} ansatz requires $757$ phases, \texttt{Diagonal (multiple)} only requires $289$ phases, and \texttt{Grid} $318$. Furthermore, the duration of \texttt{Full} is also longer, requiring $1360 \mu s$, compared to $544 \mu s$ for \texttt{Diagonal (multiple)}, and $593 \mu s$ for \texttt{Grid}.

A \emph{medium} penalty value of $\beta=10^{-4}$ already induces substantial sparsity for the simpler ansatz with little fidelity loss. For instance on the Fourier-5 states on $d=32$ \texttt{Full} reaches $4.1 \pm 8.1 \times 10^{-4}$ infidelity with $411$ phases (a $57\%$ reduction from $967$); \texttt{Diagonal (multiple)} reaches $4.6 \pm 8.6 \times 10^{-4}$ with $262$ phases (a $40\%$ reduction from $431$), and \texttt{Grid} reaches the better $4.0 \pm 3.0 \times 10^{-5}$ with $303$ phases (a $41\%$ reduction from $431$). The behavior is similar on the Haar random state, with a reduction that is particularly pronounced on \texttt{Diagonal (adaptive)} ansatz, that for $\beta=10^{-4}$ achieves an infidelity of $3.9 \pm 2.7 \times 10^{-5}$ with 196 phases, a $70\%$ reduction compared to the 674 it required when $\beta=0$. By looking at the simulations for $d=64$ we can see that there is a consistent reduction in the number of phases on all cases, for both Fourier-5 and Haar random states, in most cases by between $50\%$ and $70\%$ across all ansatzes, with a maximum of $80\%$ on \texttt{Diagonal (adaptive)} ansatz and a minimum of $14\%$ on \texttt{Diagonal (multiple)}.

With a \emph{stronger} penalty (\eg $\beta=10^{-3},10^{-2}$), all ansatzes converge to require approximately between $100-200$ phases for both the Fourier-5 and Haar random states, with infidelity in the $10^{-4}$ to $10^{-3}$ range.

Overall, the results confirm that the scalarized objective provides an effective knob to navigate the fidelity–cost trade-off, albeit with some oscillations, and that sparse phase placements can maintain competitive, or better, fidelity at a substantially reduced cost (number of phases or ansatz duration) relative to \texttt{Full}. 


\subsection{Fidelity-Cost Trade-off}
\label{sec:results:trade-off}
An important dimension to consider is the \emph{efficiency} of the proposed sparse ansatzes in terms of the resources required to prepare a quantum state given a desired target fidelity. In this respect, a good ansatz should allow us to prepare different types of states requiring either a consistently lower number of phases or having a shorter duration, compared to the others, ideally without requiring extensive fine-tuning. We discuss the results through the lenses of the \emph{Pareto frontier}, which connects the Pareto-optimal points we defined in Section \ref{sec:ansatz_hyperparams} used to compute the hypervolume. The Pareto frontier is useful to highlight, given an ansatz, the hyperparameter configuration which required the minimum number of phases (or the shortest ansatz duration) to reach the desired fidelity.
As we will now discuss, the sparse ansatzes are associated to more efficient Pareto frontiers compared to the \texttt{Full} ansatz.

\begin{table*}[ht!]
    \centering
    \caption{Hypervolume comparing the trade–off between the infidelity of the prepared state and the number of non-zero phase angles when optimizing the hyperparameters \textbf{penalizing the number of phases}. Values are reported as the bootstrap mean hypervolume with 95\% confidence intervals. Larger values indicate a better trade-off. \textbf{Bold} values indicate better mean HV than \texttt{Full}, while the maximum mean HV for each $d$ is \underline{underlined}. For readability, we report $100 \times HV$.}
    \begin{subfigure}[c]{0.45\linewidth}
        \centering
        \setlength{\tabcolsep}{4pt}
        \renewcommand{\arraystretch}{2.5}
        \begin{tabular}{c|c|c|c|c}
        \toprule
        $d$ & \texttt{Full} 
            & \texttt{\shortstack{Diagonal\\(adaptive)}}  
            & \texttt{\shortstack{Diagonal\\(multiple)}}  
            & \texttt{Grid} \\
        \midrule
        8 &
        \shortstack{51.9\\{\scriptsize[51.5, 52.3]}} &
        \shortstack{\underline{\textbf{57.5}}\\{\scriptsize[56.9, 58.1]}} &
        \shortstack{\textbf{54.1}\\{\scriptsize[53.2, 55.1]}} &
        \shortstack{\textbf{57.0}\\{\scriptsize[56.5, 57.5]}} \\
        16 &
        \shortstack{40.2\\{\scriptsize[39.8, 40.7]}} &
        \shortstack{\underline{\textbf{46.6}}\\{\scriptsize[46.1, 47.2]}} &
        \shortstack{\textbf{43.3}\\{\scriptsize[42.5, 44.2]}} &
        \shortstack{\textbf{44.9}\\{\scriptsize[44.3, 45.8]}} \\
        32 &
        \shortstack{27.4\\{\scriptsize[27.0, 27.8]}} &
        \shortstack{\underline{\textbf{35.2}}\\{\scriptsize[34.6, 36.0]}} &
        \shortstack{\textbf{32.9}\\{\scriptsize[32.3, 33.5]}} &
        \shortstack{\textbf{32.3}\\{\scriptsize[31.8, 33.0]}} \\
        64 &
        \shortstack{19.6\\{\scriptsize[18.7, 21.0]}} &
        \shortstack{\underline{\textbf{29.9}}\\{\scriptsize[28.8, 30.9]}} &
        \shortstack{\textbf{28.1}\\{\scriptsize[27.5, 28.9]}} &
        \shortstack{\textbf{29.1}\\{\scriptsize[28.5, 29.8]}} \\
        \bottomrule
        \end{tabular}
        \caption{Target state \emph{Fourier-5}.}
    \end{subfigure}\hfill
    \begin{subfigure}[c]{0.49\linewidth}
        \centering
        \setlength{\tabcolsep}{4pt}
        \renewcommand{\arraystretch}{2.5}
        \begin{tabular}{c|c|c|c|c}
        \toprule
        $d$ & \texttt{Full} 
            & \texttt{\shortstack{Diagonal\\(adaptive)}}  
            & \texttt{\shortstack{Diagonal\\(multiple)}}  
            & \texttt{Grid} \\
        \midrule
        8 &
        \shortstack{49.1\\{\scriptsize[48.4, 49.9]}} &
        \shortstack{\underline{\textbf{53.8}}\\{\scriptsize[53.1, 54.6]}} &
        \shortstack{\textbf{52.3}\\{\scriptsize[51.1, 53.4]}} &
        \shortstack{\textbf{53.0}\\{\scriptsize[51.8, 54.1]}} \\
        16 &
        \shortstack{41.1\\{\scriptsize[40.6, 41.7]}} &
        \shortstack{\underline{\textbf{45.3}}\\{\scriptsize[44.6, 46.0]}} &
        \shortstack{\textbf{42.5}\\{\scriptsize[41.6, 43.6]}} &
        \shortstack{\textbf{44.1}\\{\scriptsize[43.5, 44.8]}} \\
        32 &
        \shortstack{32.6\\{\scriptsize[32.2, 33.0]}} &
        \shortstack{\underline{\textbf{36.2}}\\{\scriptsize[35.7, 36.7]}} &
        \shortstack{\textbf{33.7}\\{\scriptsize[33.0, 34.7]}} &
        \shortstack{\textbf{35.3}\\{\scriptsize[34.6, 36.1]}} \\
        64 &
        \shortstack{22.3\\{\scriptsize[21.3, 23.5]}} &
        \shortstack{\underline{\textbf{26.5}}\\{\scriptsize[25.1, 27.8]}} &
        \shortstack{\textbf{24.6}\\{\scriptsize[23.4, 26.3]}} &
        \shortstack{\textbf{24.8}\\{\scriptsize[23.5, 26.9]}} \\
        \bottomrule
        \end{tabular}
        \caption{Target state \emph{Haar random}.}
    \end{subfigure}
    \label{tab:table/hypervolume}
\end{table*}

\begin{figure*}[ht!]
    \centering
    \caption{Pareto frontiers showing the trade-off between the infidelity of the prepared state and the number of non-zero phase angles when optimizing the hyperparameters \textbf{penalizing the number of phases}. Each faded point represents a hyperparameter configuration, while solid lines connect the configurations that achieve the best trade-off (Pareto-optimal points) for each ansatz. 
    }
    \begin{subfigure}[c]{0.49\linewidth}
        \centering
        \includegraphics[width=1.0\linewidth]{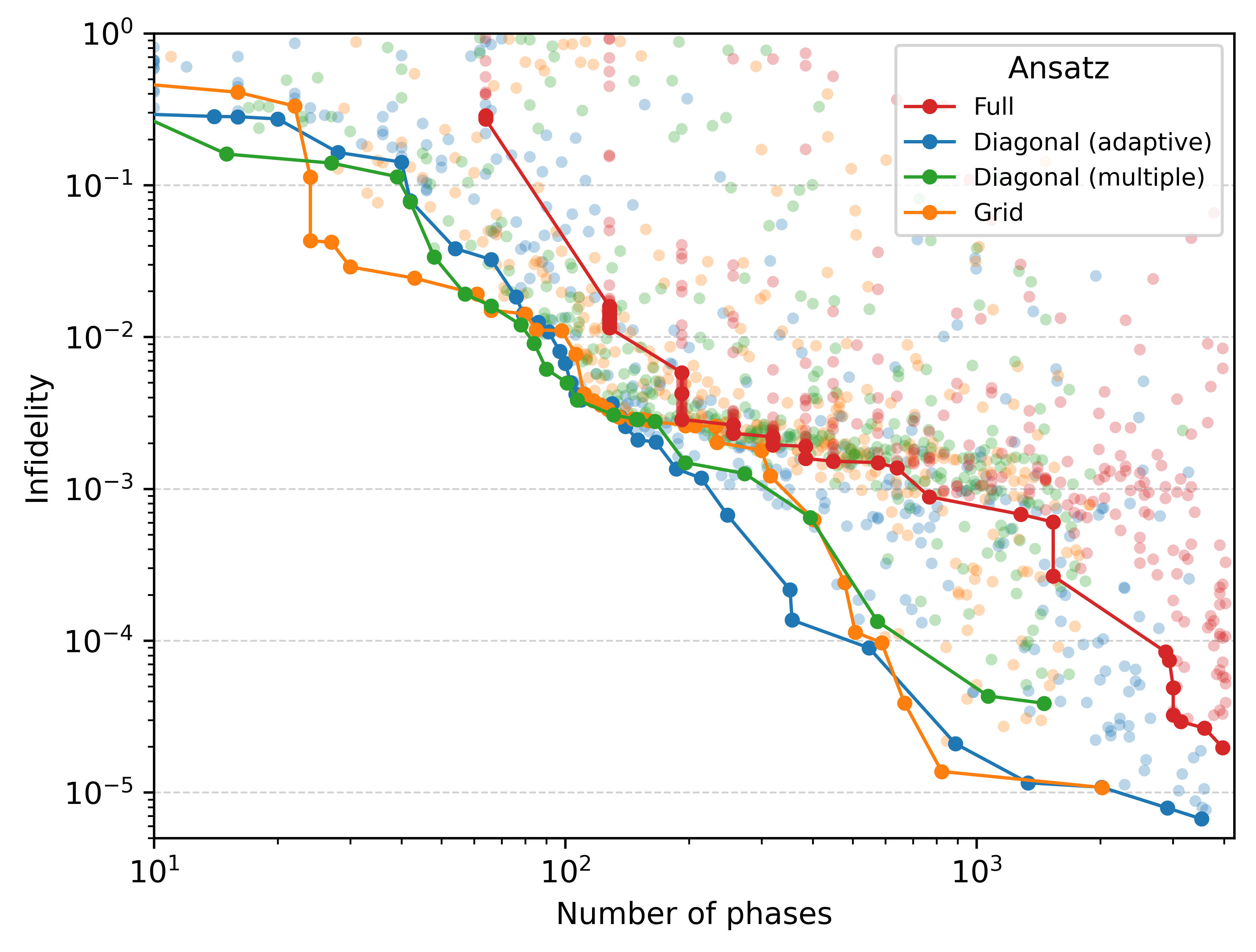}
        \caption{Target state \emph{Fourier-5} for $d=64$.}
    \end{subfigure}\hfill
    \begin{subfigure}[c]{0.49\linewidth}
        \centering
        \includegraphics[width=1.0\linewidth]{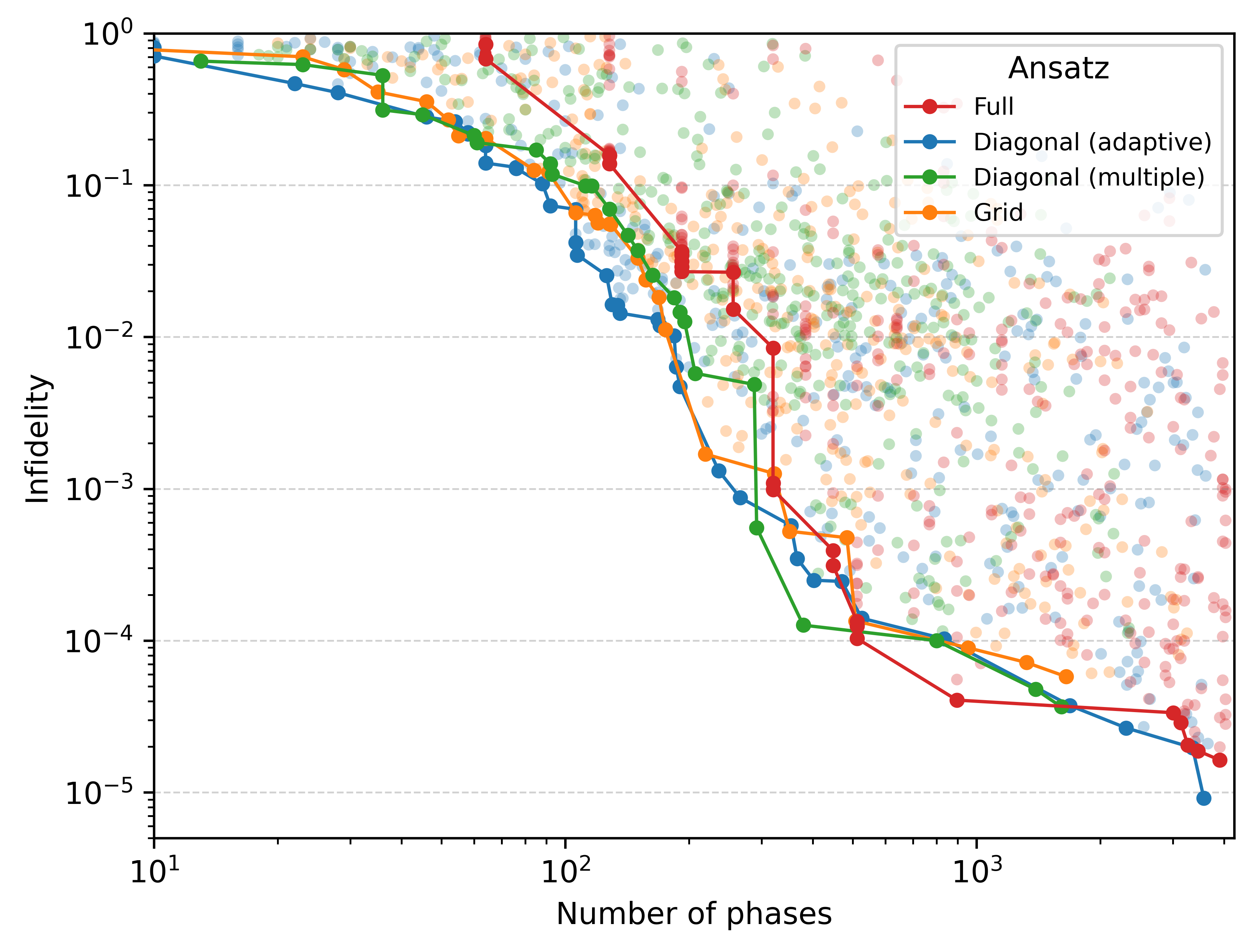}
        \caption{Target state \emph{Haar random} for $d=64$.}
    \end{subfigure}
    \label{fig:tables_phases/efficiency_scatter_phases}
\end{figure*}

We first discuss the results of the optimization penalizing the number of phases. The corresponding hypervolume metrics, together with their 95\% confidence intervals, are reported in Table \ref{tab:table/hypervolume}, while Figure \ref{fig:tables_phases/efficiency_scatter_phases} summarizes the Pareto-optimal points. The full results are reported in Appendix \ref{app:efficiency_phases}. Note that the reference point $r$ used to compute the hypervolume is defined as having the worst infidelity as well as the worst cost over all qudit sizes $d$, which ensures the result is comparable across that dimension as well. Furthermore, we compute it on the logarithm of the infidelity and cost metrics to account for their values spanning several orders of magnitude. According to the HV indicator, all sparse ansatzes exhibit a better trade-off than \texttt{Full} for both Fourier-5 and Haar random states across all qudit sizes $d$, with \texttt{Diagonal (adaptive)} consistently achieving the highest hypervolume. The difference is particularly clear for the Fourier-5 states, where the confidence intervals remain generally well separated from \texttt{Full}, while for the Haar random states the relative differences between sparse ansatzes tend to reduce, especially for larger $d$. 
If we focus on the Fourier-5 states, we can see how the improvement of the sparse ansatzes is particularly clear for the larger $d=64$, where Figure \ref{fig:tables_phases/efficiency_scatter_phases} shows that the Pareto frontier of the \texttt{Full} ansatz exhibits a substantially higher cost and approaches that of the other ansatzes only in a limited region between infidelity $10^{-2}$ and $10^{-3}$.
For the Haar random states, the difference is still present but to a smaller extent. Again, the \texttt{Full} ansatz exhibits a worse trade-off, especially for infidelities above $10^{-3}$, but then tends to overlap more with the others for infidelities below $10^{-4}$, which can be expected given that the 95\% confidence intervals of the HV indicator for \texttt{Full} are very close to, or not well separated from, those of some sparse ansatzes. 
We can also note how for both Fourier-5 and Haar random states, the absolute value of HV tends to reduce as $d$ increases, meaning that achieving good infidelities at lower cost becomes more difficult, as can be expected given the increasing qudit size. 

In the Appendix, see Table \ref{tab:tables_phases/mean_75/phase_efficiency_table}, we also report, for a given $d$ and target infidelity, the minimum number of phases required by any hyperparameter configuration encountered during their optimization. Note that sometimes, particularly for a target infidelity below $10^{-5}$ and larger qudit sizes, this target is not reached, likely due to limitations of the optimization process and the resource constraints we adopt. The results show that the simple \texttt{Diagonal (adaptive)} ansatz is almost always the one able to prepare the state with the lowest cost. The difference with respect to the \texttt{Full} ansatz varies depending on the protocol. For example, preparing a Fourier-5 state for $d=32$ and a target infidelity $\leq 10^{-4}$ requires 544 phases with the \texttt{Full} but only 348 with the \texttt{Diagonal (adaptive)} ansatz. On a larger $d=64$, the \texttt{Full} ansatz requires 2880 while the \texttt{Diagonal (adaptive)} only 548. On the Haar random state the results are similar but the difference often less pronounced, for $d=32$ and again a target infidelity $\leq 10^{-4}$, the \texttt{Full} ansatz requires 160 phases while the \texttt{Diagonal (adaptive)} requires 142. The larger $d=64$ is an example where the \texttt{Diagonal (multiple)} is more effective, requiring 799 phases compared to the 896 required by the \texttt{Full}. The outcome is, as previously discussed, less consistent when looking to the corresponding ansatz duration, as the internal distribution of the phases can either substantially reduce the gains or flip them if a lower number of phases are distributed across too many blocks making the ansatz too long. In Table \ref{tab:tables_phases/mean_75/phase_efficiency_table} we also compare with existing compilation approaches based on sequences of Givens rotations, \citet{PhysRevA.92.040303} and \citet{job2023efficient}. It should be noted that those methods focus on general unitary syntheses, which is a different type of scenario. When restricted to state preparation, both methods require only $O(d)$ Givens rotations, each of them is implemented using SNAP gates that apply a $\pi$ phase to all Fock states up to a cutoff corresponding to the level targeted by the rotation. As a consequence, when counting the number of non-zero SNAP phase angles the total cost scales as $d(d-1)$. Furthermore, achieving high fidelity typically requires slicing each Givens rotation into a sequence of smaller rotations, which allows to easily reach infidelities on the order of $10^{-8}$, but at the cost of substantially increased resource requirements. Table \ref{tab:tables_phases/mean_75/phase_efficiency_table} shows how both methods exhibit a resource requirement far beyond what is achievable with all the other ansatzes we studied, which however struggle to reach infidelities below $10^{-5}$. 

The results of the optimization penalizing the ansatz duration are reported in the Appendix (see the hypervolume metrics in Table \ref{tab:table_time/hypervolume} and the Pareto frontier in Figure \ref{fig:tables_time/efficiency_scatter_time}). The hypervolume indicator shows that, for the Fourier-5 states, all sparse ansatzes exhibit better trade-offs compared to \texttt{Full}, but to a smaller extent compared to when minimizing the number of phases. For the Haar random states, the differences are smaller and only one sparse ansatz at a time tends to outperform \texttt{Full}. The best sparse ansatz depends on the target state and $d$. On the Fourier-5 states, \texttt{Diagonal (multiple)} gives the highest hypervolume for $d=8$ and $d=16$, \texttt{Grid} for $d=32$, and \texttt{Diagonal (adaptive)} for $d=64$, while the differences on the Haar random states are quite small and the confidence intervals often overlap. The plot shows that all the ansatzes have very close Pareto frontiers in terms of their duration, with some minor exceptions.
Table \ref{tab:tables_time/mean_75/phase_efficiency_table} shows the minimal duration of the ansatz required to prepare states up to a target infidelity, again showing how penalizing the duration of the ansatz allows to reduce its duration for the Fourier-5 states, while this is less consistent for the Haar random states. These results again confirm that optimizing the duration of the ansatz is much more challenging than reducing the number of phases. However, this still produces a favorable Pareto frontier on the number of phases, see Figure \ref{fig:tables_time/efficiency_scatter_phases}, with a behavior similar to what we observed when penalizing the number of phases directly, but with a less pronounced difference. While reducing the duration of the ansatz is more difficult to achieve, it can be used as penalty criterion when the goal is to reduce the number of phases without risking to increase the duration of the ansatz.

Finally, we also run noisy simulations as described in Appendix \ref{app:photon_loss} by testing all ansatzes under photon-loss for different $T_1$ values. These simulations show that when minimizing either the number of phases or the duration of the ansatz, sparse ansatzes still achieve a better trade-off than the \texttt{Full} protocol in most cases when penalizing the number of phases, while they are less consistently effective when penalizing the duration of the ansatz, consistently with what observed in ideal conditions.

\section{Conclusions}
In this work we investigated how to reduce the resources required by the SNAP–Displacement protocol in order to provide a better trade-off between the prepared state fidelity and the cost required to prepare it, both in terms of the number of non-zero phase angles that need to be applied and in terms of the duration of the ansatz. By viewing the learnable phases as a block–Fock state grid, we proposed several sparse phase ansatzes and combined them with a two–stage optimization procedure that jointly tunes gate parameters and ansatz hyperparameters, including a cost penalty term that allows to control how much fidelity to sacrifice in order to reduce the resource cost.

Our results on several target states, up to relatively large qudit dimensions, show that the scalarized loss combining fidelity and cost allows one to smoothly transition from a high-fidelity, high-cost region to a lower-fidelity, lower-cost one, providing a useful tool for practitioners that aim to target one of these operating regimes. Furthermore, by comparing the hypervolume of the corresponding Pareto frontiers, we observe that the sparse ansatzes we propose exhibit a better trade-off than the original SD protocol in most conditions when the aim is to minimize the number of phases. When the goal is instead to minimize the duration of the ansatz, the sparse ansatzes still compare favorably in a number of cases, especially on the Fourier-5 states, but the advantage is more limited and no clear winner emerges under noise. The results remain broadly consistent under photon-loss noise, indicating that the proposed ansatzes are suitable across a wide range of conditions.

Overall, these results indicate that multi-objective optimization coupled with the introduction of regular structures controlling which SNAP phases to learn, is a viable route to make bosonic state preparation more cost efficient, reducing the number of control operations that are required.

\begin{acknowledgments}
We acknowledge the financial support from ICSC - ``National Research Centre in High Performance Computing, Big Data and Quantum Computing'' (Italy), funded by European Union – NextGenerationEU. We acknowledge ISCRA for awarding this project access to the LEONARDO supercomputer, owned by the EuroHPC Joint Undertaking, hosted by CINECA (Italy). This work has also received funding from the European Union's Horizon Europe Research and Innovation Programme under grant agreement No 101070125 (NGI Enrichers).
This work was supported by the U.S. Department of Energy, Office of Science, National Quantum Information Science Research Centers, Superconducting Quantum Materials and Systems Center (SQMS), under Contract No. 89243024CSC000002.

\end{acknowledgments}

\FloatBarrier
\bibliography{apssamp}

\clearpage
\appendix

\section{Fock and superposition state construction}
\label{app:single_parameter_ansatzes}
Here we list the optimized sequence for preparing various useful Fock and superposition states starting from the ground state using single-parameter SNAP gates in Table~\ref{tab:fock_prep}. The sequence has the structure
\begin{equation}
    \ket{\psi_f} = D(\alpha_d) \cdot \prod_{j=0}^{d-1} \left[ S_\pi(j) \cdot D(\alpha_j) \right]
\end{equation}
where $d$ is the dimension of the target state and $S_\pi(j)$ denotes a $\pi$ phase applied only on the state $\ket{j}$ \cite{Roy:2024BW}. Interestingly, the displacement parameters mostly alternate between positive and negative values for the Fock state preparation.

\begin{table*}[t]
\centering
\setlength{\tabcolsep}{4pt}
\renewcommand{\arraystretch}{1.5}
\caption{Optimized displacement parameters \(\alpha_j\) and resulting fidelities for different target Fock states and equal superposition states. Normalization constants for superposition states are omitted for brevity.}
\label{tab:fock_prep}
\scriptsize
\begin{tabular}{c|ccccccccccc|c|c}
\hline
\textbf{State} 
& \(\alpha_0\) & \(\alpha_1\) & \(\alpha_2\) & \(\alpha_3\) & \(\alpha_4\) 
& \(\alpha_5\) & \(\alpha_6\) & \(\alpha_7\) & \(\alpha_8\) & \(\alpha_9\) & \(\alpha_{10}\)
& \textbf{Fidelity} & \textbf{Infidelity} \\
\hline
\ket{1}  
& 1.143 & -0.580 & -- & -- & -- & -- & -- & -- & -- & -- & -- 
& 0.9814 & 0.0186 \\

\ket{2}  
& 0.497 & -1.133 & 0.432 & -- & -- & -- & -- & -- & -- & -- & -- 
& 0.9821 & 0.0179 \\

\ket{3}  
& 0.531 & -0.559 & 0.946 & -0.358 & -- & -- & -- & -- & -- & -- & -- 
& 0.9842 & 0.0158 \\

\ket{4}  
& 0.930 & -1.446 & 0.639 & 0.974 & -0.310 & -- & -- & -- & -- & -- & -- 
& 0.9910 & 0.0090 \\

\ket{5}  
& 0.567 & -0.921 & 0.443 & -0.385 & 0.735 & -0.278 & -- & -- & -- & -- & -- 
& 0.9874 & 0.0126 \\

\ket{6}  
& 0.559 & -0.630 & 0.769 & -0.348 & 0.343 & -0.673 & 0.254 & -- & -- & -- & -- 
& 0.9881 & 0.0119 \\

\ket{7}  
& 0.603 & -0.611 & 0.681 & -0.663 & 0.068 & -0.532 & 0.582 & -0.211 & -- & -- & -- 
& 0.9915 & 0.0085 \\

\ket{8}  
& 0.718 & -0.691 & 0.454 & -0.582 & 0.586 & -0.059 & 0.481 & -0.544 & 0.200 & -- & -- 
& 0.9921 & 0.0079 \\

\ket{9}  
& 0.586 & -0.792 & 0.524 & -0.367 & 0.508 & -0.533 & 0.056 & -0.440 & 0.513 & -0.190 & -- 
& 0.9924 & 0.0076 \\

\ket{10} 
& 0.617 & -0.645 & 0.635 & -0.443 & 0.306 & -0.457 & 0.493 & -0.055 & 0.405 & -0.488 & 0.182 
& 0.9924 & 0.0076 \\
\hline
$\sum_{j=0}^1|j\rangle$
& 0.561 & -0.243 & -- & -- & -- & -- & -- & -- & -- & -- & --
& 0.9995 & 0.0005 \\

$\sum_{j=0}^2|j\rangle$
& -0.140 & -0.968 & 0.218 & -- & -- & -- & -- & -- & -- & -- & --
& 0.9991 & 0.0009 \\

$\sum_{j=0}^3|j\rangle$
& -1.038 & -0.433 & 0.423 & 0.229 & -- & -- & -- & -- & -- & -- & --
& 0.9987 & 0.0013 \\

$\sum_{j=0}^4|j\rangle$
& 0.904 & -0.719 & -0.719 & -0.299 & 0.672 & -- & -- & -- & -- & -- & --
& 0.9873 & 0.0127 \\

$\sum_{j=0}^5|j\rangle$
& -1.205 & -0.591 & 0.327 & -0.304 & 0.345 & 0.219 & -- & -- & -- & -- & --
& 0.9987 & 0.0013 \\

$\sum_{j=0}^6|j\rangle$
& -0.805 & -0.658 & 0.072 & -0.029 & 0.514 & 0.035 & 0.496 & -- & -- & -- & --
& 0.9770 & 0.0230 \\

$\sum_{j=0}^7|j\rangle$
& -1.467 & -0.511 & 0.318 & -0.404 & 0.241 & -0.227 & 0.297 & 0.209 & -- & -- & --
& 0.9987 & 0.0013 \\

$\sum_{j=0}^8|j\rangle$
& 1.071 & -1.541 & 0.287 & -0.467 & -0.039 & 0.063 & 0.440 & 0.037 & 0.447 & -- & --
& 0.9821 & 0.0179 \\

$\sum_{j=0}^9|j\rangle$
& -1.572 & -0.614 & 0.309 & -0.360 & 0.227 & -0.316 & 0.194 & -0.182 & 0.269 & 0.201 & --
& 0.9987 & 0.0013 \\

$\sum_{j=0}^{10}|j\rangle$
& -1.285 & -0.677 & 0.220 & -0.239 & 0.276 & -0.328 & 0.007 & 0.014 & 0.375 & 0.038 & 0.427 & 0.9847 & 0.0153 \\

\hline

\end{tabular}
\end{table*}

\section{Sparse Ansatzes}
\label{app:sparse_ansatzes}
The sparse ansatzes proposed in this study are defined by one or more continuous lines on a grid $B \times d$ of points $(b, f)$, with $b$ denoting the block index and $f$ the Fock-level index, corresponding to the phase-parameter matrix $\Theta$ represented in 2D Euclidean coordinates.

The sparse ansatzes are defined from lines in Cartesian coordinates:
\begin{equation}
    f = m b + c,
\end{equation}
where $m \in \mathbb{R}$ is the slope and $c \in \mathbb{R}$ is the vertical intercept. The number of lines, as well as their slopes and intercepts, will change depending on the ansatz. Given the set of lines, a phase at position $(b, f)$ is learnable only if its distance from the closest line is within a width hyperparameter $w$.

The perpendicular distance from point $(b, f)$ to the line $f = m b + c$ is:
\begin{equation}
\mathrm{dist}(b, f) = \frac{|m b - f + c|}{\sqrt{m^2 + 1}},
\end{equation}
which is the Euclidean shortest distance from the point to the line, measured along the perpendicular passing through the point.

Let $w > 0$ be a width parameter that controls the thickness of the line. We define a binary mask $M \in \{0, 1\}^{B \times d}$ as:
\begin{equation}
    M_{b,f} =
    \begin{cases}
    1, & \text{if } \mathrm{dist}(b,f) \le \frac{w}{2}, \\
    0, & \text{otherwise.}
    \end{cases}
\end{equation}
This selects all points within a band of width $w$ centered on the line. All phases with $M_{b,f}=1$ are learnable parameters while the remaining ones are not learnable and are set to zero, \idest $M_{b,f}=0 \implies \theta_{b,f}=0$.

\subsection{\texttt{Diagonal (adaptive)}}
The \texttt{Diagonal (adaptive)} ansatz is based on a single line oriented along the main diagonal of the $(b, f)$ grid.
The line parameters are defined as $m = \frac{d-1}{B-1}$ and $c = 0$, so that it spans the range from $(0, 0)$ to $(B-1, d-1)$.

\subsection{\texttt{Diagonal (multiple)}}
\label{sec:app:diagonal_multiple}
The \texttt{Diagonal (multiple)} ansatz extends \texttt{Diagonal (adaptive)} by placing multiple parallel diagonal bands. Each band corresponds to a line of slope $m$, and the lines are evenly spaced along the anti-diagonal. The slope is the same as in the \texttt{Diagonal (adaptive)} case:
\begin{equation}
m = \frac{d - 1}{B - 1},
\end{equation}
so that all lines share the orientation of the main diagonal from $(0,0)$ to $(B-1,d-1)$.

Each diagonal line is centered at a point $(b_i, f_i)$ lying on the anti-diagonal from $(0, d-1)$ to $(B-1, 0)$. We evenly divide this range into $m_{\text{diag}}$ intervals and place the $i$-th diagonal at the center of the $i$-th interval. Specifically, we define a normalized coordinate:
\begin{equation}
t_i = \frac{i + 0.5}{m_{\text{diag}}},
\end{equation}
where $i = 0, \dots, m_{\text{diag}} - 1$. The corresponding center point $(b_i, f_i)$ is:
\begin{equation}
b_i = t_i (B - 1), \quad f_i = (1 - t_i) (d - 1).
\end{equation}

The vertical intercept $c_i$ of the $i$-th line is computed via the point–slope form of a line, to ensure it passes through $(b_i, f_i)$:
\begin{equation}
c_i = f_i - m b_i.
\end{equation}
The learnable phases are then defined based on the union (logical \texttt{OR}) of the masks $M$ associated with each line.

\subsection{\texttt{Grid}}

The \texttt{Grid} ansatz extends \texttt{Diagonal (multiple)} by introducing an additional family of bands with negative slope (parallel to the anti-diagonal), forming a grid-like pattern.

As in the previous ansatzes, the positive-slope diagonal bands are centered at positions along the anti-diagonal, using the parametrization described for \texttt{Diagonal (multiple)} in Section~\ref{sec:app:diagonal_multiple}.

In addition, the \texttt{Grid} ansatz introduces a set of $m_{\text{anti}}$ anti-diagonal bands with slope $-m$. The positions of these lines are aligned along the main diagonal and use the normalized parameter:
$$t'_j = \frac{j+0.5}{m_{\text{anti}}}.$$

The corresponding center point and intercept are:
\begin{equation}
    b'_j = t'_j(B - 1), \quad
    f'_j = t'_j(d - 1), \quad
    c'_j = f'_j + m b'_j,
\end{equation}
with $j=0,\dots,m_{\text{anti}}-1$. 

The learnable phases are then defined based on the union (logical \texttt{OR}) of the masks $M$ associated with each diagonal and anti-diagonal line.

\section{Efficient Parallel Simulation}
\label{app:efficient_parallel_simulation}
Computing the prepared state, given an ansatz, for small Hilbert spaces face a bottleneck on GPUs due to the large number of small matrix multiplications required. GPUs are typically optimized for large matrix multiplications that can be parallelized across their cores. To run multiple repetitions in parallel efficiently, we stack the unitaries as 3D tensors of shape $(R, \tilde{d}, \tilde{d})$, where $R$ is the number of repetitions, and the target state as a 2D tensor of shape $(R, \tilde{d})$. Each batch of unitaries is then applied at once using batched matrix multiplications, ensuring that all repetitions evolve independently. This approach enables much more efficient GPU utilization. To ensure that the gradient computation is also vectorized, we define a global loss function as the \emph{sum} of the losses across repetitions (as defined in Eq. \ref{eq:gate_param_loss}), rather than their average, so that the effective learning rate remains independent of $R$. Note that the $\ell_1$ and $\ell_2$ regularization terms on the gate parameters must be computed separately for each repetition, ensuring that each repetition is penalized only for its own gate parameters.

\begin{table*}[]
\centering
\caption{Hyperparameter list, ranges, and sampling distributions for all ansatzes.}
\label{tab:hyperparameters}
\begin{tabular}{|c|c|c|c|c|}
\toprule
\textbf{Ansatz} & \textbf{Description} & \textbf{Symbol} & \textbf{Range} & \textbf{Distribution} \\
\midrule
\multirow{3}{*}{\shortstack{Common Hyperparameters for \\ Gate Parameter Learning \\ (see Section \ref{sec:gate_params})}} 
    & learning rate &  & $[10^{-5}, 10^{-1}]$ & Log-uniform \\ 
    & regularization weight &  $\lambda_1$ & $[10^{-8}, 10^{-2}]$ & Log-uniform \\ 
    & regularization weight &  $\lambda_2$ & $[10^{-8}, 10^{-2}]$ & Log-uniform \\ 
\midrule
\midrule
\multirow{1}{*}{\texttt{Full}} 
    & number of blocks &  $B$ & $[1, d]$ & Log-uniform \\
\midrule
\multirow{3}{*}{\texttt{Diagonal (adaptive)}}
    & number of blocks &  $B$ & $[2, d]$ & Log-uniform \\ 
    & band width &  $w$ & $[0.1, \text{min}(d, B)]$ & Log-uniform \\ 
    & cutoff line interpolation & $\tau$ & $[0.0, 1.0]$ & Uniform \\
\midrule
\multirow{4}{*}{\texttt{Diagonal (multiple)}}
    & number of blocks &  $B$ & $[2, d]$ & Log-uniform \\ 
    & band width  & $w$ & $[0.1, \text{min}(d, w_{\text{max}})]$ & Log-uniform \\ 
    & number of diagonals &  $m_{\text{diag}}$ & $[2, \text{min}(d, B)]$ & Log-uniform \\ 
    & cutoff line interpolation &  $\tau$ & $[0.0, 1.0]$ & Uniform \\
\midrule
\multirow{5}{*}{\texttt{Grid}}
    & number of blocks &  $B$ & $[2, d]$ & Log-uniform \\ 
    & band width &  $w$ & $[0.1, \text{min}(d, w_{\text{max\_diag}},w_{\text{max\_anti}})]$  & Log-uniform \\ 
    & number of diagonals &  $m_{\text{diag}}$ & $[1, \text{min}(d, B)]$ & Log-uniform \\ 
    & number of anti-diagonals &  $m_{\text{anti}}$ & $[1, \text{min}(d, B)]$   & Log-uniform \\ 
    & cutoff line interpolation & $\tau$ & $[0.0, 1.0]$ & Uniform \\
\bottomrule
\end{tabular}
\end{table*}

\section{Hyperparameter Optimization}
\label{app:ansatz_hyperparams}
The full list of hyperparameters, including their names, ranges, and sampling distributions, is provided in Table \ref{tab:hyperparameters}. The table presents two sets of hyperparameters: those related to the optimization of the gate parameters, as described in Section \ref{sec:gate_params} and common to all ansatzes, and the ansatz-specific hyperparameters described in Section \ref{sec:sparse_ansatzes} and Appendix \ref{app:sparse_ansatzes}. All are optimized according to the process in Section \ref{sec:ansatz_hyperparams}. Note that most hyperparameters are sampled from a log-uniform distribution to favor exploration of small values, consistent with our goal of minimizing resource requirements.

The fundamental hyperparameter is the line width $w>0$ (for diagonal or anti-diagonal bands) composing the ansatz. Recall that a phase $\theta_{b,f}$ is learnable if its perpendicular distance to at least one line is $\le w/2$. Smaller $w$ promotes sparsity by selecting fewer learnable phases, whereas larger $w$ increases their number, potentially covering the entire grid for sufficiently large $w$. To improve the efficiency of the hyperparameter optimization process, we avoid configurations with an excessively high number of learnable phases and cap the maximum $w$ based on the number of lines, thereby preventing their overlap. In our simulations, we consider $w \in [0.5, w_{\text{max}}]$, where $w_{\text{max}}$ is determined by the geometric spacing between bands.

The upper bound $w_{\text{max}}$ depends on the number of lines $m$ (either diagonals $m_{\text{diag}}$ or anti-diagonals $m_{\text{anti}}$). Each line is centered at a point $(b_i, f_i)$ as described in Appendix \ref{app:sparse_ansatzes}. The Euclidean distance between consecutive line centers is
\begin{equation}
\Delta = \frac{1}{m} \sqrt{(B - 1)^2 + (d - 1)^2}.
\end{equation}
To limit overlap, we set the maximum width to a fraction $\alpha \in [0,1]$ of this spacing:
\begin{equation}
w_{\text{max}} = \alpha \Delta.
\end{equation}
We use $\alpha = 0.9$ to minimize interference between neighboring bands while maintaining good coverage of the available phases.

\subsection{Pruning Mask}
To further reduce the number of learnable phases, we introduce an optional masking rule that removes all learnable phases below a tunable threshold line. The purpose of this rule is to avoid placing learnable phases in regions that are unlikely to influence the quantum state, such as high Fock states during early blocks because it is likely that the displacement gates have not populated them yet. 
We define a cutoff line that is parallel to the main diagonal and intersects the anti-diagonal at a tunable position. 
All learnable phases below the cutoff line are removed.

More formally, the main diagonal is defined by the line $f = m b$, and the cutoff line is parallel to it:
\begin{equation}
    f_c = m b + c,
\end{equation}
where $c$ is a vertical offset controlling how aggressively the mask excludes phases for high Fock states in early blocks.
To ensure that $c$ is geometrically meaningful, we define it as a linear interpolation between two key points:
\begin{itemize}
    \item The \emph{intersection} of the main diagonal and anti-diagonal (the \emph{center} of the phase grid):
    \begin{equation}
    b_0 = \frac{d - 1}{2m} = \frac{B - 1}{2}, \qquad f_0 = \frac{d - 1}{2};
    \end{equation}
    \item The \emph{bottom-left corner} of the grid: $(0, d - 1)$.
\end{itemize}

Let $\tau \in [0, 1]$ be a coefficient that determines the interpolation between these points. The corresponding cutoff line passes through the point:
\begin{equation}
    b_\tau = (1 - \tau) \frac{B - 1}{2}, \qquad
    f_\tau = (1 - \tau) \frac{d - 1}{2} + \tau (d - 1),
\end{equation}
and, being parallel to the main diagonal, we obtain the cutoff line:
\begin{equation}
    f_c = mb + (f_\tau-mb_\tau) = m b + \tau (d - 1).
\end{equation}

All phases that lie below the cutoff line, \idest $f_c \geq m b + \tau (d - 1)$ are applied to high Fock states on the initial blocks, and should be removed. This mechanism provides a smooth interpolation between no masking ($\tau = 1$) and the removal of learnable phases below the main diagonal ($\tau = 0$). Given the narrow range of this hyperparameter, and since very small changes are likely to yield similar selections of learnable phases, we sample it from a uniform distribution.

\section{Phase Penalty Results}
\label{app:phase_penalty}
This section reports the full results of the hyperparameter optimization of all ansatzes and qudit sizes $d=8,16,32$ and $64$. The results of the hyperparameter optimization penalizing the number of phases are reported in Table \ref{tab:tables_phases/mean_75/fourier_5/initial_0-best_scalarized} (Fourier-5) and Table \ref{tab:tables_phases/mean_75/random_gaussian/initial_0-best_scalarized} (Haar random). The results of the hyperparameter optimization penalizing the number of phases are reported in Table \ref{tab:tables_time/mean_75/fourier_5/initial_0-best_scalarized} (Fourier-5) and Table \ref{tab:tables_time/mean_75/random_gaussian/initial_0-best_scalarized} (Haar random).

\begin{figure*}[ht!]
    \centering
    \caption{Infidelity and duration of the ansatz for the best hyperparameter configuration $h$ obtained when \textbf{penalizing the number of phases}, for each penalty coefficient $\beta$. 
        }
    \begin{subfigure}[c]{0.49\linewidth}
        \centering
        \includegraphics[width=0.9\linewidth]{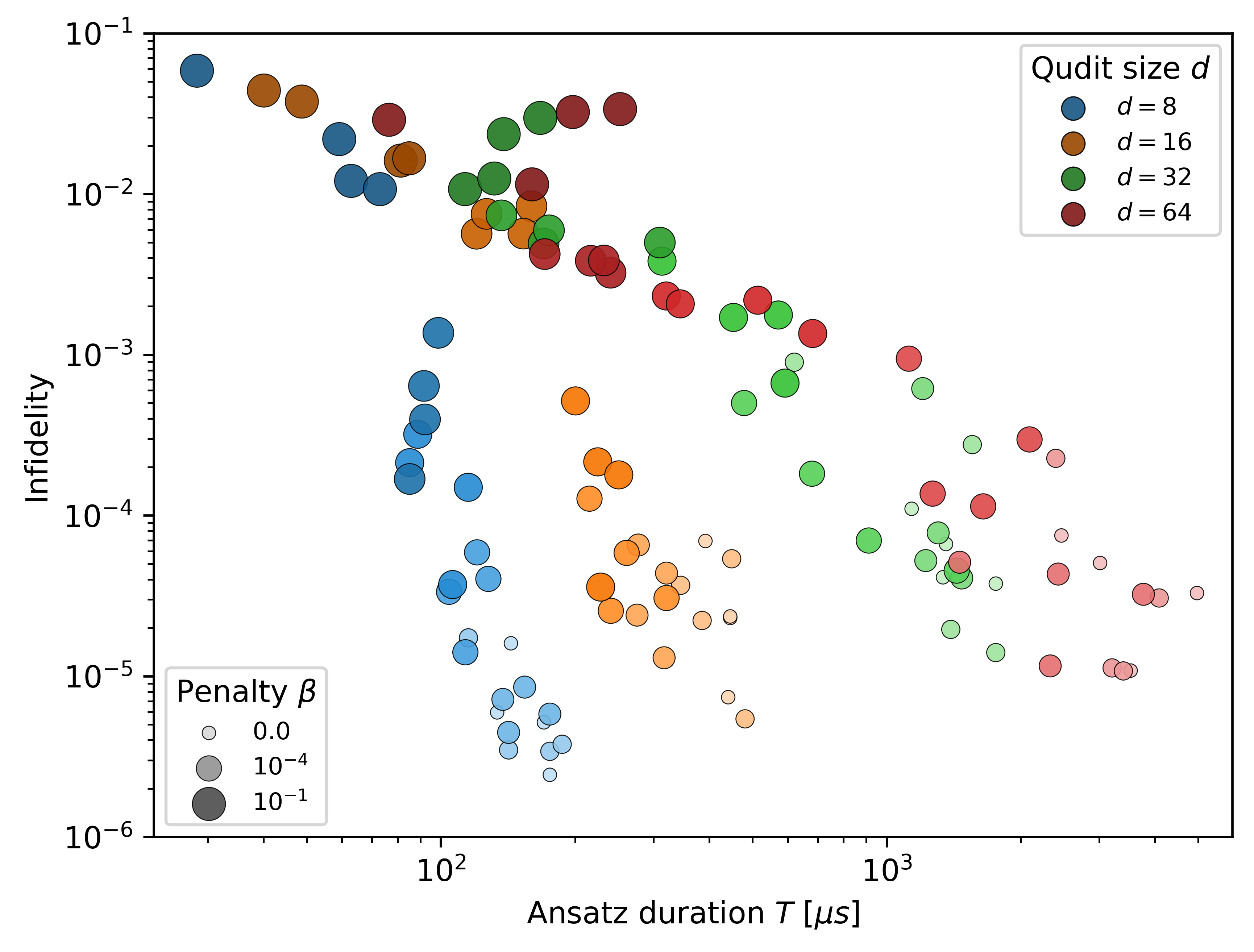}
        \caption{Target state \emph{Fourier-5}.}
        \label{fig:tables_phases/mean_75/fourier_5/initial_0/pareto_d=8_16_32_64_no_arch_time}
    \end{subfigure}\hfill
    \begin{subfigure}[c]{0.49\linewidth}
        \centering
        \includegraphics[width=0.9\linewidth]{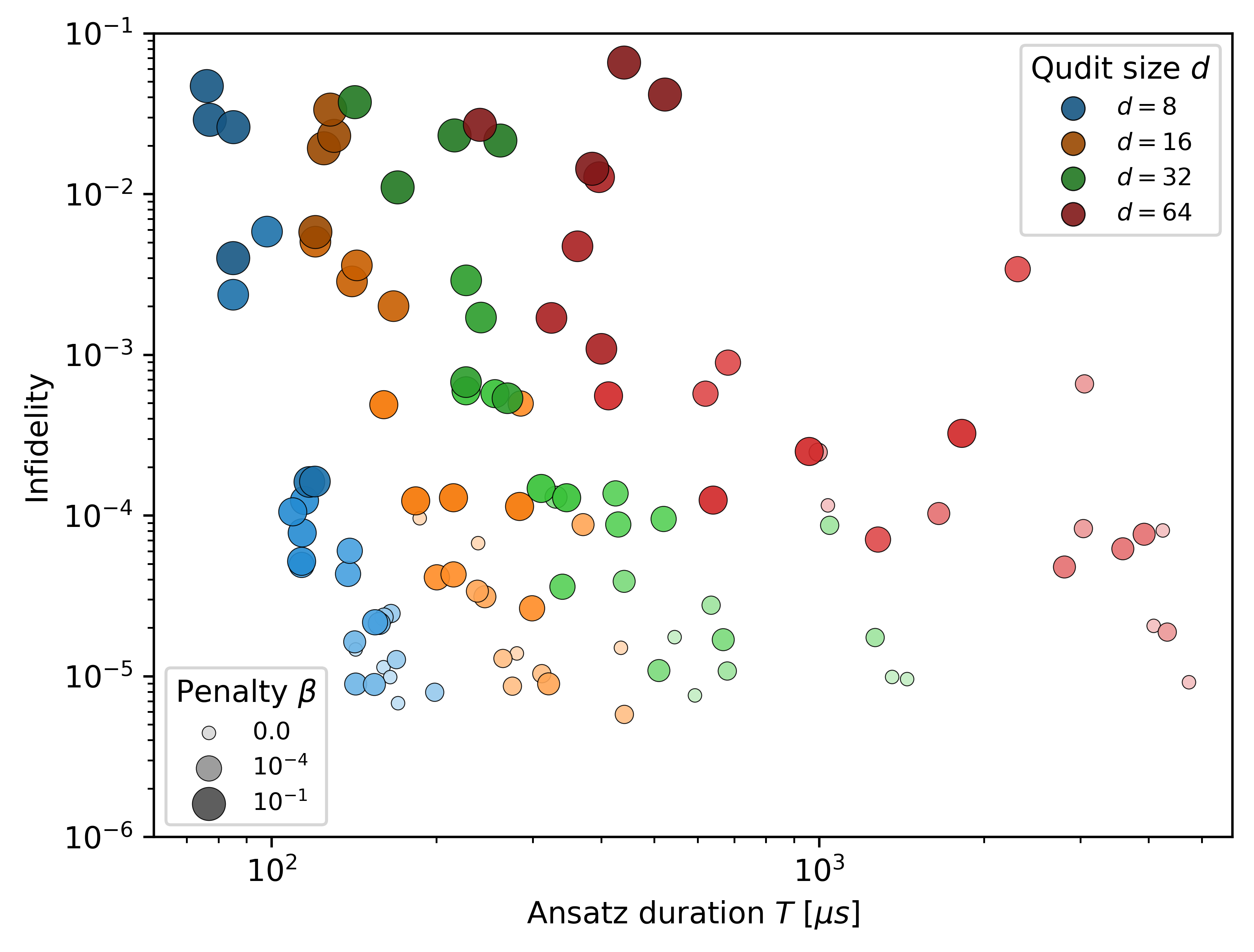}
        \caption{Target state \emph{Haar random}.}
        \label{fig:tables_phases/mean_75/random_gaussian/initial_0/pareto_d=8_16_32_64_no_arch_time}
    \end{subfigure}
    \label{fig:tables_phases/pareto_d=8_16_32_64_no_arch_time}
\end{figure*}

\begin{figure*}[ht!]
    \centering
    \caption{Infidelity and number of non-zero phase angles for the best hyperparameter configuration $h$ obtained when \textbf{penalizing the duration of the ansatz}, for each penalty coefficient $\beta$. 
        }
    \begin{subfigure}[c]{0.49\linewidth}
        \centering
        \includegraphics[width=0.9\linewidth]{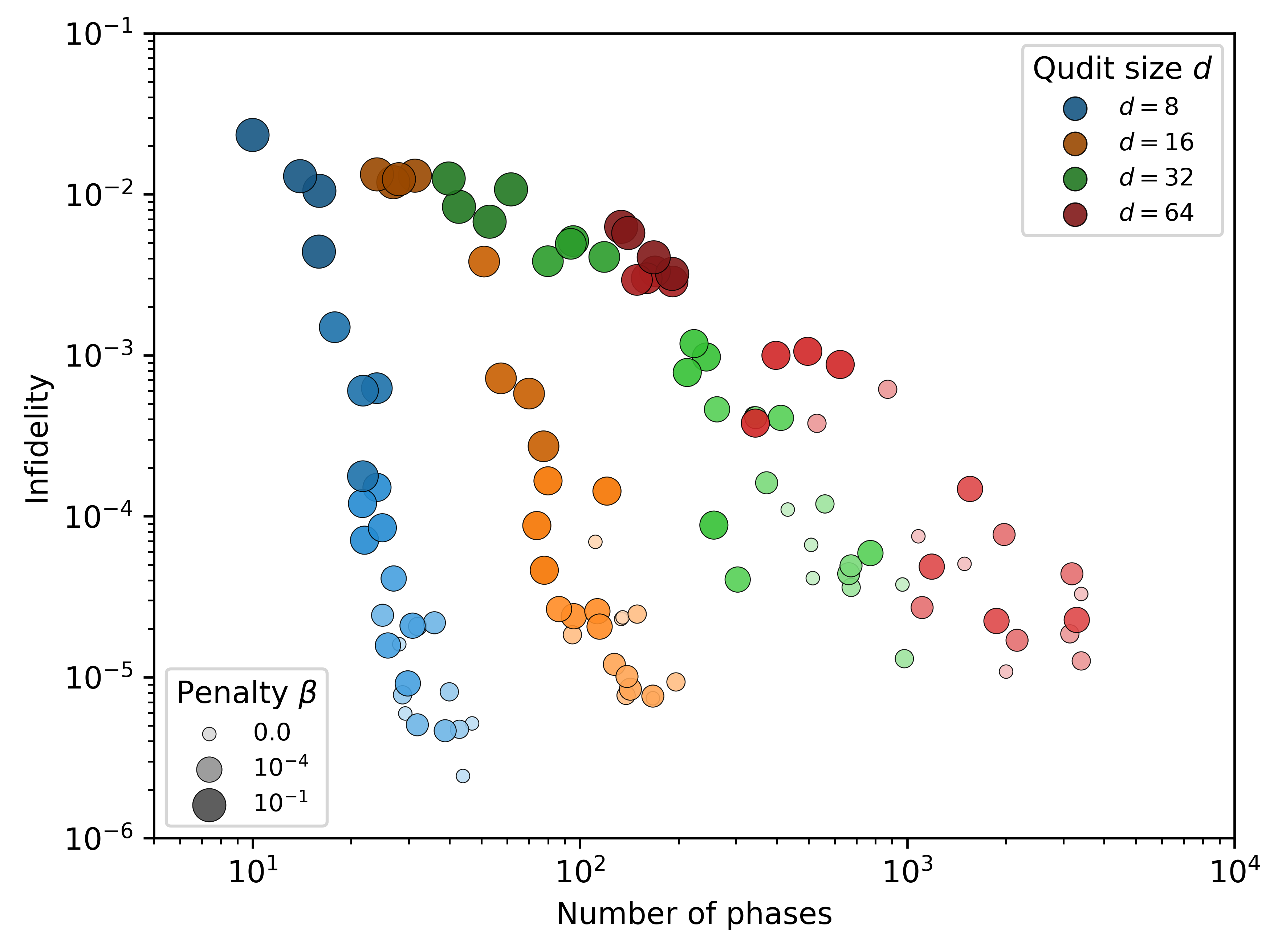}
        \caption{Target state \emph{Fourier-5}.}
        \label{fig:tables_time/mean_75/fourier_5/initial_0/pareto_d=8_16_32_64_no_arch_phases}
    \end{subfigure}\hfill
    \begin{subfigure}[c]{0.49\linewidth}
        \centering
        \includegraphics[width=0.9\linewidth]{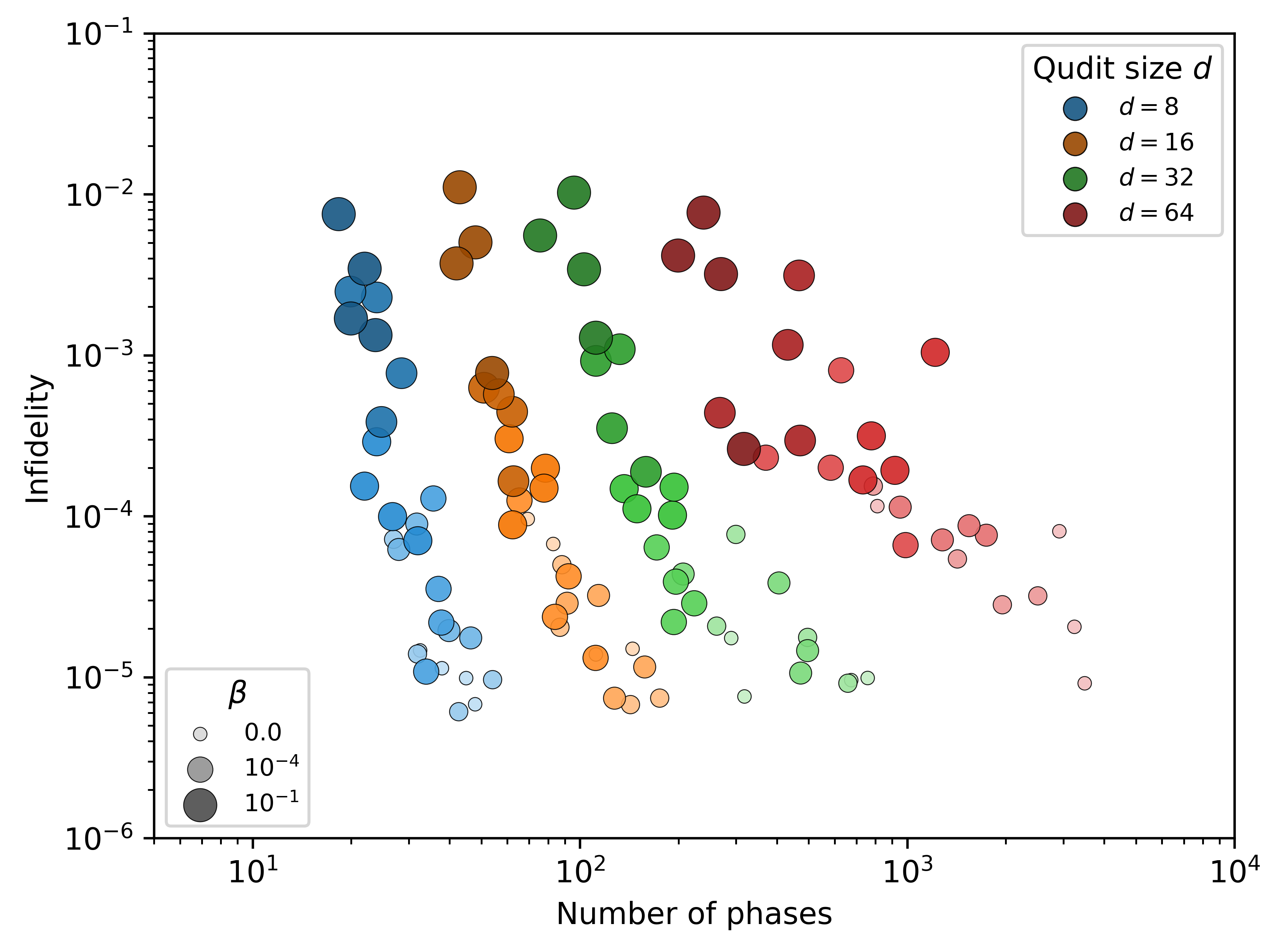}
        \caption{Target state \emph{Haar random}.}
        \label{fig:tables_time/mean_75/random_gaussian/initial_0/pareto_d=8_16_32_64_no_arch}
    \end{subfigure}
    \label{fig:tables_time/pareto_d=8_16_32_64_no_arch_phases}
\end{figure*}

We also report a visualization of the trade-off across the other metric, that is, the ansatz duration when the optimal hyperparameters are selected by penalizing the number of phases (see Figure \ref{fig:tables_time/pareto_d=8_16_32_64_no_arch_phases}) and the number of phases when the hyperparameters are selected by optimizing the duration of the ansatz (see Figure \ref{fig:tables_phases/pareto_d=8_16_32_64_no_arch_time}). The results indicate that penalizing the duration of the ansatz allows to control smoothly the number of phases as well, exhibiting a similar pattern compared to the simulation that penalized the number of phases directly. The only difference is that the number of phases can be reduced to a lesser extent when the penalty coefficient $\beta$ is high. On the opposite front, penalizing the number of phases is a poor proxy for the duration of the ansatz in many regions, in particular when the penalty coefficient $\beta$ is high. This result can be expected given that the duration of a SNAP gate is a function of the square root of the phases it needs to apply and therefore the same total number of phases will result in a longer overall ansatz if they are split across more gates, \idest blocks.

\begin{table*}
    \centering
    \footnotesize
    \caption{Results of the best hyperparameter configuration $h$ obtained for the preparation of the \textbf{Fourier-5} states, when \textbf{penalizing the number of phases}, for each penalty coefficient $\beta$. The table reports the infidelity, number of non-zero phase angles and duration for the different ansatzes we compare. 
        }    \input{tables_phases/mean_75/fourier_5/initial_0/results_best_scalarized}
    \label{tab:tables_phases/mean_75/fourier_5/initial_0-best_scalarized}
\end{table*}

\begin{table*}
    \centering
    \footnotesize
    \caption{Results of the best hyperparameter configuration $h$ obtained for the preparation of the \textbf{Haar random} states, when \textbf{penalizing the number of phases}, for each penalty coefficient $\beta$. The table reports the infidelity, number of non-zero phase angles and duration for the different ansatzes we compare. 
        }     \input{tables_phases/mean_75/random_gaussian/initial_0/results_best_scalarized}
    \label{tab:tables_phases/mean_75/random_gaussian/initial_0-best_scalarized}
\end{table*}

\begin{table*}
    \centering
    \footnotesize
    \caption{Results of the best hyperparameter configuration $h$ obtained for the preparation of the \textbf{Fourier-5} states, when \textbf{penalizing the duration of the ansatz}, for each penalty coefficient $\beta$. The table reports the infidelity, number of non-zero phase angles and duration for the different ansatzes we compare. 
        }     \input{tables_time/mean_75/fourier_5/initial_0/results_best_scalarized}
    \label{tab:tables_time/mean_75/fourier_5/initial_0-best_scalarized}
\end{table*}

\begin{table*}
    \centering
    \footnotesize
    \caption{Results of the best hyperparameter configuration $h$ obtained for the preparation of the \textbf{Haar random} states, when \textbf{penalizing the duration of the ansatz}, for each penalty coefficient $\beta$. The table reports the infidelity, number of non-zero phase angles and duration for the different ansatzes we compare. 
        }     \input{tables_time/mean_75/random_gaussian/initial_0/results_best_scalarized}
    \label{tab:tables_time/mean_75/random_gaussian/initial_0-best_scalarized}
\end{table*}

\clearpage
\section{Fidelity-cost trade-off when optimizing the number of phases}
\label{app:efficiency_phases}
\begin{figure*}[ht!]
    \centering
    \caption{Pareto frontiers showing the trade-off between the infidelity of the prepared state and the number of non-zero phase angles when optimizing the hyperparameters \textbf{penalizing the number of phases}. Each faded point represents a hyperparameter configuration, while solid lines connect the configurations that achieve the best trade-off (Pareto-optimal points) for each ansatz. 
        }
    \begin{subfigure}[c]{0.49\linewidth}
        \centering
        \includegraphics[width=1.0\linewidth]{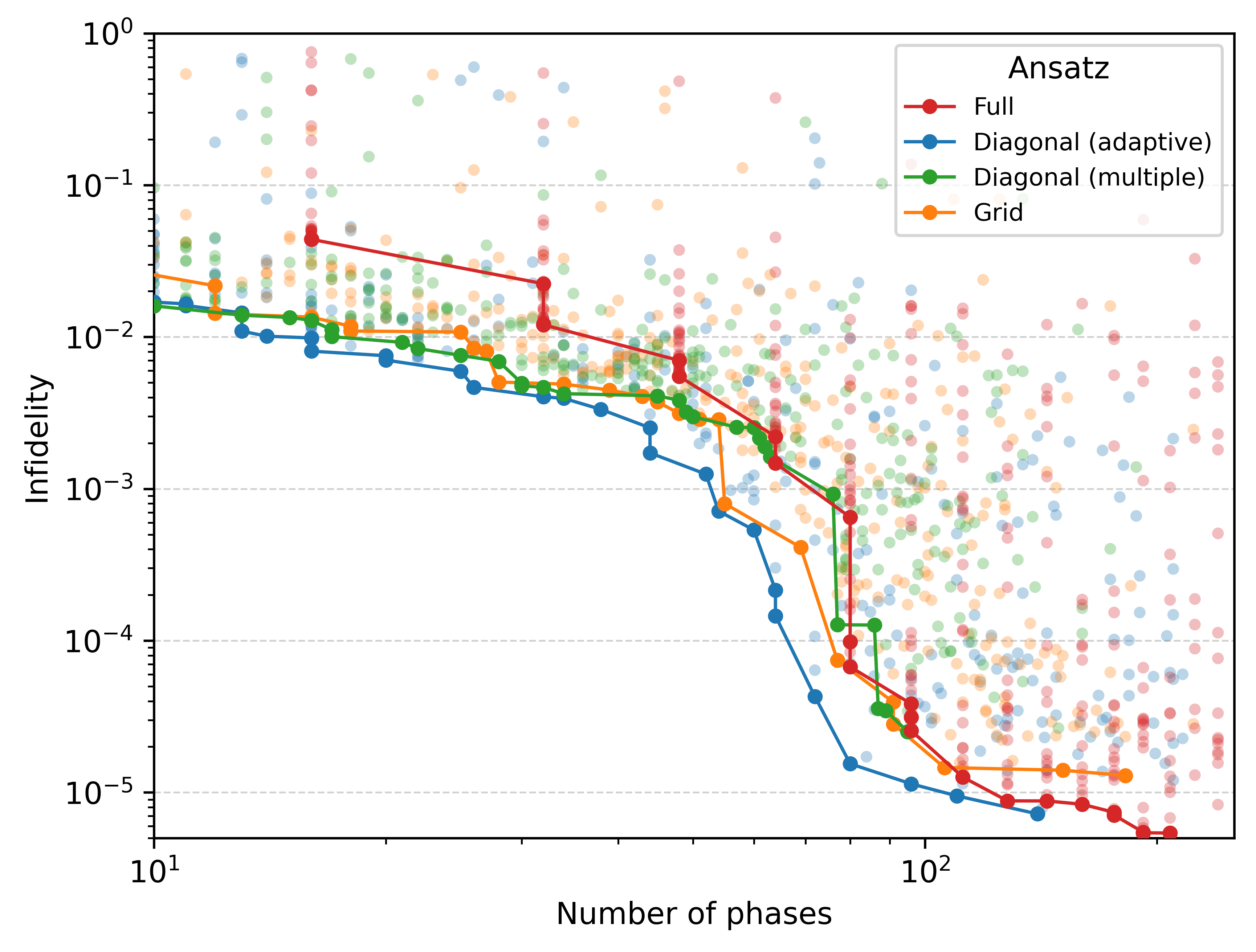}
        \caption{Target state \emph{Fourier-5} for $d=16$.}
    \end{subfigure}\hfill
    \begin{subfigure}[c]{0.49\linewidth}
        \centering
        \includegraphics[width=1.0\linewidth]{tables_phases/mean_75/fourier_5/initial_0/efficiency_scatter_phases/fig5a.png}
        \caption{Target state \emph{Fourier-5} for $d=64$.}
    \end{subfigure}\hfill
    \begin{subfigure}[c]{0.49\linewidth}
        \centering
        \includegraphics[width=1.0\linewidth]{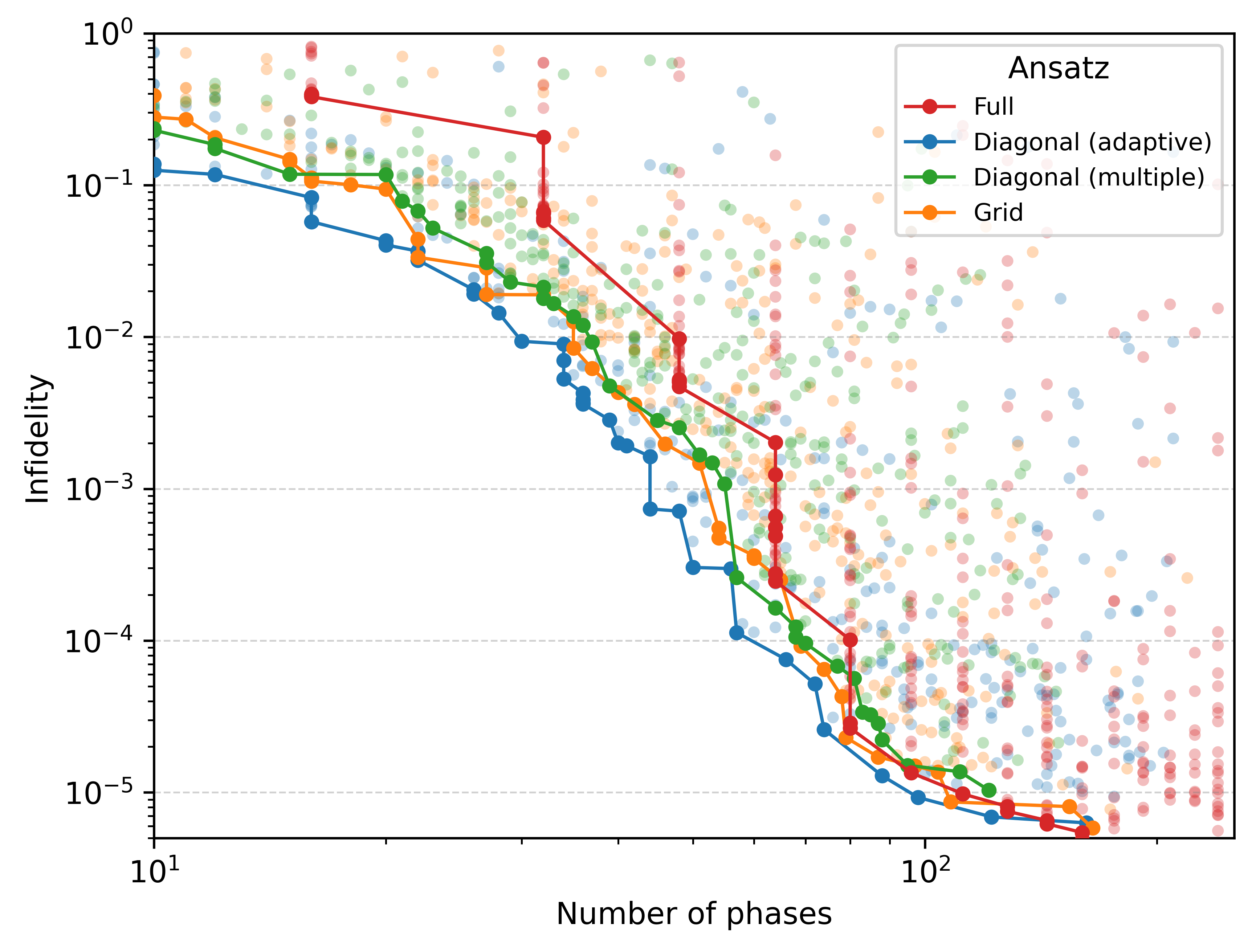}
        \caption{Target state \emph{Haar random} for $d=16$.}
    \end{subfigure}\hfill
    \begin{subfigure}[c]{0.49\linewidth}
        \centering
        \includegraphics[width=1.0\linewidth]{tables_phases/mean_75/random_gaussian/initial_0/efficiency_scatter_phases/fig5b.png}
        \caption{Target state \emph{Haar random} for $d=64$.}
    \end{subfigure}
    \label{app:fig:tables_phases/efficiency_scatter_phases}
\end{figure*}

This section reports the full results on the efficiency of the state preparation for all ansatzes and qudit sizes $d=8,16,32$ and $64$, when minimizing the number of phases. The Pareto frontiers are summarized in Figure \ref{app:fig:tables_phases/efficiency_scatter_phases}, and the number of phases required to prepare a state given a target infidelity are summarized in Table \ref{tab:tables_phases/mean_75/phase_efficiency_table}. 

In this section we also report the results of two existing methods, \citet{PhysRevA.92.040303} and \citet{job2023efficient}. However, since our task is state preparation rather than full unitary synthesis, we only compile the sequence required to prepare the target state. The direct compilation method of \citet{job2023efficient} provides a closed-form expression for the displacement parameter, while the fidelity can be controlled by slicing the rotation angle of each Givens rotation into multiple successive rotations, each applying a smaller angle. Accordingly, we search for the optimal number of slices by exploring all values from $1$ to $10$, and then from $11$ to $100$ with a step of $2$. Note that, since this method is deterministic, we report the results of only one repetition. 
\citet{PhysRevA.92.040303} originally proposed compiling a unitary as a sequence of Givens rotations, but relies on numerical optimization to identify the optimal displacement parameter. Since this optimization only requires identifying a single parameter at a time, namely that of the displacements, we use the \texttt{L-BFGS} optimizer from PyTorch. The maximum number of iterations is set to $100$ and the learning rate to $1.0$. We use the same number of repetitions as for the sparse ansatzes we propose, initializing the displacement parameter from small random values around zero. Here as well, we slice the rotation angle to improve the approximation and use the same set of values as for \citet{job2023efficient}. We find this strategy to be highly effective, as observed in \citet{job2023efficient} one can reach easily infidelities in the order of $10^{-8}$, but is also a very inefficient strategy, as it results in extremely long ansatzes, far beyond what can be achieved with even the \texttt{Full} ansatz. It should be noted, however, that these methods address a different trade-off, since they are designed for the compilation of general unitaries and prepare their components independently. Moreover, \citet{job2023efficient} is a direct compilation method that does not require any optimization and runs very fast, while our approach requires optimization but can produce much shorter ansatzes. Which method is preferable therefore depends on the use case.

\begin{table*}[ht]
    \centering
    \footnotesize
    \caption{Infidelity and resource requirements of the ansatz that reaches a target infidelity with the minimum number of phases, when optimizing the hyperparameters \textbf{penalizing the number of phases}. \textbf{Bold} values indicate sparse ansatzes that use fewer phases than \texttt{Full}, while the minimum number for each target infidelity is \underline{underlined}. A dash (-) indicates that none of the hyperparameter configurations achieved the target infidelity.
    }
    \begin{subfigure}[c]{\linewidth}
        \centering
        
                \input{tables_phases/mean_75/fourier_5/initial_0/min_phases_table_phases}
                \caption{Target state \emph{Fourier-5}.}
            \end{subfigure}\hfill
    \begin{subfigure}[c]{\linewidth}
        \centering
        
                \input{tables_phases/mean_75/random_gaussian/initial_0/min_phases_table_phases}

                \caption{Target state \emph{Haar random}.}
            \end{subfigure}\hfill  
    \label{tab:tables_phases/mean_75/phase_efficiency_table}
\end{table*}

\clearpage
\section{Fidelity-cost trade-off when optimizing the duration of the ansatz}
\label{app:efficiency_time}
\begin{figure*}[ht!]
    \centering
    \caption{Pareto frontiers showing the trade-off between the infidelity of the prepared state and the duration of the ansatz when optimizing the hyperparameters \textbf{penalizing the duration of the ansatz}. Each faded point represents a hyperparameter configuration, while solid lines connect the configurations that achieve the best trade-off (Pareto-optimal points) for each ansatz. 
        }
    \begin{subfigure}[c]{0.49\linewidth}
        \centering
        \includegraphics[width=1.0\linewidth]{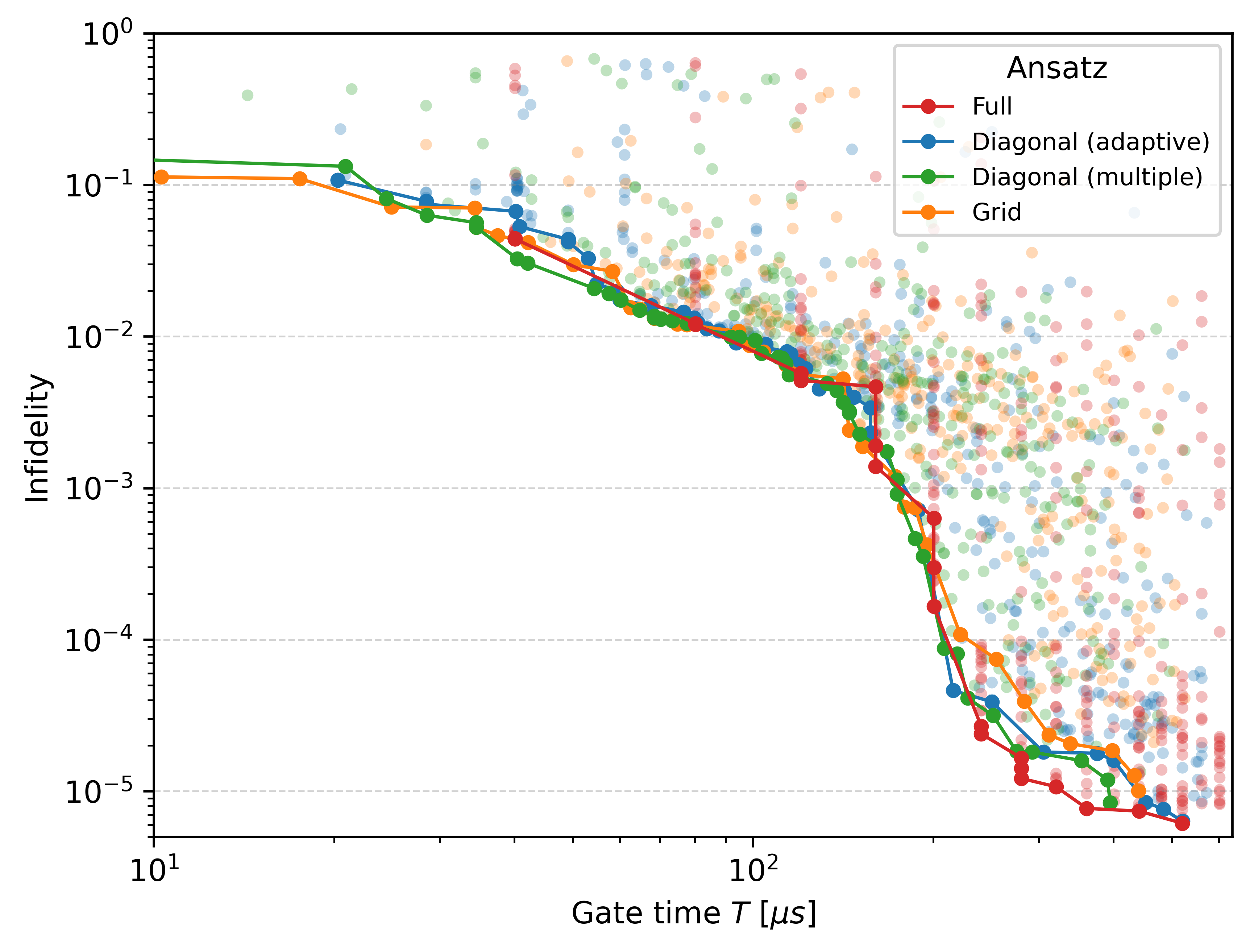}
        \caption{Target state \emph{Fourier-5} for $d=16$}
    \end{subfigure}\hfill
    \begin{subfigure}[c]{0.49\linewidth}
        \centering
        \includegraphics[width=1.0\linewidth]{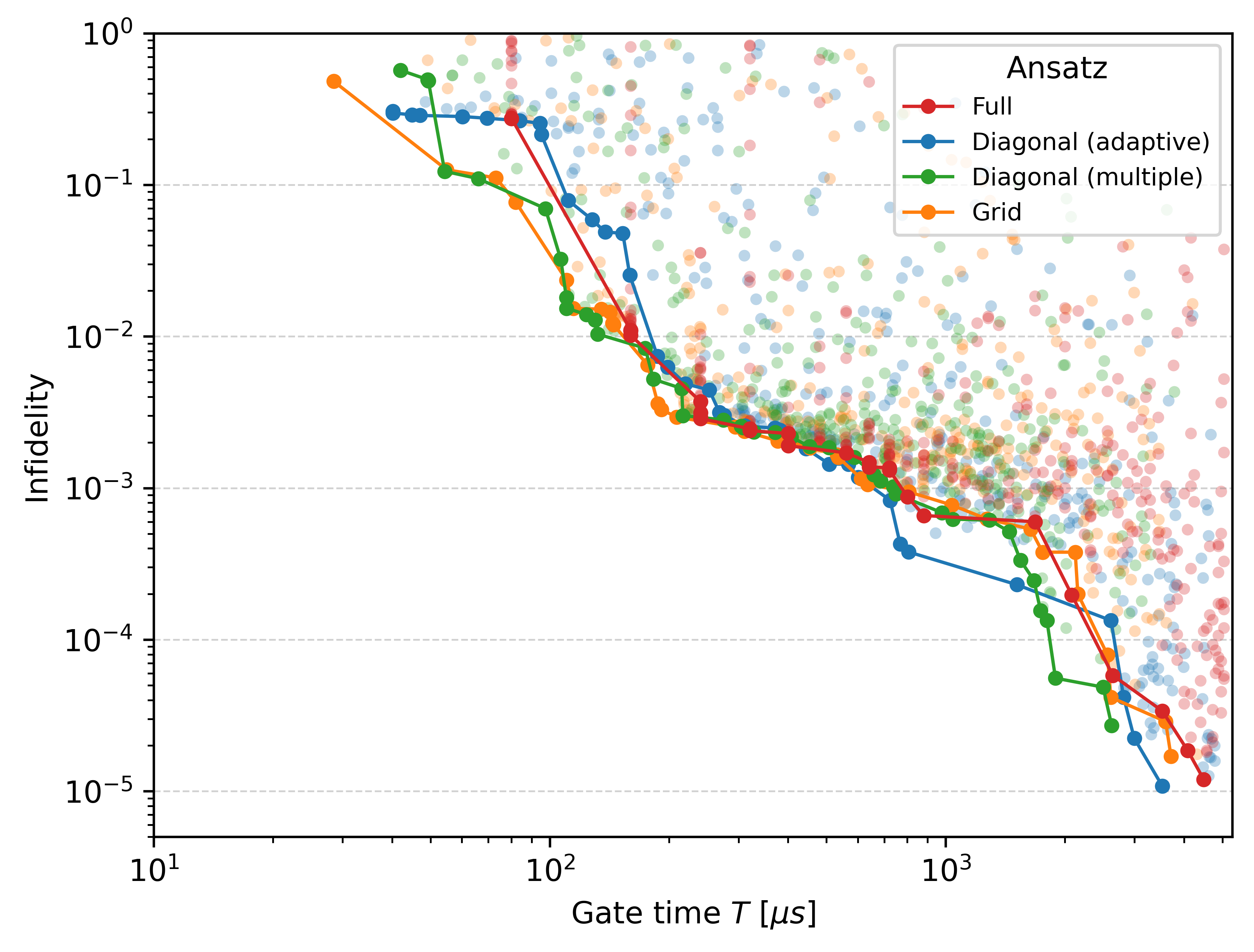}
        \caption{Target state \emph{Fourier-5} for $d=64$}
    \end{subfigure}\hfill    
    \begin{subfigure}[c]{0.49\linewidth}
        \centering
        \includegraphics[width=1.0\linewidth]{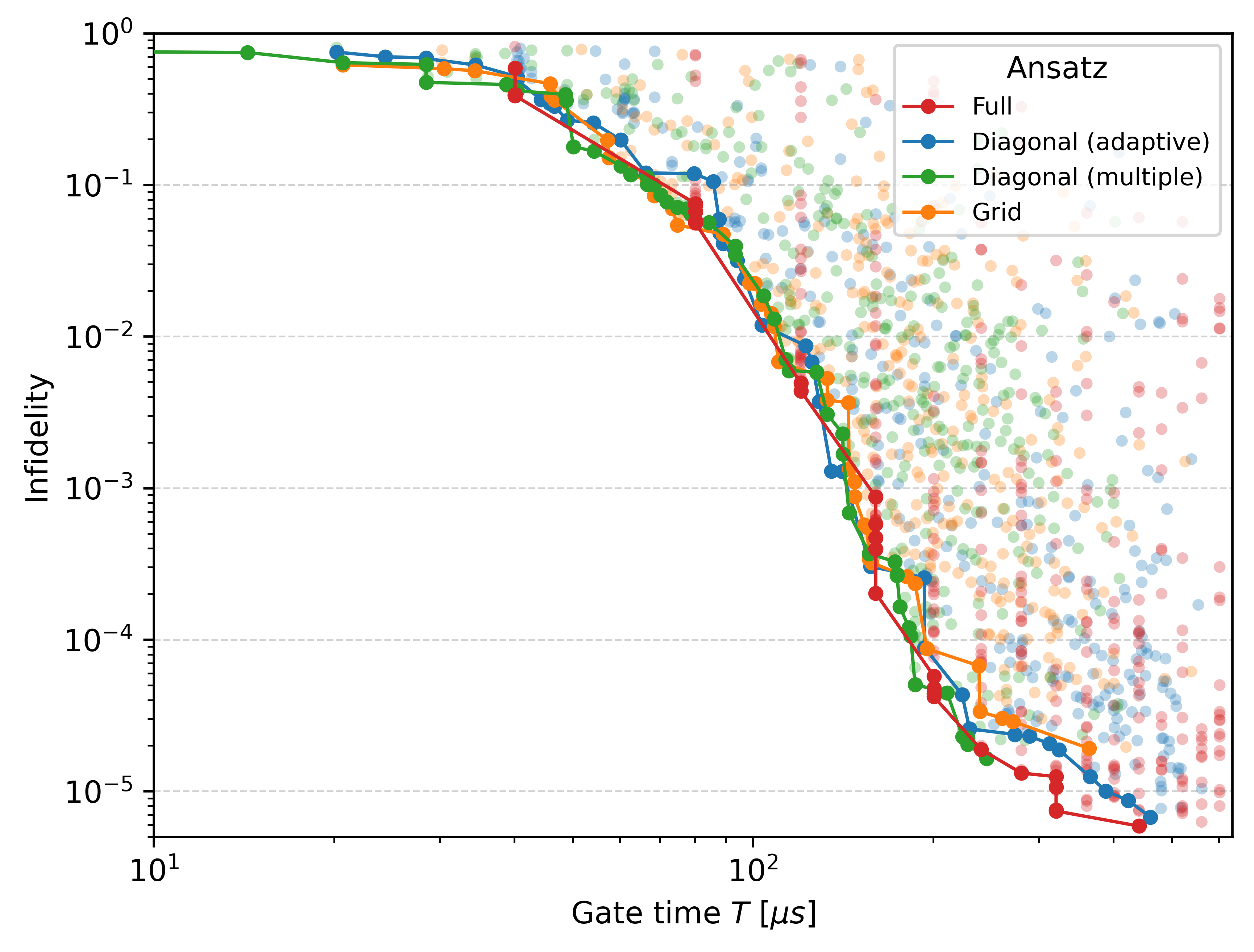}
        \caption{Target state \emph{Haar random} for $d=16$}
    \end{subfigure}\hfill
    \begin{subfigure}[c]{0.49\linewidth}
        \centering
        \includegraphics[width=1.0\linewidth]{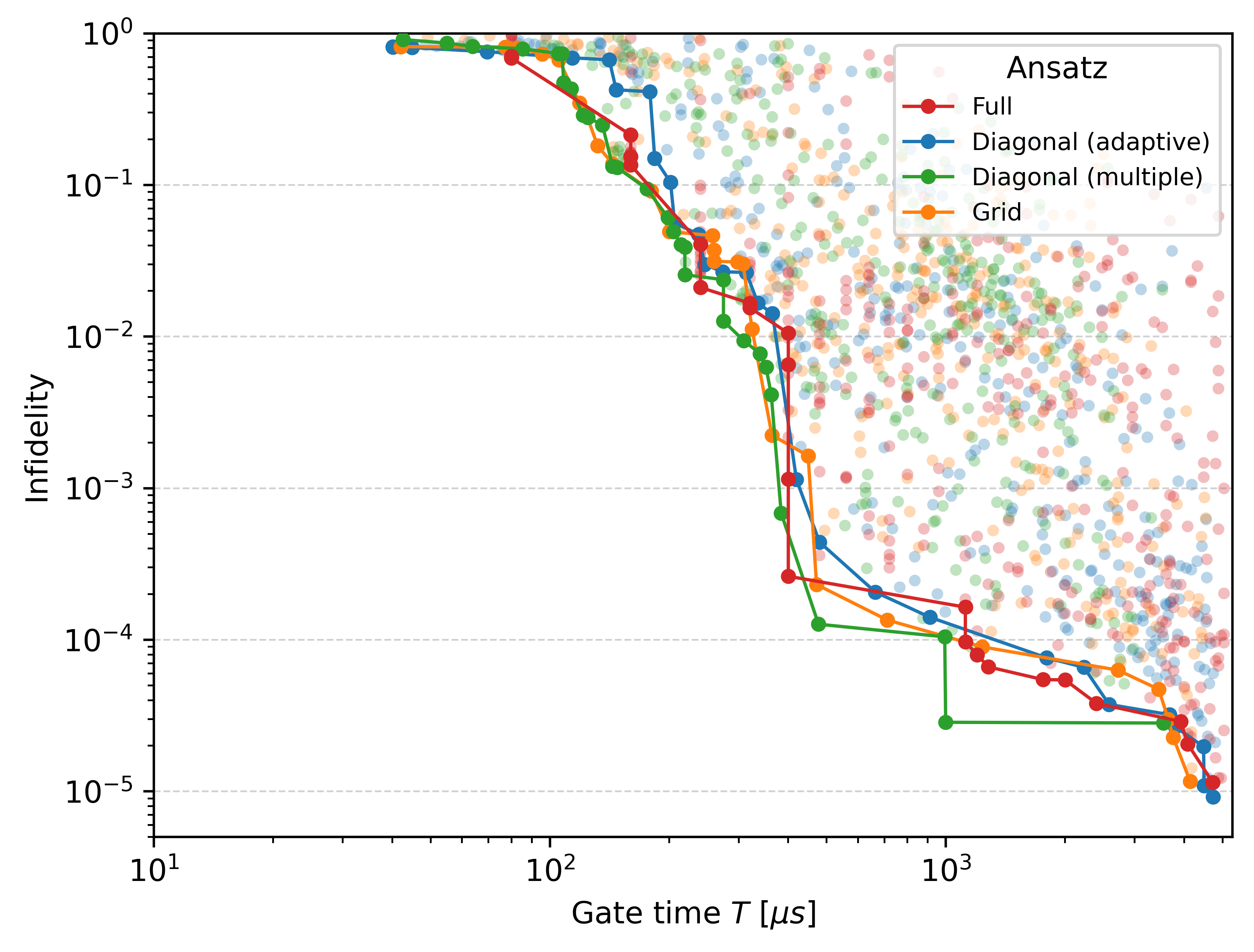}
        \caption{Target state \emph{Haar random} $d=64$}
    \end{subfigure}
    \label{fig:tables_time/efficiency_scatter_time}
\end{figure*}

\begin{figure*}[ht!]
    \centering
    \caption{Pareto frontiers showing the trade-off between the infidelity of the prepared state and the number of non-zero phase angles when optimizing the hyperparameters \textbf{penalizing the duration of the ansatz}. Each faded point represents a hyperparameter configuration, while solid lines connect the configurations that achieve the best trade-off (Pareto-optimal points) for each ansatz. 
        }
    \begin{subfigure}[c]{0.49\linewidth}
        \centering
        \includegraphics[width=1.0\linewidth]{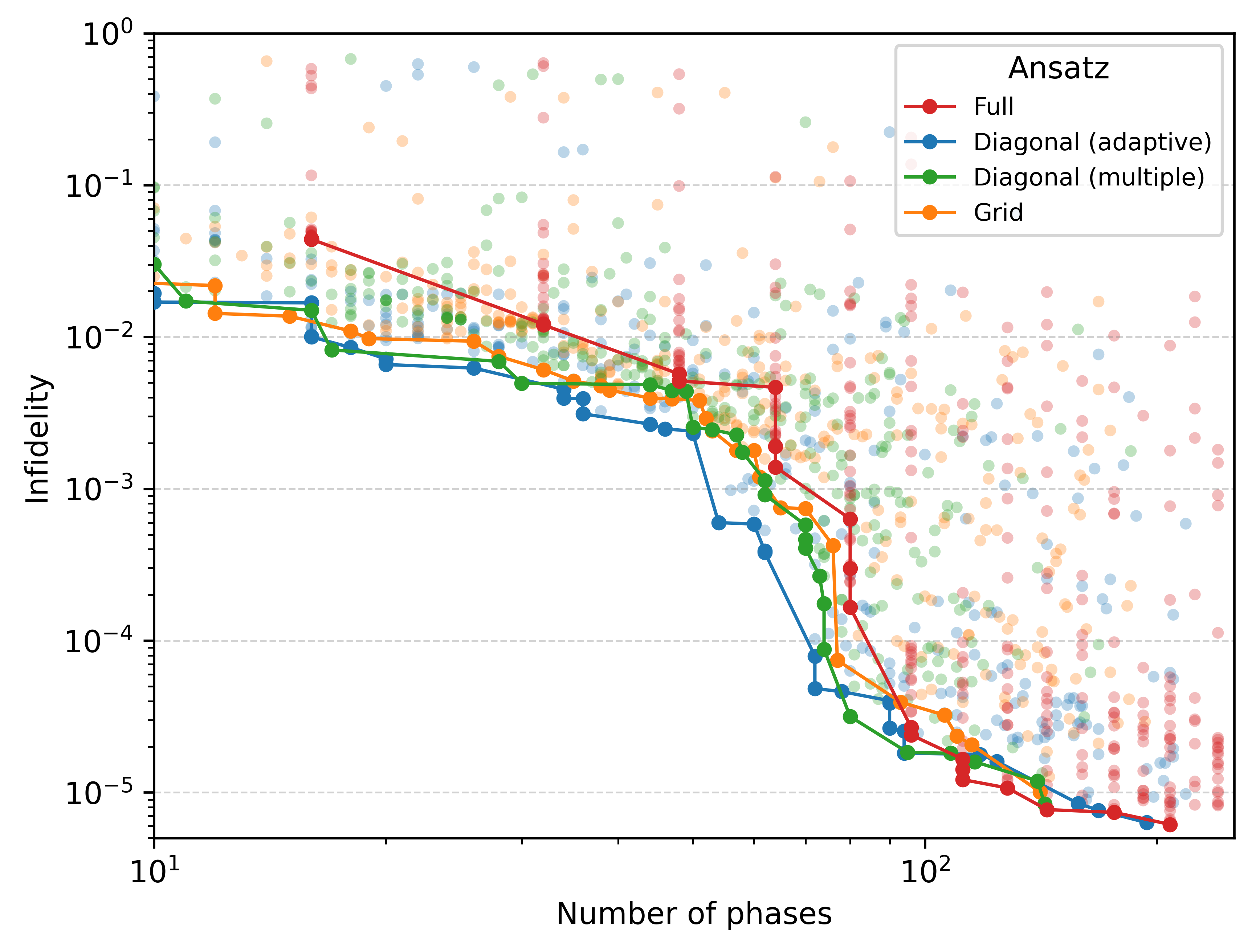}
        \caption{Target state \emph{Fourier-5} for $d=16$}
    \end{subfigure}\hfill
    \begin{subfigure}[c]{0.49\linewidth}
        \centering
        \includegraphics[width=1.0\linewidth]{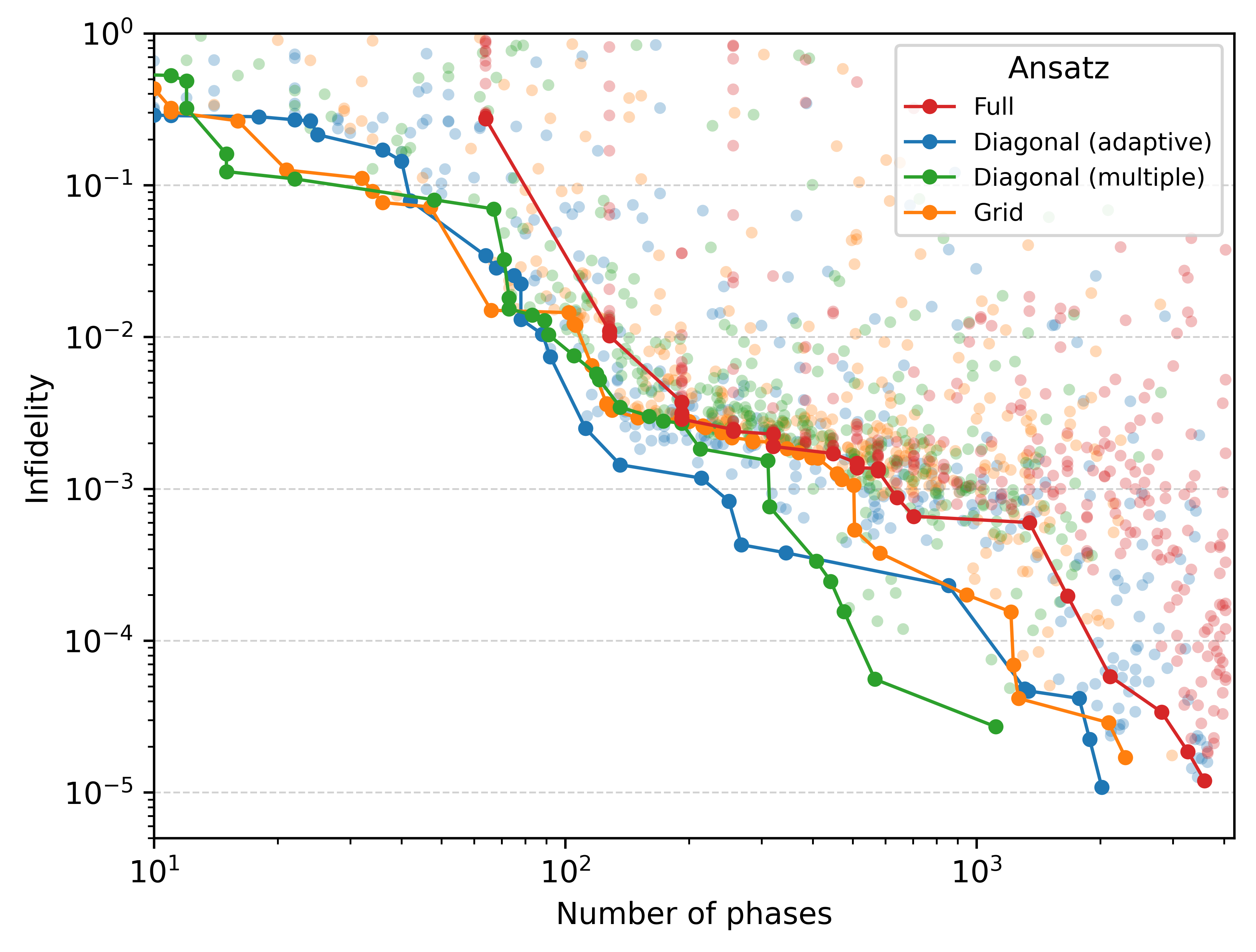}
        \caption{Target state \emph{Fourier-5} for $d=64$}
    \end{subfigure}\hfill    
    \begin{subfigure}[c]{0.49\linewidth}
        \centering
        \includegraphics[width=1.0\linewidth]{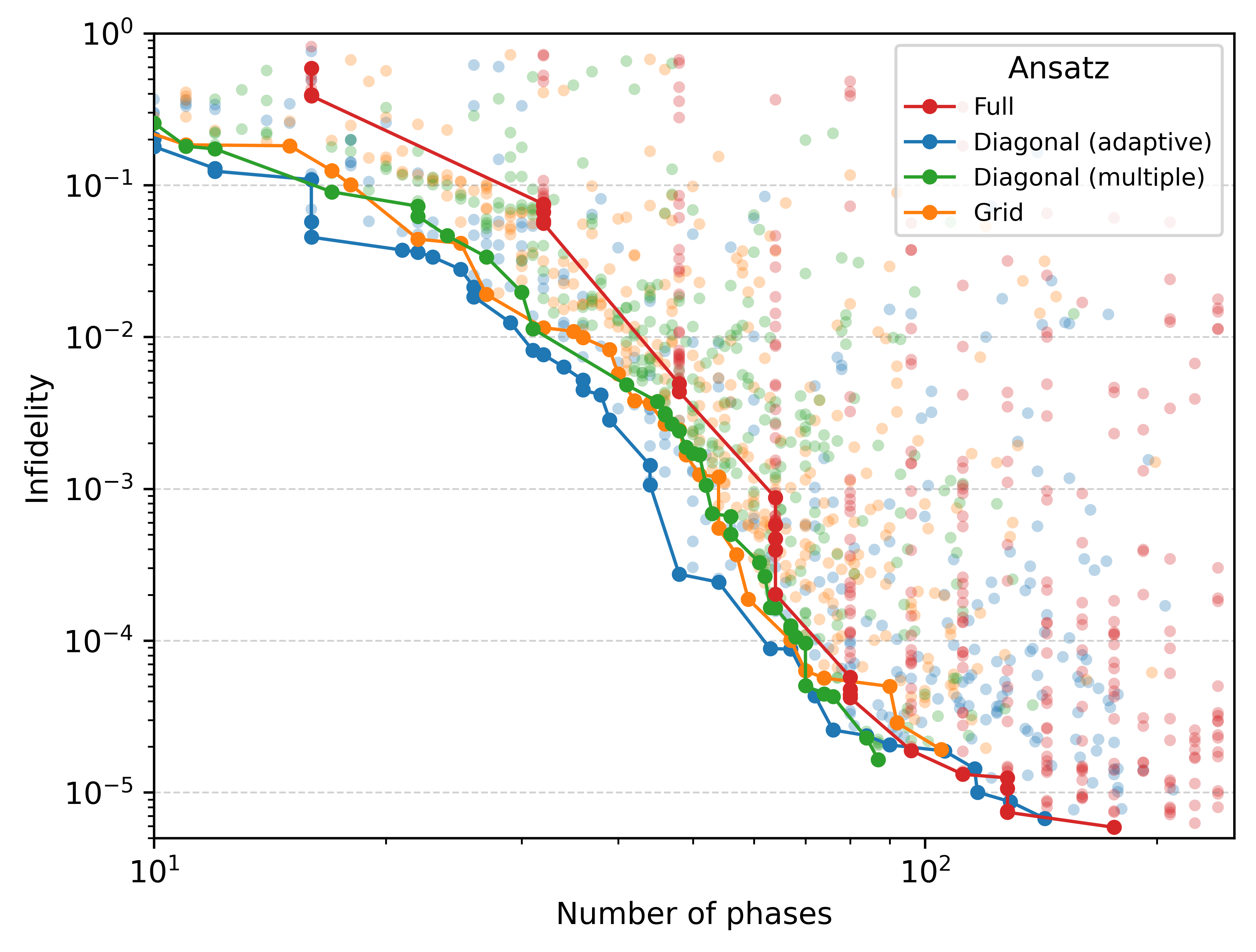}
        \caption{Target state \emph{Haar random} for $d=16$}
    \end{subfigure}\hfill
    \begin{subfigure}[c]{0.49\linewidth}
        \centering
        \includegraphics[width=1.0\linewidth]{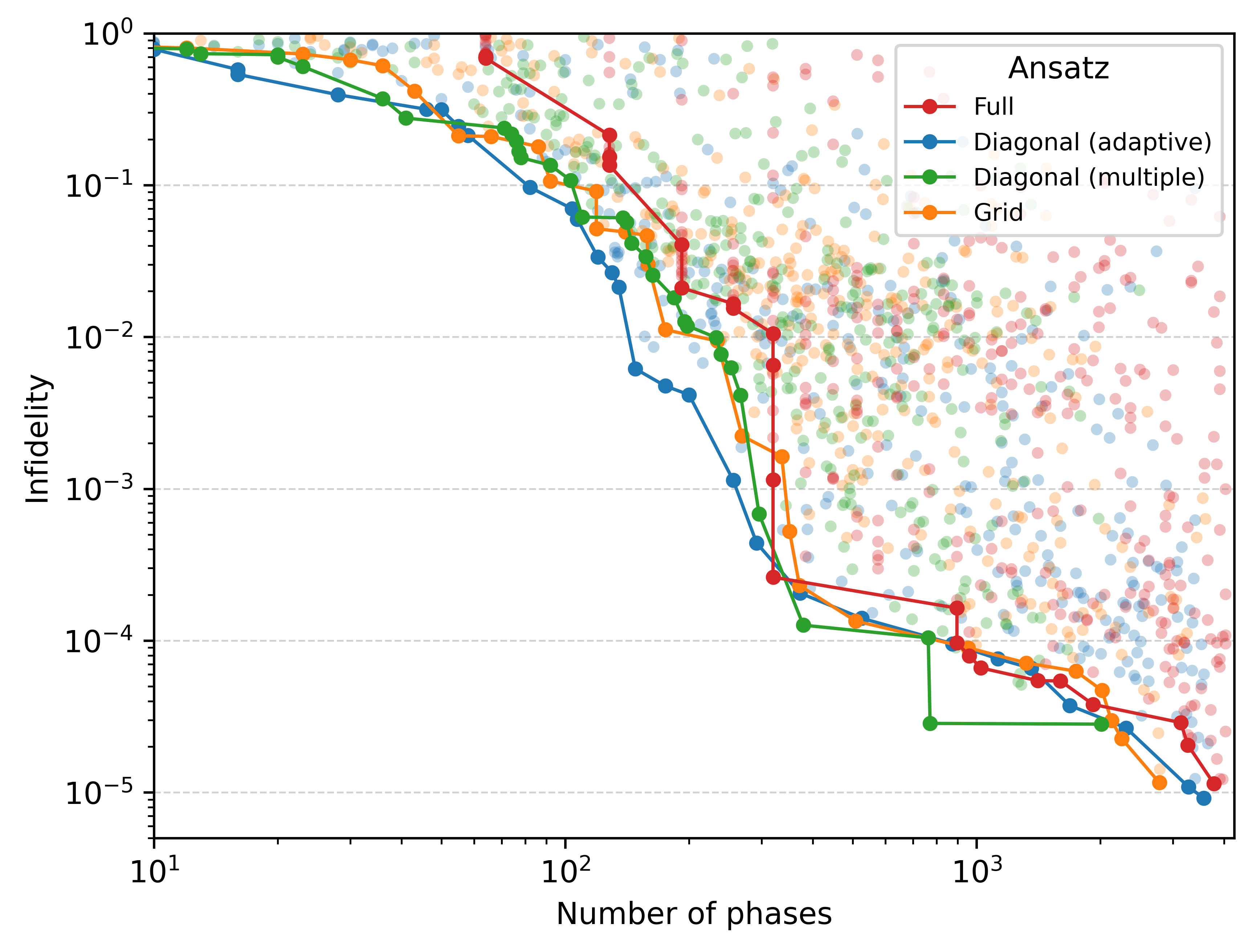}
        \caption{Target state \emph{Haar random} $d=64$}
    \end{subfigure}
    \label{fig:tables_time/efficiency_scatter_phases}
\end{figure*}

This section reports the full results on the efficiency of the state preparation for all ansatzes and qudit sizes $d=8,16,32$ and $64$, when minimizing the duration of the ansatz. The Pareto frontiers are shown in Figure \ref{fig:tables_time/efficiency_scatter_time} and the corresponding hypervolume is reported in Table \ref{tab:table_time/hypervolume}. The duration of the ansatzes required to prepare a state given a target infidelity are summarized in Table \ref{tab:tables_time/mean_75/phase_efficiency_table}. Lastly, the Pareto frontiers with respect to the number of phases (rather than the duration of the ansatz) are shown in Figure \ref{fig:tables_time/efficiency_scatter_phases}. While we do not conduct a detailed analysis of the optimal hyperparametes values, we can note that the pruning mask coefficient $\tau$ is generally below $0.8$, indicating only limited masking, when the optimization is carried out without penalizing the number of phases. When the penalization is introduced, the optimal value for $\tau$ rapidly drops to values in the range of $0.4-0.2$. While the behavior is not clearly monotonic, as a number of other hyperparameter influence the number of phases, those values indicate a rather strong pruning, which is consistent with the observation that applying phases to high Fock levels at early blocks is rather inefficient because the displacement gates have likely not yet populated them.

\begin{table*}[ht!]
    \centering
    \caption{Hypervolume comparing the trade–off between the infidelity of the prepared state and the duration of the ansatz when optimizing the hyperparameters \textbf{penalizing the duration of the ansatz}. Values are reported as the bootstrap mean hypervolume with 95\% confidence intervals. Larger values indicate a better trade-off. \textbf{Bold} values indicate better mean HV than \texttt{Full}, while the maximum mean HV for each $d$ is \underline{underlined}. For readability, we report $100 \times HV$.}
    \begin{subfigure}[c]{0.49\linewidth}
        \centering
        \setlength{\tabcolsep}{4pt}
        \renewcommand{\arraystretch}{2.5}
        \begin{tabular}{c|c|c|c|c}
        \toprule
        $d$ & \texttt{Full} 
            & \texttt{\shortstack{Diagonal\\(adaptive)}}  
            & \texttt{\shortstack{Diagonal\\(multiple)}}  
            & \texttt{Grid} \\
        \midrule
        8 &
        \shortstack{39.3\\{\scriptsize[38.9, 39.7]}} &
        \shortstack{\textbf{41.3}\\{\scriptsize[40.9, 41.7]}} &
        \shortstack{\underline{\textbf{43.0}}\\{\scriptsize[42.4, 43.7]}} &
        \shortstack{\textbf{40.4}\\{\scriptsize[40.0, 40.9]}} \\
        16 &
        \shortstack{32.5\\{\scriptsize[32.2, 32.8]}} &
        \shortstack{\textbf{33.5}\\{\scriptsize[33.0, 34.0]}} &
        \shortstack{\underline{\textbf{36.8}}\\{\scriptsize[36.3, 37.3]}} &
        \shortstack{\textbf{33.8}\\{\scriptsize[33.1, 34.4]}} \\
        32 &
        \shortstack{23.4\\{\scriptsize[23.1, 23.8]}} &
        \shortstack{\textbf{24.0}\\{\scriptsize[23.5, 24.5]}} &
        \shortstack{\textbf{23.7}\\{\scriptsize[22.9, 24.6]}} &
        \shortstack{\underline{\textbf{25.4}}\\{\scriptsize[25.0, 25.8]}} \\
        64 &
        \shortstack{19.6\\{\scriptsize[18.8, 20.9]}} &
        \shortstack{\underline{\textbf{22.0}}\\{\scriptsize[20.8, 23.9]}} &
        \shortstack{\textbf{21.7}\\{\scriptsize[21.3, 22.9]}} &
        \shortstack{\textbf{21.4}\\{\scriptsize[21.0, 22.1]}} \\
        \bottomrule
        \end{tabular}
        \caption{Target state \emph{Fourier-5}.}
    \end{subfigure}\hfill
    \begin{subfigure}[c]{0.49\linewidth}
        \centering
        \setlength{\tabcolsep}{4pt}
        \renewcommand{\arraystretch}{2.5}
        \begin{tabular}{c|c|c|c|c}
        \toprule
        $d$ & \texttt{Full} 
            & \texttt{\shortstack{Diagonal\\(adaptive)}}  
            & \texttt{\shortstack{Diagonal\\(multiple)}}  
            & \texttt{Grid} \\
        \midrule
        8 &
        \shortstack{39.0\\{\scriptsize[38.7, 39.4]}} &
        \shortstack{38.8\\{\scriptsize[38.3, 39.4]}} &
        \shortstack{\textbf{39.1}\\{\scriptsize[38.3, 40.1]}} &
        \shortstack{\underline{\textbf{40.7}}\\{\scriptsize[40.1, 41.2]}} \\
        16 &
        \shortstack{33.6\\{\scriptsize[33.3, 33.9]}} &
        \shortstack{\underline{\textbf{33.8}}\\{\scriptsize[33.4, 34.1]}} &
        \shortstack{33.1\\{\scriptsize[32.7, 33.6]}} &
        \shortstack{31.8\\{\scriptsize[31.3, 32.4]}} \\
        32 &
        \shortstack{28.6\\{\scriptsize[28.3, 29.0]}} &
        \shortstack{28.5\\{\scriptsize[28.2, 28.9]}} &
        \shortstack{28.5\\{\scriptsize[27.9, 29.3]}} &
        \shortstack{\underline{\textbf{29.8}}\\{\scriptsize[29.1, 30.5]}} \\
        64 &
        \shortstack{22.0\\{\scriptsize[21.1, 23.0]}} &
        \shortstack{20.9\\{\scriptsize[20.2, 21.9]}} &
        \shortstack{\underline{\textbf{22.5}}\\{\scriptsize[21.5, 23.4]}} &
        \shortstack{21.6\\{\scriptsize[20.6, 22.9]}} \\
        \bottomrule
        \end{tabular}
        \caption{Target state \emph{Haar random}.}
    \end{subfigure}
    \label{tab:table_time/hypervolume}
\end{table*}

\begin{table*}[ht]
    \centering
    \footnotesize
    \caption{Infidelity and resource requirements of the ansatz that reaches a target infidelity with the shortest duration, when optimizing the hyperparameters \textbf{penalizing the duration of the ansatz}. \textbf{Bold} values indicate sparse ansatzes that have a shorter duration than \texttt{Full}, while the minimum duration for each target infidelity is \underline{underlined}. A dash (-) indicates that none of the hyperparameter configurations achieved the target infidelity.
    }
    \begin{subfigure}[c]{\linewidth}
        \centering
        
                \input{tables_time/mean_75/fourier_5/initial_0/min_phases_table_time}
                \caption{Target state \emph{Fourier-5}.}
            \end{subfigure}\hfill
    \begin{subfigure}[c]{\linewidth}
        \centering
        
                \input{tables_time/mean_75/random_gaussian/initial_0/min_phases_table_time}
                \caption{Target state \emph{Haar random}.}
            \end{subfigure}\hfill  
    \label{tab:tables_time/mean_75/phase_efficiency_table}
\end{table*}

\clearpage
\section{Photon-Loss Noise}
\label{app:photon_loss}

This appendix presents additional results for simulations performed in the presence of photon-loss noise, where the SD gate parameters are optimized directly under \emph{noisy} conditions rather than for an idealized, noiseless model. 

Because quantum computing is an interdisciplinary field that brings together researchers with diverse backgrounds, we describe our noise model and numerical implementation in some detail to make the assumptions and approximations transparent, as well as our work more accessible. Section~\ref{sec:photonloss_model} introduces the photon-loss channel and the numerically stable formulation used in our simulations and Section~\ref{sec:time_estimation} explains how we estimate the duration of the ansatz.

\subsection{Photon-Loss Model}
\label{sec:photonloss_model}
Photon-loss is modeled as a trace-preserving quantum channel acting after each block of the SNAP displacement protocol. The channel is represented using a Kraus decomposition of the form:
\begin{equation}
\mathcal{E}(\rho) = \sum_{\ell=0}^{\ell_{\max}} K_\ell   \rho   K_\ell^\dagger
\end{equation}
where $K_\ell$ are the Kraus operators describing the loss of $\ell$ photons. 
Each Kraus operator is given by:
\begin{equation}
\label{eq:photon_loss:canonical}
K_\ell = \sqrt{\frac{(1 - \eta)^\ell}{\ell!}} \cdot \eta^{\hat{n}/2} \cdot \hat{a}^\ell,
\end{equation}
where $\eta = \exp( -\frac{T_{\text{block}}}{T_1} ) \in [0,1]$ is the photon survival probability over the duration of one block, with $T_1$ the photon-number (energy) lifetime of the cavity, $T_{\text{block}}$ is the duration of the block, $\hat{a}$ is the annihilation operator, and $\hat{n} = \hat{a}^\dagger \hat{a}$ is the number operator. Recall that we run our simulations in a truncated Hilbert space of size $\tilde d$, corresponding to the controllable qudit levels $d$ plus the bumper states. In this truncated model, $\hat a^\ell=0$ for all $\ell\ge \tilde d$, which bounds the maximum number of photons that can be lost to $\ell=\ell_{\max}=\tilde d-1$.

The operators $K_\ell$ are applied directly to the density matrix $\rho$ at each block of the ansatz:
\begin{align*}
L(\alpha, \vec{\theta}) &= D(\alpha)   S(\vec{\theta})   D^\dagger(\alpha), \\
\rho_{b+1} &= \mathcal{E}\left( L(\alpha_b, \vec{\theta}_b)   \rho_b   L^\dagger(\alpha_b, \vec{\theta}_b) \right),
\quad \text{for } b = 1, \dots, B.
\end{align*}

This canonical form has the major drawback of requiring to multiply numbers which tend to differ by tens of orders of magnitude, \idest a very small scalar coefficient and a very large $\hat{a}^\ell$ which for $\ell>10$ and $d>64$ easily differ by more than 50 orders of magnitude. Due to this, large numerical errors accumulate resulting in the trace of the final density matrix contracting by around $10^{-7}$ at each block. Since for smaller qudit sizes we can achieve infidelities on the order of $10^{-6}$, we found this numerical error too large. We rely therefore on an equivalent but much more numerically stable formulation. 

Using the action of $\hat a^\ell$ on number states and the fact that $\eta^{\hat n/2}$ is diagonal, one can write each Kraus operator directly in the Fock basis without forming powers of $\hat a$:
\begin{equation}
K_\ell  |m\rangle
= \sqrt{\binom{m}{\ell}} (1-\eta)^{\ell/2} \eta^{(m-\ell)/2} |m-\ell\rangle
\quad (\text{for } m\ge \ell),
\end{equation}
and $K_\ell|m\rangle=0$ for $m<\ell$. Equivalently, the only nonzero matrix elements of $K_\ell$ lie on the $\ell$-th subdiagonal:
\begin{equation}
\big(K_\ell\big)_{m-\ell, m}
= \sqrt{\binom{m}{\ell}} (1-\eta)^{\ell/2} \eta^{(m-\ell)/2},
\qquad m=\ell,\dots,d-1,
\end{equation}
with all other entries equal to zero. This representation is algebraically equivalent to the canonical form of Eq. \ref{eq:photon_loss:canonical}. 
For further improved numerical stability, the $K_\ell$ terms are computed in the logarithmic domain transforming products into summations. Overall, this formulation preserves the trace up to machine precision, which for double-precision floating-point arithmetic is on the order of $10^{-15}$.

\subsection{Gate Duration Estimation}
\label{sec:time_estimation}
The simulation of photon-loss noise requires to estimate the duration of each block, $T_{\text{block}}$. The duration of a displacement gate is very short compared to the cavity lifetime and can be safely assumed to be $T_{D}\simeq 0.1~\mu\mathrm s$ \cite{fosel2020efficient}. The duration of a SNAP gate instead depends strongly on the number of phases that need to be applied because the available drive power must be distributed across all frequencies. For current implementations the SNAP duration scales approximately as
$T_S \propto \sqrt{d}$, with a recent study reporting $T_S \approx 10\sqrt{d} \ \mu\text{s}$ \cite{bornman2025benchmarking}, which is the value we will use. Since in our ansatzes the number of phases optimized by the SNAP gates is not fixed to $d$ but rather changes between architectures and blocks, the duration of the SNAP gates is computed at each block based on how many phases should be applied.

\subsection{Results}
In this section we report the results obtained when learning the gate parameters under noisy conditions. Due to the very large computational cost of running the numerical simulations, in particular the hyperparameter optimization phase, we choose to rely on the optimal hyperparameters identified in noiseless conditions. Note, however, that as the underlying conditions shift, \idest when we add photon-loss noise, it is likely that the optimal hyperparameters for a given ansatz will change as well.  Running a separate hyperparameter optimization for each value of $T_1$ would be unfeasible, and moreover our goal is to develop a strategy that is easy to deploy, reliable, and robust to changing conditions. Indeed, in real experiments the noise parameters will shift over time. A comparison performed by optimizing the hyperparameters on small qudit sizes indicated that in that region the conclusions remain consistent.
The results when using the hyperparameter configuration penalizing the number of phases are reported in Table \ref{photon_loss/tables_phases/mean_75/fourier_5/initial_0/photon_loss_table} (Fourier-5) and Table \ref{photon_loss/tables_phases/mean_75/random_gaussian/initial_0/photon_loss_table} (Haar random). The results when using the hyperparameter configuration penalizing the duration of the ansatz are instead reported in Table \ref{photon_loss/tables_time/mean_75/fourier_5/initial_0/photon_loss_table} (Fourier-5) and Table \ref{photon_loss/tables_time/mean_75/random_gaussian/initial_0/photon_loss_table} (Haar random). 

As expected, these results exhibit a higher infidelity compared to those obtained in noiseless conditions, especially for low values of $T_1$ and for large qudit sizes, \idest $d = 64$. 
Given that our main goal is to assess whether the sparse ansatzes allow us to obtain a better trade-off between prepared-state infidelity and required resources also in noisy conditions, in Table \ref{tab:photon_loss/hypervolume} we report the hypervolume computed when minimizing the number of phases, by collecting the results obtained with the optimal hyperparameters corresponding to different cost penalties $\beta$. Note that these results are based on a more limited set of data points compared to those that use all individual hyperparameter tuning trials, and are therefore noisier. 
The results indicate again that the sparse ansatzes exhibit a better trade-off compared to \texttt{Full} in most settings, however, this behavior depends strongly on the target-state family. For the Fourier-5 states, excluding the very small $d=8$ case, the sparse ansatz with the highest mean HV is \texttt{Grid} at $d=32$ and $d=64$, and also always well separated from the other sparse ansatzes in terms of its 95\% confidence intervals. For $d=16$, instead, the highest mean HV is given by \texttt{Diagonal (multiple)}, while \texttt{Grid} is slightly behind but with overlapping confidence intervals. For the Haar random states, as observed for the ideal noiseless setting, the picture becomes progressively less clear as $d$ increases. \texttt{Diagonal (adaptive)} is best at $d=8$, \texttt{Grid} is most often best at $d=16$, while at $d=32$ and especially $d=64$ the differences between sparse ansatzes become less consistent and their confidence intervals overlap more frequently. This confirms that it is more difficult, as could be expected, to consistently reduce the number of phases required to prepare Haar random target states, and the presence of noise further increases this challenge.
Overall, excluding the few ties with \texttt{Full} and counting across all combinations of target state, $d$, and $T_1$, \texttt{Diagonal (adaptive)} exhibits a better trade-off than \texttt{Full} in 59\% of cases, \texttt{Diagonal (multiple)} in 78\%, and \texttt{Grid} in 81\%. Including ties, \texttt{Diagonal (adaptive)} is at least as efficient as \texttt{Full} in 63\% of cases, \texttt{Diagonal (multiple)} in 88\%, and \texttt{Grid} in 84\% of cases.

The hypervolume computed when minimizing the duration of the ansatz is reported in Table \ref{tab:photon_loss_time/hypervolume}, and again the results are broadly consistent in indicating that this is a more challenging optimization problem. The sparse ansatzes provide a competitive trade-off most of the time, but with some exceptions and without a clear winner among the sparse ansatzes. In particular, excluding the few ties with \texttt{Full}, \texttt{Diagonal (adaptive)} exhibits a better trade-off than \texttt{Full} in 28\% of cases, \texttt{Diagonal (multiple)} in 62\%, and \texttt{Grid} in 62\%. The results also show clearly how strong the impact of a small $T_1$ is in degrading fidelity, especially for the Haar random states.

Overall, these results indicate that, even under photon-loss noise, sparse ansatzes provide a competitive and often better trade-off compared to \texttt{Full}, especially when penalizing the number of phases, and can therefore be a useful tool to minimize the number of control operations that need to be applied and to reduce the complexity of the SNAP pulses, thereby helping to reduce the error introduced by control operations.

\begin{table*}[ht!]
    \centering
    \caption{Hypervolume comparing the trade–off between the infidelity of the prepared state and the number of non-zero phase angles when optimizing the hyperparameters \textbf{penalizing the number of phases}. Values are reported as the bootstrap mean hypervolume with 95\% confidence intervals. Larger values indicate a better trade-off. \textbf{Bold} values indicate better mean HV than \texttt{Full}, while the maximum mean HV for each $d$ and $T_1$ is \underline{underlined}. For readability, we report $100 \times HV$.}
    \begin{subfigure}[c]{0.49\linewidth}
        \centering
        \footnotesize
        \setlength{\tabcolsep}{4pt}
        \renewcommand{\arraystretch}{2.2}
        \begin{tabular}{ll|c|c|c|c}
        \toprule
        $d$ & \shortstack{$T_1$\\ {[ms]}} & \texttt{Full} 
            & \texttt{\shortstack{Diagonal\\(adaptive)}}  
            & \texttt{\shortstack{Diagonal\\(multiple)}}  
            & \texttt{Grid} \\
            \midrule
            \multirow[c]{4}{*}{8} & $10^{-1} $ & \shortstack{\underline{27.0}\\{\scriptsize[26.5, 27.6]}} & \shortstack{20.5\\{\scriptsize[19.7, 21.5]}} & \shortstack{26.1\\{\scriptsize[25.6, 26.9]}} & \shortstack{26.6\\{\scriptsize[26.2, 27.0]}} \\
             & $10^{0} $ & \shortstack{\underline{39.2}\\{\scriptsize[38.7, 39.7]}} & \shortstack{33.7\\{\scriptsize[33.1, 34.3]}} & \shortstack{38.8\\{\scriptsize[38.4, 39.2]}} & \shortstack{39.0\\{\scriptsize[38.6, 39.5]}} \\
             & $10^{1} $ & \shortstack{58.2\\{\scriptsize[57.8, 58.8]}} & \shortstack{55.2\\{\scriptsize[54.7, 55.7]}} & \shortstack{58.2\\{\scriptsize[57.4, 58.8]}} & \shortstack{\underline{\textbf{60.1}}\\{\scriptsize[59.5, 60.7]}} \\
             & $10^{2} $ & \shortstack{77.0\\{\scriptsize[76.4, 77.5]}} & \shortstack{\textbf{77.1}\\{\scriptsize[76.2, 78.1]}} & \shortstack{74.9\\{\scriptsize[74.1, 75.8]}} & \shortstack{\underline{\textbf{79.3}}\\{\scriptsize[78.6, 80.0]}} \\
            \midrule
            \multirow[c]{4}{*}{16} & $10^{-1} $ & \shortstack{18.5\\{\scriptsize[18.5, 18.5]}} & \shortstack{18.5\\{\scriptsize[18.5, 18.5]}} & \shortstack{\underline{\textbf{22.8}}\\{\scriptsize[22.6, 23.1]}} & \shortstack{\textbf{22.8}\\{\scriptsize[22.7, 22.8]}} \\
             & $10^{0} $ & \shortstack{29.5\\{\scriptsize[29.4, 29.6]}} & \shortstack{\textbf{30.3}\\{\scriptsize[30.0, 30.5]}} & \shortstack{\underline{\textbf{35.1}}\\{\scriptsize[34.7, 35.5]}} & \shortstack{\textbf{35.1}\\{\scriptsize[34.9, 35.3]}} \\
             & $10^{1} $ & \shortstack{41.9\\{\scriptsize[41.6, 42.2]}} & \shortstack{\textbf{45.7}\\{\scriptsize[45.3, 46.1]}} & \shortstack{\underline{\textbf{49.1}}\\{\scriptsize[48.7, 49.5]}} & \shortstack{\textbf{48.5}\\{\scriptsize[48.1, 48.9]}} \\
             & $10^{2} $ & \shortstack{56.6\\{\scriptsize[56.3, 56.9]}} & \shortstack{\textbf{61.9}\\{\scriptsize[61.5, 62.4]}} & \shortstack{\underline{\textbf{64.5}}\\{\scriptsize[64.1, 64.9]}} & \shortstack{\textbf{63.5}\\{\scriptsize[63.1, 64.0]}} \\
            \midrule
            \multirow[c]{4}{*}{32} & $10^{-1} $ & \shortstack{8.8\\{\scriptsize[8.8, 8.8]}} & \shortstack{\textbf{9.6}\\{\scriptsize[9.6, 9.7]}} & \shortstack{\textbf{10.1}\\{\scriptsize[10.1, 10.2]}} & \shortstack{\underline{\textbf{12.9}}\\{\scriptsize[12.8, 13.0]}} \\
             & $10^{0} $ & \shortstack{17.0\\{\scriptsize[17.0, 17.1]}} & \shortstack{\textbf{17.5}\\{\scriptsize[17.4, 17.6]}} & \shortstack{\textbf{19.0}\\{\scriptsize[18.8, 19.1]}} & \shortstack{\underline{\textbf{24.7}}\\{\scriptsize[24.5, 24.9]}} \\
             & $10^{1} $ & \shortstack{27.7\\{\scriptsize[27.6, 27.8]}} & \shortstack{\textbf{30.2}\\{\scriptsize[29.8, 30.6]}} & \shortstack{\textbf{30.5}\\{\scriptsize[30.1, 30.9]}} & \shortstack{\underline{\textbf{37.8}}\\{\scriptsize[37.5, 38.1]}} \\
             & $10^{2} $ & \shortstack{38.2\\{\scriptsize[38.0, 38.3]}} & \shortstack{\textbf{42.0}\\{\scriptsize[41.7, 42.3]}} & \shortstack{\textbf{40.8}\\{\scriptsize[40.4, 41.2]}} & \shortstack{\underline{\textbf{48.7}}\\{\scriptsize[48.4, 48.9]}} \\
            \midrule
            \multirow[c]{4}{*}{64} & $10^{-1} $ & \shortstack{4.4\\{\scriptsize[4.3, 4.4]}} & \shortstack{\textbf{4.8}\\{\scriptsize[4.8, 4.9]}} & \shortstack{\textbf{4.8}\\{\scriptsize[4.8, 4.9]}} & \shortstack{\underline{\textbf{7.0}}\\{\scriptsize[6.6, 7.1]}} \\
             & $10^{0} $ & \shortstack{9.2\\{\scriptsize[8.9, 9.4]}} & \shortstack{8.6\\{\scriptsize[8.2, 8.9]}} & \shortstack{\textbf{9.3}\\{\scriptsize[8.9, 9.7]}} & \shortstack{\underline{\textbf{14.6}}\\{\scriptsize[13.6, 15.0]}} \\
             & $10^{1} $ & \shortstack{17.7\\{\scriptsize[16.6, 18.6]}} & \shortstack{\textbf{18.8}\\{\scriptsize[17.2, 19.2]}} & \shortstack{\textbf{19.6}\\{\scriptsize[18.9, 20.3]}} & \shortstack{\underline{\textbf{26.4}}\\{\scriptsize[24.2, 27.5]}} \\
             & $10^{2} $ & \shortstack{28.0\\{\scriptsize[27.6, 28.4]}} & \shortstack{\textbf{28.8}\\{\scriptsize[25.0, 29.9]}} & \shortstack{\textbf{30.6}\\{\scriptsize[30.0, 31.3]}} & \shortstack{\underline{\textbf{33.8}}\\{\scriptsize[32.5, 34.9]}} \\
            \bottomrule
            \end{tabular}
         \caption{Target state \emph{Fourier-5}.}
    \end{subfigure}\hfill
    \begin{subfigure}[c]{0.49\linewidth}
        \centering
        \footnotesize
        \setlength{\tabcolsep}{4pt}
        \renewcommand{\arraystretch}{2.2}
        \begin{tabular}{ll|c|c|c|c}
        \toprule
        $d$ & \shortstack{$T_1$\\ {[ms]}} & \texttt{Full} 
            & \texttt{\shortstack{Diagonal\\(adaptive)}}  
            & \texttt{\shortstack{Diagonal\\(multiple)}}  
            & \texttt{Grid} \\
        \midrule
        \multirow[c]{4}{*}{8} & $10^{-1} $ & \shortstack{8.2\\{\scriptsize[8.1, 8.2]}} & \shortstack{\underline{\textbf{9.3}}\\{\scriptsize[9.2, 9.4]}} & \shortstack{\textbf{8.9}\\{\scriptsize[8.9, 8.9]}} & \shortstack{\textbf{9.1}\\{\scriptsize[8.9, 9.2]}} \\
         & $10^{0} $ & \shortstack{23.9\\{\scriptsize[23.7, 24.2]}} & \shortstack{\underline{\textbf{27.5}}\\{\scriptsize[27.1, 27.9]}} & \shortstack{\textbf{25.5}\\{\scriptsize[25.0, 25.9]}} & \shortstack{\textbf{25.6}\\{\scriptsize[25.3, 25.9]}} \\
         & $10^{1} $ & \shortstack{52.4\\{\scriptsize[52.1, 52.7]}} & \shortstack{\underline{\textbf{57.4}}\\{\scriptsize[56.7, 58.1]}} & \shortstack{\textbf{55.4}\\{\scriptsize[54.8, 56.0]}} & \shortstack{\textbf{55.7}\\{\scriptsize[55.2, 56.2]}} \\
         & $10^{2} $ & \shortstack{80.1\\{\scriptsize[79.4, 80.7]}} & \shortstack{\underline{\textbf{85.7}}\\{\scriptsize[84.5, 86.9]}} & \shortstack{\textbf{84.5}\\{\scriptsize[83.7, 85.3]}} & \shortstack{\textbf{85.2}\\{\scriptsize[84.5, 85.9]}} \\
        \midrule
        \multirow[c]{4}{*}{16} & $10^{-1} $ & \shortstack{2.9\\{\scriptsize[2.8, 2.9]}} & \shortstack{2.8\\{\scriptsize[2.8, 2.9]}} & \shortstack{\textbf{3.5}\\{\scriptsize[3.5, 3.6]}} & \shortstack{\underline{\textbf{3.9}}\\{\scriptsize[3.9, 3.9]}} \\
         & $10^{0} $ & \shortstack{12.5\\{\scriptsize[12.3, 12.6]}} & \shortstack{11.8\\{\scriptsize[11.7, 12.0]}} & \shortstack{\textbf{13.9}\\{\scriptsize[13.8, 14.1]}} & \shortstack{\underline{\textbf{14.4}}\\{\scriptsize[14.2, 14.6]}} \\
         & $10^{1} $ & \shortstack{34.2\\{\scriptsize[33.9, 34.4]}} & \shortstack{\textbf{35.5}\\{\scriptsize[35.2, 35.8]}} & \shortstack{\textbf{37.1}\\{\scriptsize[36.8, 37.5]}} & \shortstack{\underline{\textbf{37.8}}\\{\scriptsize[37.3, 38.2]}} \\
         & $10^{2} $ & \shortstack{57.9\\{\scriptsize[57.4, 58.3]}} & \shortstack{\textbf{60.8}\\{\scriptsize[60.2, 61.4]}} & \shortstack{\underline{\textbf{61.5}}\\{\scriptsize[61.0, 62.0]}} & \shortstack{\textbf{60.4}\\{\scriptsize[59.7, 61.1]}} \\
        \midrule
        \multirow[c]{4}{*}{32} & $10^{-1} $ & \shortstack{1.2\\{\scriptsize[1.2, 1.2]}} & \shortstack{1.1\\{\scriptsize[1.1, 1.1]}} & \shortstack{\textbf{1.3}\\{\scriptsize[1.3, 1.3]}} & \shortstack{\underline{\textbf{1.4}}\\{\scriptsize[1.3, 1.4]}} \\
         & $10^{0} $ & \shortstack{\underline{5.0}\\{\scriptsize[5.0, 5.0]}} & \shortstack{3.0\\{\scriptsize[3.0, 3.0]}} & \shortstack{\underline{5.0}\\{\scriptsize[5.0, 5.1]}} & \shortstack{\underline{5.0}\\{\scriptsize[4.9, 5.0]}} \\
         & $10^{1} $ & \shortstack{18.9\\{\scriptsize[18.7, 19.1]}} & \shortstack{18.2\\{\scriptsize[18.1, 18.3]}} & \shortstack{\textbf{19.0}\\{\scriptsize[18.8, 19.2]}} & \shortstack{\underline{\textbf{19.2}}\\{\scriptsize[19.0, 19.4]}} \\
         & $10^{2} $ & \shortstack{37.6\\{\scriptsize[37.1, 38.1]}} & \shortstack{\underline{\textbf{40.3}}\\{\scriptsize[40.1, 40.5]}} & \shortstack{\textbf{38.8}\\{\scriptsize[38.6, 39.0]}} & \shortstack{\textbf{38.8}\\{\scriptsize[38.4, 39.2]}} \\
        \midrule
        \multirow[c]{4}{*}{64} & $10^{-1} $ & \shortstack{0.9\\{\scriptsize[0.7, 0.9]}} & \shortstack{\textbf{1.0}\\{\scriptsize[0.9, 1.0]}} & \shortstack{\textbf{1.0}\\{\scriptsize[0.8, 1.0]}} & \shortstack{\underline{\textbf{1.2}}\\{\scriptsize[1.2, 1.2]}} \\
         & $10^{0} $ & \shortstack{\underline{1.9}\\{\scriptsize[1.8, 1.9]}} & \shortstack{1.2\\{\scriptsize[1.2, 1.3]}} & \shortstack{\underline{1.9}\\{\scriptsize[1.8, 1.9]}} & \shortstack{1.8\\{\scriptsize[1.6, 1.9]}} \\
         & $10^{1} $ & \shortstack{\underline{8.4}\\{\scriptsize[8.2, 8.6]}} & \shortstack{6.6\\{\scriptsize[6.4, 6.8]}} & \shortstack{7.9\\{\scriptsize[7.8, 8.1]}} & \shortstack{7.2\\{\scriptsize[7.1, 7.4]}} \\
         & $10^{2} $ & \shortstack{20.9\\{\scriptsize[19.8, 21.7]}} & \shortstack{20.7\\{\scriptsize[19.2, 22.2]}} & \shortstack{\underline{\textbf{21.1}}\\{\scriptsize[20.2, 21.9]}} & \shortstack{18.7\\{\scriptsize[18.1, 19.2]}} \\
        \bottomrule
        \end{tabular}
        \caption{Target state \emph{Haar random}.}
    \end{subfigure}
    \label{tab:photon_loss/hypervolume}    
\end{table*}

\begin{table*}[ht!]
    \centering
    \caption{Hypervolume comparing the trade–off between the infidelity of the prepared state and the duration of the ansatz when optimizing the hyperparameters \textbf{penalizing the duration of the ansatz}. Values are reported as the bootstrap mean hypervolume with 95\% confidence intervals. Larger values indicate a better trade-off. \textbf{Bold} values indicate better mean HV than \texttt{Full}, while the maximum mean HV for each $d$ and $T_1$ is \underline{underlined}. For readability, we report $100 \times HV$.}
    \begin{subfigure}[c]{0.49\linewidth}
        \centering
        \footnotesize
        \setlength{\tabcolsep}{4pt}
        \renewcommand{\arraystretch}{2.2}
        \begin{tabular}{ll|c|c|c|c}
        \toprule
        $d$ & \shortstack{$T_1$\\ {[ms]}} & \texttt{Full} 
            & \texttt{\shortstack{Diagonal\\(adaptive)}}  
            & \texttt{\shortstack{Diagonal\\(multiple)}}  
            & \texttt{Grid} \\
            \midrule
            \multirow[c]{4}{*}{8} & $ 10^{-1} $ & \shortstack{20.4\\{\scriptsize[19.6, 21.2]}} & \shortstack{\textbf{20.8}\\{\scriptsize[19.0, 21.9]}} & \shortstack{\textbf{22.1}\\{\scriptsize[21.6, 22.6]}} & \shortstack{\underline{\textbf{23.0}}\\{\scriptsize[21.9, 23.9]}} \\
             & $ 10^{0} $ & \shortstack{35.0\\{\scriptsize[34.5, 35.5]}} & \shortstack{31.8\\{\scriptsize[31.2, 32.7]}} & \shortstack{\textbf{36.3}\\{\scriptsize[35.8, 36.7]}} & \shortstack{\underline{\textbf{37.4}}\\{\scriptsize[36.7, 38.1]}} \\
             & $ 10^{1} $ & \shortstack{59.0\\{\scriptsize[58.5, 59.6]}} & \shortstack{55.5\\{\scriptsize[54.8, 56.3]}} & \shortstack{\textbf{59.5}\\{\scriptsize[58.7, 60.3]}} & \shortstack{\underline{\textbf{60.2}}\\{\scriptsize[59.0, 61.4]}} \\
             & $ 10^{2} $ & \shortstack{\underline{83.2}\\{\scriptsize[82.6, 83.7]}} & \shortstack{77.2\\{\scriptsize[76.1, 78.2]}} & \shortstack{80.9\\{\scriptsize[79.6, 82.1]}} & \shortstack{80.7\\{\scriptsize[79.6, 81.9]}} \\
            \midrule
            \multirow[c]{4}{*}{16} & $ 10^{-1} $ & \shortstack{14.8\\{\scriptsize[14.7, 14.8]}} & \shortstack{13.7\\{\scriptsize[13.6, 13.7]}} & \shortstack{14.1\\{\scriptsize[14.0, 14.1]}} & \shortstack{\underline{\textbf{15.0}}\\{\scriptsize[15.0, 15.0]}} \\
             & $ 10^{0} $ & \shortstack{28.2\\{\scriptsize[28.0, 28.3]}} & \shortstack{26.4\\{\scriptsize[26.2, 26.5]}} & \shortstack{26.9\\{\scriptsize[26.7, 27.0]}} & \shortstack{\underline{\textbf{28.6}}\\{\scriptsize[28.5, 28.7]}} \\
             & $ 10^{1} $ & \shortstack{\underline{44.8}\\{\scriptsize[44.5, 45.1]}} & \shortstack{41.6\\{\scriptsize[41.3, 42.0]}} & \shortstack{42.6\\{\scriptsize[42.2, 43.0]}} & \shortstack{44.7\\{\scriptsize[44.4, 45.1]}} \\
             & $ 10^{2} $ & \shortstack{\underline{63.4}\\{\scriptsize[63.1, 63.7]}} & \shortstack{59.3\\{\scriptsize[58.8, 59.7]}} & \shortstack{61.5\\{\scriptsize[61.1, 61.9]}} & \shortstack{61.2\\{\scriptsize[60.9, 61.6]}} \\
            \midrule
            \multirow[c]{4}{*}{32} & $ 10^{-1} $ & \shortstack{5.6\\{\scriptsize[5.6, 5.6]}} & \shortstack{\underline{\textbf{6.9}}\\{\scriptsize[6.9, 6.9]}} & \shortstack{\textbf{6.4}\\{\scriptsize[6.4, 6.4]}} & \shortstack{\textbf{6.5}\\{\scriptsize[6.4, 6.5]}} \\
             & $ 10^{0} $ & \shortstack{16.0\\{\scriptsize[16.0, 16.0]}} & \shortstack{\textbf{16.8}\\{\scriptsize[16.6, 16.9]}} & \shortstack{\underline{\textbf{18.6}}\\{\scriptsize[18.5, 18.7]}} & \shortstack{\textbf{18.5}\\{\scriptsize[18.4, 18.6]}} \\
             & $ 10^{1} $ & \shortstack{29.4\\{\scriptsize[29.3, 29.5]}} & \shortstack{\textbf{33.8}\\{\scriptsize[33.6, 33.9]}} & \shortstack{\textbf{33.9}\\{\scriptsize[33.8, 34.1]}} & \shortstack{\underline{\textbf{34.0}}\\{\scriptsize[33.9, 34.2]}} \\
             & $ 10^{2} $ & \shortstack{42.7\\{\scriptsize[42.4, 42.9]}} & \shortstack{\textbf{46.2}\\{\scriptsize[45.9, 46.6]}} & \shortstack{\underline{\textbf{47.3}}\\{\scriptsize[47.0, 47.6]}} & \shortstack{\textbf{46.9}\\{\scriptsize[46.4, 47.5]}} \\
            \midrule
            \multirow[c]{4}{*}{64} & $ 10^{-1} $ & \shortstack{1.1\\{\scriptsize[1.1, 1.1]}} & \shortstack{\underline{\textbf{1.5}}\\{\scriptsize[1.2, 1.6]}} & \shortstack{\textbf{1.2}\\{\scriptsize[1.2, 1.2]}} & \shortstack{1.1\\{\scriptsize[1.1, 1.1]}} \\
             & $ 10^{0} $ & \shortstack{7.5\\{\scriptsize[7.1, 7.8]}} & \shortstack{6.7\\{\scriptsize[6.1, 7.3]}} & \shortstack{\underline{\textbf{8.2}}\\{\scriptsize[8.0, 8.4]}} & \shortstack{\textbf{8.2}\\{\scriptsize[7.7, 8.4]}} \\
             & $ 10^{1} $ & \shortstack{20.3\\{\scriptsize[19.9, 20.7]}} & \shortstack{15.7\\{\scriptsize[14.4, 16.4]}} & \shortstack{\underline{\textbf{21.9}}\\{\scriptsize[21.6, 22.2]}} & \shortstack{\textbf{20.8}\\{\scriptsize[20.3, 21.3]}} \\
             & $ 10^{2} $ & \shortstack{32.9\\{\scriptsize[31.6, 34.1]}} & \shortstack{29.5\\{\scriptsize[23.3, 32.1]}} & \shortstack{\underline{\textbf{33.7}}\\{\scriptsize[32.3, 34.6]}} & \shortstack{\textbf{33.4}\\{\scriptsize[29.4, 34.9]}} \\
            \bottomrule
        \end{tabular}
         \caption{Target state \emph{Fourier-5}.}
    \end{subfigure}\hfill
    \begin{subfigure}[c]{0.49\linewidth}
        \centering
        \footnotesize
        \setlength{\tabcolsep}{4pt}
        \renewcommand{\arraystretch}{2.2}
        \begin{tabular}{ll|c|c|c|c}
        \toprule
        $d$ & \shortstack{$T_1$\\ {[ms]}} & \texttt{Full} 
            & \texttt{\shortstack{Diagonal\\(adaptive)}}  
            & \texttt{\shortstack{Diagonal\\(multiple)}}  
            & \texttt{Grid} \\
            \midrule
            \multirow[c]{4}{*}{8} & $ 10^{-1} $ & \shortstack{9.2\\{\scriptsize[9.2, 9.2]}} & \shortstack{9.1\\{\scriptsize[9.1, 9.2]}} & \shortstack{\textbf{\underline{9.5}}\\{\scriptsize[9.5, 9.5]}} & \shortstack{\underline{\textbf{9.5}}\\{\scriptsize[9.5, 9.5]}} \\
             & $ 10^{0} $ & \shortstack{27.0\\{\scriptsize[26.6, 27.3]}} & \shortstack{\textbf{27.2}\\{\scriptsize[26.9, 27.7]}} & \shortstack{\underline{\textbf{27.7}}\\{\scriptsize[27.1, 28.3]}} & \shortstack{\textbf{27.2}\\{\scriptsize[26.8, 27.7]}} \\
             & $ 10^{1} $ & \shortstack{59.3\\{\scriptsize[58.8, 59.8]}} & \shortstack{57.8\\{\scriptsize[57.2, 58.4]}} & \shortstack{\underline{\textbf{60.1}}\\{\scriptsize[59.3, 60.8]}} & \shortstack{\textbf{59.7}\\{\scriptsize[59.1, 60.3]}} \\
             & $ 10^{2} $ & \shortstack{88.5\\{\scriptsize[87.3, 89.6]}} & \shortstack{86.5\\{\scriptsize[84.8, 88.7]}} & \shortstack{\underline{\textbf{91.0}}\\{\scriptsize[89.8, 92.1]}} & \shortstack{\textbf{89.1}\\{\scriptsize[88.0, 90.1]}} \\
            \midrule
            \multirow[c]{4}{*}{16} & $ 10^{-1} $ & \shortstack{\underline{3.1}\\{\scriptsize[3.0, 3.1]}} & \shortstack{3.0\\{\scriptsize[3.0, 3.0]}} & \shortstack{\underline{3.1}\\{\scriptsize[3.1, 3.1]}} & \shortstack{2.9\\{\scriptsize[2.9, 2.9]}} \\
             & $ 10^{0} $ & \shortstack{\underline{13.5}\\{\scriptsize[13.4, 13.7]}} & \shortstack{13.3\\{\scriptsize[13.1, 13.5]}} & \shortstack{12.9\\{\scriptsize[12.7, 13.1]}} & \shortstack{12.9\\{\scriptsize[12.7, 13.1]}} \\
             & $ 10^{1} $ & \shortstack{37.0\\{\scriptsize[36.8, 37.3]}} & \shortstack{\underline{\textbf{37.3}}\\{\scriptsize[36.9, 37.9]}} & \shortstack{35.8\\{\scriptsize[35.5, 36.1]}} & \shortstack{34.9\\{\scriptsize[34.7, 35.2]}} \\
             & $ 10^{2} $ & \shortstack{63.3\\{\scriptsize[63.0, 63.6]}} & \shortstack{\underline{\textbf{64.2}}\\{\scriptsize[63.5, 64.9]}} & \shortstack{60.8\\{\scriptsize[60.5, 61.2]}} & \shortstack{59.8\\{\scriptsize[59.4, 60.2]}} \\
            \midrule
            \multirow[c]{4}{*}{32} & $ 10^{-1} $ & \shortstack{1.4\\{\scriptsize[1.4, 1.4]}} & \shortstack{1.1\\{\scriptsize[1.1, 1.1]}} & \shortstack{\underline{\textbf{1.5}}\\{\scriptsize[1.5, 1.5]}} & \shortstack{\underline{\textbf{1.5}}\\{\scriptsize[1.5, 1.5]}} \\
             & $ 10^{0} $ & \shortstack{5.7\\{\scriptsize[5.7, 5.7]}} & \shortstack{3.2\\{\scriptsize[3.1, 3.2]}} & \shortstack{\underline{\textbf{6.0}}\\{\scriptsize[5.9, 6.0]}} & \shortstack{\textbf{5.9}\\{\scriptsize[5.8, 5.9]}} \\
             & $ 10^{1} $ & \shortstack{21.7\\{\scriptsize[21.5, 21.9]}} & \shortstack{19.1\\{\scriptsize[18.9, 19.3]}} & \shortstack{\underline{\textbf{22.8}}\\{\scriptsize[22.5, 23.0]}} & \shortstack{\textbf{22.0}\\{\scriptsize[21.8, 22.2]}} \\
             & $ 10^{2} $ & \shortstack{44.2\\{\scriptsize[43.9, 44.5]}} & \shortstack{40.3\\{\scriptsize[39.4, 41.0]}} & \shortstack{\underline{\textbf{45.7}}\\{\scriptsize[45.3, 46.0]}} & \shortstack{\textbf{45.1}\\{\scriptsize[44.9, 45.3]}} \\
            \midrule
            \multirow[c]{4}{*}{64} & $ 10^{-1} $ & \shortstack{\underline{1.0}\\{\scriptsize[0.8, 1.1]}} & \shortstack{0.9\\{\scriptsize[0.6, 1.0]}} & \shortstack{0.7\\{\scriptsize[0.3, 1.1]}} & \shortstack{0.5\\{\scriptsize[0.2, 0.9]}} \\
             & $ 10^{0} $ & \shortstack{\underline{2.1}\\{\scriptsize[2.0, 2.1]}} & \shortstack{1.2\\{\scriptsize[1.2, 1.2]}} & \shortstack{\underline{2.1}\\{\scriptsize[2.0, 2.1]}} & \shortstack{2.0\\{\scriptsize[1.9, 2.1]}} \\
             & $ 10^{1} $ & \shortstack{9.6\\{\scriptsize[9.5, 9.8]}} & \shortstack{7.0\\{\scriptsize[6.7, 7.2]}} & \shortstack{\underline{\textbf{9.7}}\\{\scriptsize[9.3, 10.1]}} & \shortstack{9.5\\{\scriptsize[9.2, 9.8]}} \\
             & $ 10^{2} $ & \shortstack{\underline{24.8}\\{\scriptsize[23.7, 25.5]}} & \shortstack{21.9\\{\scriptsize[20.9, 22.7]}} & \shortstack{24.3\\{\scriptsize[22.6, 26.0]}} & \shortstack{23.1\\{\scriptsize[22.1, 24.3]}} \\
            \bottomrule
        \end{tabular}
        \caption{Target state \emph{Haar random}.}
    \end{subfigure}
    \label{tab:photon_loss_time/hypervolume}  
\end{table*}

\begin{table*}[ht]
    \centering
    \footnotesize
    \caption{Infidelity for the \emph{Fourier-5} states under photon-loss noise. Gate parameters are learned to optimize fidelity accounting for photon-loss with the hyperparameter configuration $h$ obtained \textbf{when penalizing the number of phases}. Column $d$ represents the number of levels of the qudit, $\beta$ is the phase penalty used to optimize the hyperparameters. 
    }
    \input{photon_loss/tables_phases/mean_75/fourier_5/initial_0/photon_loss_table}
    \label{photon_loss/tables_phases/mean_75/fourier_5/initial_0/photon_loss_table}
\end{table*}

\begin{table*}[ht]
    \centering
    \footnotesize
    \caption{Infidelity for the \emph{Haar random} states under photon-loss noise. Gate parameters are learned to optimize fidelity accounting for photon-loss with the hyperparameter configuration $h$ obtained \textbf{when penalizing the number of phases}. Column $d$ represents the number of levels of the qudit, $\beta$ is the phase penalty used to optimize the hyperparameters.
    }
    \input{photon_loss/tables_phases/mean_75/random_gaussian/initial_0/photon_loss_table}
    \label{photon_loss/tables_phases/mean_75/random_gaussian/initial_0/photon_loss_table}
\end{table*}

\begin{table*}[ht]
    \centering
    \footnotesize
    \caption{Infidelity for the \emph{Fourier-5} states under photon-loss noise. Gate parameters are learned to optimize fidelity accounting for photon-loss with the hyperparameter configuration $h$ obtained \textbf{when penalizing the duration of the ansatz}. Column $d$ represents the number of levels of the qudit, $\beta$ is the phase penalty used to optimize the hyperparameters. 
    }
    \input{photon_loss/tables_time/mean_75/fourier_5/initial_0/photon_loss_table}
    \label{photon_loss/tables_time/mean_75/fourier_5/initial_0/photon_loss_table}
\end{table*}

\begin{table*}[ht]
    \centering
    \footnotesize
    \caption{Infidelity for the \emph{Haar random} states under photon-loss noise. Gate parameters are learned to optimize fidelity accounting for photon-loss with the hyperparameter configuration $h$ obtained \textbf{when penalizing the duration of the ansatz}. Column $d$ represents the number of levels of the qudit, $\beta$ is the phase penalty used to optimize the hyperparameters. 
    }
    \input{photon_loss/tables_time/mean_75/random_gaussian/initial_0/photon_loss_table}
    \label{photon_loss/tables_time/mean_75/random_gaussian/initial_0/photon_loss_table}
\end{table*}


\end{document}

%% file: tables_phases/mean_75/fourier_5/initial_0/results_best_scalarized.tex
\begin{tabular}{ll|ccc|ccc|ccc|ccc}
\toprule
 &  & \multicolumn{3}{c}{\texttt{Full}} & \multicolumn{3}{c}{\texttt{Diagonal (adaptive)}} & \multicolumn{3}{c}{\texttt{Diagonal (multiple)}} & \multicolumn{3}{c}{\texttt{Grid}} \\
 \multicolumn{1}{c}{d} & \multicolumn{1}{c}{$\beta$} \vline & Infidelity & Phases & T $[\mu s]$ & Infidelity & Phases & T $[\mu s]$ & Infidelity & Phases & T $[\mu s]$ & Infidelity & Phases & T $[\mu s]$ \\
\midrule
\multirow[c]{7}{*}{8} & $0.0$ & $ 5.2 \pm 2.6 \times 10^{-6} $ & 46 & 170 & $ 2.4 \pm 1.3 \times 10^{-6} $ & 43 & 175 & $ 1.6 \pm 1.1 \times 10^{-5} $ & 28 & 143 & $ 6.0 \pm 4.0 \times 10^{-6} $ & 29 & 133 \\
 & $ 10^{-6} $ & $ 3.5 \pm 2.4 \times 10^{-6} $ & 39 & 141 & $ 3.4 \pm 2.0 \times 10^{-6} $ & 43 & 175 & $ 1.7 \pm 1.6 \times 10^{-5} $ & 21 & 115 & $ 3.8 \pm 1.9 \times 10^{-6} $ & 47 & 187 \\
 & $ 10^{-5} $ & $ 4.5 \pm 3.2 \times 10^{-6} $ & 39 & 141 & $ 5.8 \pm 3.1 \times 10^{-6} $ & 42 & 175 & $ 8.5 \pm 6.5 \times 10^{-6} $ & 33 & 154 & $ 7.2 \pm 6.3 \times 10^{-6} $ & 31 & 137 \\
 & $ 10^{-4} $ & $ 1.4 \pm 1.4 \times 10^{-5} $ & 31 & 113 & $ 3.3 \pm 5.5 \times 10^{-5} $ & 21 & 104 & $ 5.9 \pm 5.6 \times 10^{-5} $ & 23 & 120 & $ 4.0 \pm 4.9 \times 10^{-5} $ & 23 & 127 \\
 & $ 10^{-3} $ & $ 2.1 \pm 2.7 \times 10^{-4} $ & 23 & 85 & $ 3.2 \pm 5.5 \times 10^{-4} $ & 15 & 88 & $ 3.7 \pm 6.0 \times 10^{-5} $ & 18 & 106 & $ 1.5 \pm 1.6 \times 10^{-4} $ & 21 & 115 \\
 & $ 10^{-2} $ & $ 1.7 \pm 2.4 \times 10^{-4} $ & 23 & 85 & $ 1.4 \pm 2.1 \times 10^{-3} $ & 13 & 98 & $ 4.0 \pm 7.4 \times 10^{-4} $ & 14 & 92 & $ 6.4 \pm 9.1 \times 10^{-4} $ & 17 & 91 \\
 & $ 10^{-1} $ & $ 5.8 \pm 2.7 \times 10^{-2} $ & 8 & 28 & $ 1.2 \pm 1.0 \times 10^{-2} $ & 8 & 62 & $ 2.2 \pm 1.6 \times 10^{-2} $ & 8 & 59 & $ 1.1 \pm 1.1 \times 10^{-2} $ & 8 & 73 \\
\midrule
\multirow[c]{7}{*}{16} & $0.0$ & $ 7.4 \pm 5.2 \times 10^{-6} $ & 167 & 441 & $ 2.3 \pm 2.0 \times 10^{-5} $ & 133 & 445 & $ 0.7 \pm 1.5 \times 10^{-4} $ & 111 & 392 & $ 2.4 \pm 2.4 \times 10^{-5} $ & 135 & 445 \\
 & $ 10^{-6} $ & $ 5.4 \pm 4.2 \times 10^{-6} $ & 190 & 481 & $ 3.7 \pm 6.8 \times 10^{-5} $ & 91 & 345 & $ 5.4 \pm 4.4 \times 10^{-5} $ & 130 & 449 & $ 2.2 \pm 1.9 \times 10^{-5} $ & 125 & 385 \\
 & $ 10^{-5} $ & $ 4.4 \pm 3.3 \times 10^{-5} $ & 127 & 320 & $ 1.3 \pm 1.4 \times 10^{-5} $ & 111 & 316 & $ 6.5 \pm 6.8 \times 10^{-5} $ & 95 & 277 & $ 2.4 \pm 3.6 \times 10^{-5} $ & 103 & 275 \\
 & $ 10^{-4} $ & $ 2.6 \pm 3.8 \times 10^{-5} $ & 95 & 240 & $ 5.8 \pm 5.1 \times 10^{-5} $ & 85 & 261 & $ 1.3 \pm 1.7 \times 10^{-4} $ & 76 & 215 & $ 3.1 \pm 3.1 \times 10^{-5} $ & 102 & 320 \\
 & $ 10^{-3} $ & $ 0.5 \pm 1.2 \times 10^{-3} $ & 79 & 200 & $ 2.2 \pm 3.7 \times 10^{-4} $ & 63 & 224 & $ 3.6 \pm 4.1 \times 10^{-5} $ & 86 & 228 & $ 1.8 \pm 2.4 \times 10^{-4} $ & 78 & 250 \\
 & $ 10^{-2} $ & $ 5.7 \pm 2.8 \times 10^{-3} $ & 47 & 120 & $ 7.5 \pm 2.9 \times 10^{-3} $ & 20 & 126 & $ 8.4 \pm 3.0 \times 10^{-3} $ & 21 & 159 & $ 5.7 \pm 3.0 \times 10^{-3} $ & 33 & 153 \\
 & $ 10^{-1} $ & $ 4.4 \pm 0.0 \times 10^{-2} $ & 15 & 40 & $ 1.6 \pm 0.5 \times 10^{-2} $ & 10 & 81 & $ 1.7 \pm 0.3 \times 10^{-2} $ & 8 & 84 & $ 3.8 \pm 1.0 \times 10^{-2} $ & 5 & 48 \\
\midrule
\multirow[c]{7}{*}{32} & $0.0$ & $ 3.8 \pm 2.6 \times 10^{-5} $ & 967 & 1756 & $ 6.6 \pm 5.0 \times 10^{-5} $ & 508 & 1358 & $ 1.1 \pm 1.0 \times 10^{-4} $ & 431 & 1137 & $ 4.1 \pm 3.6 \times 10^{-5} $ & 514 & 1337 \\
 & $ 10^{-6} $ & $ 1.4 \pm 1.0 \times 10^{-5} $ & 987 & 1756 & $ 2.0 \pm 1.9 \times 10^{-5} $ & 627 & 1392 & $ 0.9 \pm 1.1 \times 10^{-3} $ & 336 & 620 & $ 2.8 \pm 2.8 \times 10^{-4} $ & 723 & 1556 \\
 & $ 10^{-5} $ & $ 4.1 \pm 4.6 \times 10^{-5} $ & 803 & 1473 & $ 5.2 \pm 5.9 \times 10^{-5} $ & 563 & 1223 & $ 6.2 \pm 9.4 \times 10^{-4} $ & 533 & 1204 & $ 7.8 \pm 7.1 \times 10^{-5} $ & 503 & 1304 \\
 & $ 10^{-4} $ & $ 1.8 \pm 3.2 \times 10^{-4} $ & 340 & 680 & $ 0.7 \pm 2.3 \times 10^{-4} $ & 335 & 911 & $ 4.5 \pm 4.3 \times 10^{-5} $ & 650 & 1435 & $ 5.0 \pm 7.0 \times 10^{-4} $ & 248 & 478 \\
 & $ 10^{-3} $ & $ 1.7 \pm 1.2 \times 10^{-3} $ & 241 & 453 & $ 6.7 \pm 9.9 \times 10^{-4} $ & 151 & 591 & $ 1.8 \pm 1.0 \times 10^{-3} $ & 108 & 571 & $ 3.8 \pm 1.1 \times 10^{-3} $ & 88 & 313 \\
 & $ 10^{-2} $ & $ 4.9 \pm 0.4 \times 10^{-3} $ & 95 & 170 & $ 5.9 \pm 1.5 \times 10^{-3} $ & 43 & 174 & $ 5.0 \pm 0.9 \times 10^{-3} $ & 50 & 309 & $ 7.4 \pm 1.2 \times 10^{-3} $ & 49 & 136 \\
 & $ 10^{-1} $ & $ 1.1 \pm 0.1 \times 10^{-2} $ & 62 & 113 & $ 2.4 \pm 1.0 \times 10^{-2} $ & 23 & 138 & $ 3.0 \pm 0.9 \times 10^{-2} $ & 21 & 167 & $ 1.2 \pm 0.2 \times 10^{-2} $ & 15 & 131 \\
\midrule
\multirow[c]{7}{*}{64} & $0.0$ & $ 3.3 \pm 2.2 \times 10^{-5} $ & 3401 & 4966 & $ 1.1 \pm 1.0 \times 10^{-5} $ & 2005 & 3524 & $ 7.5 \pm 4.1 \times 10^{-5} $ & 1081 & 2465 & $ 5.1 \pm 2.7 \times 10^{-5} $ & 1496 & 3010 \\
 & $ 10^{-6} $ & $ 3.1 \pm 2.2 \times 10^{-5} $ & 3139 & 4085 & $ 1.1 \pm 1.0 \times 10^{-5} $ & 2264 & 3203 & $ 2.3 \pm 4.2 \times 10^{-4} $ & 1580 & 2395 & $ 1.1 \pm 0.6 \times 10^{-5} $ & 1961 & 3394 \\
 & $ 10^{-5} $ & $ 3.2 \pm 1.2 \times 10^{-5} $ & 2987 & 3764 & $ 1.2 \pm 1.0 \times 10^{-5} $ & 1318 & 2326 & $ 4.3 \pm 1.9 \times 10^{-5} $ & 1056 & 2425 & $ 5.1 \pm 5.8 \times 10^{-5} $ & 857 & 1457 \\
 & $ 10^{-4} $ & $ 9.4 \pm 6.3 \times 10^{-4} $ & 830 & 1121 & $ 1.4 \pm 1.2 \times 10^{-4} $ & 355 & 1267 & $ 3.0 \pm 1.5 \times 10^{-4} $ & 926 & 2091 & $ 1.1 \pm 0.7 \times 10^{-4} $ & 502 & 1645 \\
 & $ 10^{-3} $ & $ 2.3 \pm 0.4 \times 10^{-3} $ & 255 & 320 & $ 1.4 \pm 0.6 \times 10^{-3} $ & 185 & 682 & $ 2.1 \pm 0.9 \times 10^{-3} $ & 243 & 344 & $ 2.2 \pm 0.3 \times 10^{-3} $ & 243 & 514 \\
 & $ 10^{-2} $ & $ 3.2 \pm 0.2 \times 10^{-3} $ & 189 & 240 & $ 3.8 \pm 0.8 \times 10^{-3} $ & 108 & 216 & $ 3.9 \pm 0.7 \times 10^{-3} $ & 114 & 232 & $ 4.2 \pm 0.8 \times 10^{-3} $ & 111 & 171 \\
 & $ 10^{-1} $ & $ 1.1 \pm 0.2 \times 10^{-2} $ & 126 & 160 & $ 3.2 \pm 1.0 \times 10^{-2} $ & 66 & 197 & $ 3.4 \pm 1.9 \times 10^{-2} $ & 48 & 252 & $ 2.9 \pm 0.4 \times 10^{-2} $ & 25 & 76 \\
\bottomrule
\end{tabular}

%% file: tables_phases/mean_75/random_gaussian/initial_0/results_best_scalarized.tex
\begin{tabular}{ll|ccc|ccc|ccc|ccc}
\toprule
 &  & \multicolumn{3}{c}{\texttt{Full}} & \multicolumn{3}{c}{\texttt{Diagonal (adaptive)}} & \multicolumn{3}{c}{\texttt{Diagonal (multiple)}} & \multicolumn{3}{c}{\texttt{Grid}} \\
 \multicolumn{1}{c}{d} & \multicolumn{1}{c}{$\beta$} \vline & Infidelity & Phases & T $[\mu s]$ & Infidelity & Phases & T $[\mu s]$ & Infidelity & Phases & T $[\mu s]$ & Infidelity & Phases & T $[\mu s]$ \\
\midrule
\multirow[c]{7}{*}{8} & $0.0$ & $ 6.8 \pm 4.4 \times 10^{-6} $ & 47 & 170 & $ 1.5 \pm 1.4 \times 10^{-5} $ & 32 & 142 & $ 1.1 \pm 0.9 \times 10^{-5} $ & 37 & 160 & $ 9.9 \pm 8.3 \times 10^{-6} $ & 44 & 164 \\
 & $ 10^{-6} $ & $ 7.9 \pm 4.5 \times 10^{-6} $ & 55 & 198 & $ 1.3 \pm 1.0 \times 10^{-5} $ & 40 & 169 & $ 2.5 \pm 1.9 \times 10^{-5} $ & 37 & 165 & $ 2.3 \pm 1.5 \times 10^{-5} $ & 40 & 160 \\
 & $ 10^{-5} $ & $ 1.6 \pm 1.3 \times 10^{-5} $ & 39 & 141 & $ 8.9 \pm 8.3 \times 10^{-6} $ & 33 & 142 & $ 8.9 \pm 7.9 \times 10^{-6} $ & 33 & 154 & $ 2.1 \pm 1.5 \times 10^{-5} $ & 39 & 157 \\
 & $ 10^{-4} $ & $ 4.9 \pm 5.7 \times 10^{-5} $ & 31 & 113 & $ 4.3 \pm 4.5 \times 10^{-5} $ & 28 & 138 & $ 0.6 \pm 1.1 \times 10^{-4} $ & 27 & 139 & $ 2.2 \pm 2.1 \times 10^{-5} $ & 33 & 154 \\
 & $ 10^{-3} $ & $ 5.2 \pm 5.3 \times 10^{-5} $ & 31 & 113 & $ 7.8 \pm 7.6 \times 10^{-5} $ & 25 & 113 & $ 1.2 \pm 2.2 \times 10^{-4} $ & 21 & 114 & $ 1.1 \pm 1.8 \times 10^{-4} $ & 24 & 109 \\
 & $ 10^{-2} $ & $ 2.4 \pm 4.3 \times 10^{-3} $ & 23 & 85 & $ 5.8 \pm 9.3 \times 10^{-3} $ & 15 & 98 & $ 1.6 \pm 3.1 \times 10^{-4} $ & 22 & 117 & $ 1.6 \pm 2.1 \times 10^{-4} $ & 21 & 120 \\
 & $ 10^{-1} $ & $ 4.0 \pm 6.6 \times 10^{-3} $ & 23 & 85 & $ 2.9 \pm 3.5 \times 10^{-2} $ & 9 & 77 & $ 4.7 \pm 3.9 \times 10^{-2} $ & 9 & 76 & $ 2.6 \pm 2.7 \times 10^{-2} $ & 12 & 85 \\
\midrule
\multirow[c]{7}{*}{16} & $0.0$ & $ 1.4 \pm 1.2 \times 10^{-5} $ & 111 & 280 & $ 1.5 \pm 1.0 \times 10^{-5} $ & 144 & 434 & $ 9.6 \pm 7.8 \times 10^{-5} $ & 69 & 186 & $ 0.7 \pm 1.0 \times 10^{-4} $ & 82 & 238 \\
 & $ 10^{-6} $ & $ 5.8 \pm 4.4 \times 10^{-6} $ & 175 & 441 & $ 1.3 \pm 1.2 \times 10^{-5} $ & 87 & 264 & $ 1.0 \pm 1.1 \times 10^{-5} $ & 120 & 311 & $ 8.7 \pm 8.5 \times 10^{-6} $ & 107 & 275 \\
 & $ 10^{-5} $ & $ 9.0 \pm 7.2 \times 10^{-6} $ & 127 & 320 & $ 3.1 \pm 3.0 \times 10^{-5} $ & 74 & 245 & $ 3.4 \pm 2.7 \times 10^{-5} $ & 77 & 237 & $ 8.8 \pm 8.4 \times 10^{-5} $ & 100 & 370 \\
 & $ 10^{-4} $ & $ 4.1 \pm 5.0 \times 10^{-5} $ & 76 & 200 & $ 2.6 \pm 2.2 \times 10^{-5} $ & 82 & 299 & $ 5.0 \pm 7.6 \times 10^{-4} $ & 59 & 285 & $ 4.3 \pm 5.4 \times 10^{-5} $ & 77 & 215 \\
 & $ 10^{-3} $ & $ 4.9 \pm 8.0 \times 10^{-4} $ & 64 & 160 & $ 1.3 \pm 2.0 \times 10^{-4} $ & 57 & 215 & $ 1.2 \pm 1.7 \times 10^{-4} $ & 67 & 183 & $ 1.1 \pm 1.4 \times 10^{-4} $ & 74 & 283 \\
 & $ 10^{-2} $ & $ 5.0 \pm 5.9 \times 10^{-3} $ & 48 & 120 & $ 2.0 \pm 2.8 \times 10^{-3} $ & 39 & 167 & $ 2.9 \pm 3.2 \times 10^{-3} $ & 48 & 140 & $ 3.6 \pm 3.4 \times 10^{-3} $ & 41 & 143 \\
 & $ 10^{-1} $ & $ 5.8 \pm 6.6 \times 10^{-3} $ & 47 & 120 & $ 1.9 \pm 1.3 \times 10^{-2} $ & 26 & 124 & $ 2.3 \pm 1.7 \times 10^{-2} $ & 28 & 130 & $ 3.3 \pm 1.9 \times 10^{-2} $ & 21 & 128 \\
\midrule
\multirow[c]{7}{*}{32} & $0.0$ & $ 1.0 \pm 1.2 \times 10^{-5} $ & 757 & 1360 & $ 9.6 \pm 8.8 \times 10^{-6} $ & 674 & 1448 & $ 1.7 \pm 2.4 \times 10^{-5} $ & 289 & 544 & $ 7.6 \pm 8.2 \times 10^{-6} $ & 318 & 593 \\
 & $ 10^{-6} $ & $ 1.1 \pm 1.1 \times 10^{-5} $ & 371 & 680 & $ 2.8 \pm 2.7 \times 10^{-5} $ & 256 & 635 & $ 8.7 \pm 6.9 \times 10^{-5} $ & 361 & 1045 & $ 1.7 \pm 1.3 \times 10^{-5} $ & 524 & 1266 \\
 & $ 10^{-5} $ & $ 1.1 \pm 1.3 \times 10^{-5} $ & 287 & 510 & $ 1.7 \pm 1.9 \times 10^{-5} $ & 280 & 668 & $ 1.3 \pm 1.6 \times 10^{-4} $ & 181 & 330 & $ 3.9 \pm 4.3 \times 10^{-5} $ & 237 & 440 \\
 & $ 10^{-4} $ & $ 3.6 \pm 3.3 \times 10^{-5} $ & 180 & 340 & $ 9.5 \pm 9.4 \times 10^{-5} $ & 207 & 520 & $ 0.9 \pm 1.4 \times 10^{-4} $ & 205 & 429 & $ 1.4 \pm 1.8 \times 10^{-4} $ & 169 & 425 \\
 & $ 10^{-3} $ & $ 6.0 \pm 8.1 \times 10^{-4} $ & 127 & 226 & $ 1.5 \pm 2.2 \times 10^{-4} $ & 121 & 310 & $ 1.3 \pm 1.7 \times 10^{-4} $ & 144 & 346 & $ 0.6 \pm 1.2 \times 10^{-3} $ & 130 & 256 \\
 & $ 10^{-2} $ & $ 0.7 \pm 1.2 \times 10^{-3} $ & 127 & 226 & $ 2.9 \pm 3.7 \times 10^{-3} $ & 73 & 226 & $ 1.7 \pm 2.9 \times 10^{-3} $ & 119 & 241 & $ 5.4 \pm 5.8 \times 10^{-4} $ & 118 & 269 \\
 & $ 10^{-1} $ & $ 1.1 \pm 1.2 \times 10^{-2} $ & 95 & 170 & $ 2.3 \pm 1.3 \times 10^{-2} $ & 51 & 215 & $ 3.7 \pm 3.7 \times 10^{-2} $ & 69 & 141 & $ 2.2 \pm 2.1 \times 10^{-2} $ & 75 & 261 \\
\midrule
\multirow[c]{7}{*}{64} & $0.0$ & $ 2.1 \pm 1.3 \times 10^{-5} $ & 3242 & 4085 & $ 9.2 \pm 4.4 \times 10^{-6} $ & 3485 & 4738 & $ 1.2 \pm 1.0 \times 10^{-4} $ & 810 & 1038 & $ 8.1 \pm 5.3 \times 10^{-5} $ & 2916 & 4244 \\
 & $ 10^{-6} $ & $ 1.9 \pm 1.0 \times 10^{-5} $ & 3251 & 4325 & $ 6.6 \pm 4.7 \times 10^{-4} $ & 1943 & 3053 & $ 2.5 \pm 3.0 \times 10^{-4} $ & 744 & 996 & $ 8.3 \pm 4.9 \times 10^{-5} $ & 1704 & 3039 \\
 & $ 10^{-5} $ & $ 7.6 \pm 3.8 \times 10^{-5} $ & 2838 & 3924 & $ 1.0 \pm 0.7 \times 10^{-4} $ & 812 & 1654 & $ 4.8 \pm 4.1 \times 10^{-5} $ & 1221 & 2807 & $ 6.2 \pm 1.7 \times 10^{-5} $ & 2059 & 3588 \\
 & $ 10^{-4} $ & $ 7.1 \pm 5.3 \times 10^{-5} $ & 1020 & 1281 & $ 5.7 \pm 9.2 \times 10^{-4} $ & 330 & 620 & $ 3.4 \pm 1.6 \times 10^{-3} $ & 562 & 2306 & $ 8.9 \pm 6.9 \times 10^{-4} $ & 465 & 681 \\
 & $ 10^{-3} $ & $ 1.2 \pm 1.1 \times 10^{-4} $ & 503 & 640 & $ 2.5 \pm 2.6 \times 10^{-4} $ & 401 & 959 & $ 5.5 \pm 6.4 \times 10^{-4} $ & 289 & 412 & $ 3.2 \pm 2.6 \times 10^{-4} $ & 568 & 1823 \\
 & $ 10^{-2} $ & $ 1.1 \pm 1.3 \times 10^{-3} $ & 318 & 400 & $ 4.7 \pm 7.9 \times 10^{-3} $ & 185 & 362 & $ 1.3 \pm 1.4 \times 10^{-2} $ & 281 & 396 & $ 1.7 \pm 2.0 \times 10^{-3} $ & 213 & 324 \\
 & $ 10^{-1} $ & $ 2.7 \pm 1.8 \times 10^{-2} $ & 191 & 240 & $ 1.4 \pm 1.5 \times 10^{-2} $ & 136 & 385 & $ 4.2 \pm 1.3 \times 10^{-2} $ & 175 & 523 & $ 6.6 \pm 1.7 \times 10^{-2} $ & 105 & 440 \\
\bottomrule
\end{tabular}

%% file: tables_time/mean_75/fourier_5/initial_0/results_best_scalarized.tex
\begin{tabular}{ll|ccc|ccc|ccc|ccc}
\toprule
 &  & \multicolumn{3}{c}{\texttt{Full}} & \multicolumn{3}{c}{\texttt{Diagonal (adaptive)}} & \multicolumn{3}{c}{\texttt{Diagonal (multiple)}} & \multicolumn{3}{c}{\texttt{Grid}} \\
 \multicolumn{1}{c}{d} & \multicolumn{1}{c}{$\beta$} \vline & Infidelity & Phases & T $[\mu s]$ & Infidelity & Phases & T $[\mu s]$ & Infidelity & Phases & T $[\mu s]$ & Infidelity & Phases & T $[\mu s]$ \\
\midrule
\multirow[c]{7}{*}{8} & $0.0$ & $ 5.2 \pm 2.6 \times 10^{-6} $ & 46 & 170 & $ 2.4 \pm 1.3 \times 10^{-6} $ & 43 & 175 & $ 1.6 \pm 1.1 \times 10^{-5} $ & 28 & 143 & $ 6.0 \pm 4.0 \times 10^{-6} $ & 29 & 133 \\
 & $ 10^{-6} $ & $ 8.1 \pm 4.8 \times 10^{-6} $ & 39 & 141 & $ 4.7 \pm 2.4 \times 10^{-6} $ & 42 & 173 & $ 7.8 \pm 4.4 \times 10^{-6} $ & 28 & 131 & $ 2.1 \pm 1.4 \times 10^{-5} $ & 31 & 153 \\
 & $ 10^{-5} $ & $ 4.6 \pm 2.8 \times 10^{-6} $ & 38 & 141 & $ 2.2 \pm 1.3 \times 10^{-5} $ & 35 & 158 & $ 5.1 \pm 2.7 \times 10^{-6} $ & 31 & 148 & $ 2.4 \pm 3.4 \times 10^{-5} $ & 24 & 111 \\
 & $ 10^{-4} $ & $ 2.1 \pm 1.8 \times 10^{-5} $ & 30 & 113 & $ 9.1 \pm 9.9 \times 10^{-6} $ & 29 & 122 & $ 1.6 \pm 1.4 \times 10^{-5} $ & 25 & 123 & $ 4.1 \pm 5.4 \times 10^{-5} $ & 26 & 103 \\
 & $ 10^{-3} $ & $ 1.5 \pm 2.0 \times 10^{-4} $ & 24 & 85 & $ 0.7 \pm 1.1 \times 10^{-4} $ & 21 & 104 & $ 0.8 \pm 1.0 \times 10^{-4} $ & 24 & 98 & $ 1.2 \pm 1.0 \times 10^{-4} $ & 21 & 115 \\
 & $ 10^{-2} $ & $ 0.6 \pm 1.6 \times 10^{-3} $ & 23 & 85 & $ 1.5 \pm 2.5 \times 10^{-3} $ & 17 & 73 & $ 1.8 \pm 2.7 \times 10^{-4} $ & 21 & 91 & $ 6.0 \pm 7.1 \times 10^{-4} $ & 21 & 81 \\
 & $ 10^{-1} $ & $ 1.1 \pm 0.9 \times 10^{-2} $ & 16 & 56 & $ 4.4 \pm 5.2 \times 10^{-3} $ & 15 & 68 & $ 2.3 \pm 1.3 \times 10^{-2} $ & 10 & 42 & $ 1.3 \pm 1.1 \times 10^{-2} $ & 13 & 52 \\
\midrule
\multirow[c]{7}{*}{16} & $0.0$ & $ 7.4 \pm 5.2 \times 10^{-6} $ & 167 & 441 & $ 2.3 \pm 2.0 \times 10^{-5} $ & 133 & 445 & $ 0.7 \pm 1.5 \times 10^{-4} $ & 111 & 392 & $ 2.4 \pm 2.4 \times 10^{-5} $ & 135 & 445 \\
 & $ 10^{-6} $ & $ 7.7 \pm 4.7 \times 10^{-6} $ & 138 & 360 & $ 9.3 \pm 6.1 \times 10^{-6} $ & 196 & 544 & $ 1.8 \pm 1.8 \times 10^{-5} $ & 94 & 275 & $ 2.5 \pm 1.5 \times 10^{-5} $ & 149 & 465 \\
 & $ 10^{-5} $ & $ 1.2 \pm 1.2 \times 10^{-5} $ & 127 & 320 & $ 7.6 \pm 6.7 \times 10^{-6} $ & 167 & 484 & $ 8.4 \pm 6.0 \times 10^{-6} $ & 142 & 395 & $ 1.0 \pm 0.9 \times 10^{-5} $ & 139 & 440 \\
 & $ 10^{-4} $ & $ 2.4 \pm 3.0 \times 10^{-5} $ & 95 & 240 & $ 2.6 \pm 2.5 \times 10^{-5} $ & 86 & 328 & $ 2.6 \pm 2.2 \times 10^{-5} $ & 113 & 323 & $ 2.1 \pm 1.8 \times 10^{-5} $ & 114 & 338 \\
 & $ 10^{-3} $ & $ 1.7 \pm 2.5 \times 10^{-4} $ & 79 & 200 & $ 4.6 \pm 3.8 \times 10^{-5} $ & 77 & 215 & $ 8.7 \pm 9.3 \times 10^{-5} $ & 73 & 208 & $ 1.4 \pm 1.3 \times 10^{-4} $ & 121 & 317 \\
 & $ 10^{-2} $ & $ 2.7 \pm 8.3 \times 10^{-4} $ & 77 & 200 & $ 0.7 \pm 1.5 \times 10^{-3} $ & 57 & 188 & $ 5.8 \pm 8.4 \times 10^{-4} $ & 69 & 201 & $ 3.8 \pm 2.8 \times 10^{-3} $ & 51 & 175 \\
 & $ 10^{-1} $ & $ 1.3 \pm 0.1 \times 10^{-2} $ & 31 & 80 & $ 1.2 \pm 0.2 \times 10^{-2} $ & 26 & 90 & $ 1.3 \pm 0.1 \times 10^{-2} $ & 24 & 68 & $ 1.2 \pm 0.1 \times 10^{-2} $ & 28 & 74 \\
\midrule
\multirow[c]{7}{*}{32} & $0.0$ & $ 3.8 \pm 2.6 \times 10^{-5} $ & 967 & 1756 & $ 6.6 \pm 5.0 \times 10^{-5} $ & 508 & 1358 & $ 1.1 \pm 1.0 \times 10^{-4} $ & 431 & 1137 & $ 4.1 \pm 3.6 \times 10^{-5} $ & 514 & 1337 \\
 & $ 10^{-6} $ & $ 1.3 \pm 0.9 \times 10^{-5} $ & 980 & 1756 & $ 3.6 \pm 4.6 \times 10^{-5} $ & 674 & 1349 & $ 4.2 \pm 8.4 \times 10^{-4} $ & 339 & 616 & $ 1.2 \pm 1.0 \times 10^{-4} $ & 560 & 1334 \\
 & $ 10^{-5} $ & $ 4.4 \pm 4.7 \times 10^{-5} $ & 662 & 1246 & $ 4.9 \pm 4.9 \times 10^{-5} $ & 673 & 1423 & $ 4.1 \pm 6.9 \times 10^{-4} $ & 344 & 620 & $ 1.6 \pm 1.6 \times 10^{-4} $ & 371 & 1072 \\
 & $ 10^{-4} $ & $ 4.1 \pm 8.1 \times 10^{-4} $ & 411 & 736 & $ 5.9 \pm 5.3 \times 10^{-5} $ & 771 & 1582 & $ 4.6 \pm 8.6 \times 10^{-4} $ & 262 & 579 & $ 4.0 \pm 3.0 \times 10^{-5} $ & 303 & 555 \\
 & $ 10^{-3} $ & $ 1.0 \pm 1.1 \times 10^{-3} $ & 243 & 453 & $ 0.8 \pm 1.1 \times 10^{-3} $ & 212 & 496 & $ 1.2 \pm 1.1 \times 10^{-3} $ & 223 & 508 & $ 8.8 \pm 9.3 \times 10^{-5} $ & 256 & 509 \\
 & $ 10^{-2} $ & $ 5.1 \pm 0.6 \times 10^{-3} $ & 95 & 170 & $ 3.8 \pm 1.1 \times 10^{-3} $ & 79 & 251 & $ 4.1 \pm 0.9 \times 10^{-3} $ & 118 & 218 & $ 4.9 \pm 0.4 \times 10^{-3} $ & 93 & 168 \\
 & $ 10^{-1} $ & $ 1.1 \pm 0.1 \times 10^{-2} $ & 61 & 113 & $ 8.4 \pm 1.1 \times 10^{-3} $ & 42 & 110 & $ 1.3 \pm 0.1 \times 10^{-2} $ & 39 & 85 & $ 6.8 \pm 0.4 \times 10^{-3} $ & 53 & 116 \\
\midrule
\multirow[c]{7}{*}{64} & $0.0$ & $ 3.3 \pm 2.2 \times 10^{-5} $ & 3401 & 4966 & $ 1.1 \pm 1.0 \times 10^{-5} $ & 2005 & 3524 & $ 7.5 \pm 4.1 \times 10^{-5} $ & 1081 & 2465 & $ 5.1 \pm 2.7 \times 10^{-5} $ & 1496 & 3010 \\
 & $ 10^{-6} $ & $ 1.9 \pm 1.8 \times 10^{-5} $ & 3140 & 4085 & $ 1.3 \pm 1.1 \times 10^{-5} $ & 3402 & 4618 & $ 6.1 \pm 7.5 \times 10^{-4} $ & 871 & 1290 & $ 3.8 \pm 0.7 \times 10^{-4} $ & 530 & 1757 \\
 & $ 10^{-5} $ & $ 4.4 \pm 2.6 \times 10^{-5} $ & 3186 & 4165 & $ 7.7 \pm 4.9 \times 10^{-5} $ & 1977 & 3330 & $ 2.7 \pm 0.5 \times 10^{-5} $ & 1110 & 2627 & $ 1.7 \pm 0.8 \times 10^{-5} $ & 2165 & 3708 \\
 & $ 10^{-4} $ & $ 2.3 \pm 1.6 \times 10^{-5} $ & 3296 & 4165 & $ 2.2 \pm 1.4 \times 10^{-5} $ & 1874 & 2994 & $ 4.9 \pm 3.2 \times 10^{-5} $ & 1188 & 2498 & $ 1.5 \pm 0.5 \times 10^{-4} $ & 1555 & 3466 \\
 & $ 10^{-3} $ & $ 8.8 \pm 6.4 \times 10^{-4} $ & 624 & 801 & $ 3.8 \pm 4.7 \times 10^{-4} $ & 343 & 806 & $ 1.0 \pm 0.3 \times 10^{-3} $ & 397 & 754 & $ 1.1 \pm 0.5 \times 10^{-3} $ & 497 & 634 \\
 & $ 10^{-2} $ & $ 2.9 \pm 0.3 \times 10^{-3} $ & 192 & 240 & $ 3.3 \pm 0.2 \times 10^{-3} $ & 169 & 299 & $ 3.0 \pm 0.1 \times 10^{-3} $ & 160 & 216 & $ 2.9 \pm 0.3 \times 10^{-3} $ & 149 & 209 \\
 & $ 10^{-1} $ & $ 3.2 \pm 0.2 \times 10^{-3} $ & 191 & 240 & $ 6.3 \pm 1.3 \times 10^{-3} $ & 133 & 198 & $ 5.7 \pm 0.5 \times 10^{-3} $ & 140 & 209 & $ 4.1 \pm 2.9 \times 10^{-3} $ & 168 & 224 \\
\bottomrule
\end{tabular}

%% file: tables_time/mean_75/random_gaussian/initial_0/results_best_scalarized.tex
\begin{tabular}{ll|ccc|ccc|ccc|ccc}
\toprule
 &  & \multicolumn{3}{c}{\texttt{Full}} & \multicolumn{3}{c}{\texttt{Diagonal (adaptive)}} & \multicolumn{3}{c}{\texttt{Diagonal (multiple)}} & \multicolumn{3}{c}{\texttt{Grid}} \\
 \multicolumn{1}{c}{d} & \multicolumn{1}{c}{$\beta$} \vline & Infidelity & Phases & T $[\mu s]$ & Infidelity & Phases & T $[\mu s]$ & Infidelity & Phases & T $[\mu s]$ & Infidelity & Phases & T $[\mu s]$ \\
\midrule
\multirow[c]{7}{*}{8} & $0.0$ & $ 6.8 \pm 4.4 \times 10^{-6} $ & 47 & 170 & $ 1.5 \pm 1.4 \times 10^{-5} $ & 32 & 142 & $ 1.1 \pm 0.9 \times 10^{-5} $ & 37 & 160 & $ 9.9 \pm 8.3 \times 10^{-6} $ & 44 & 164 \\
 & $ 10^{-6} $ & $ 9.6 \pm 5.7 \times 10^{-6} $ & 54 & 198 & $ 6.1 \pm 3.7 \times 10^{-6} $ & 42 & 173 & $ 1.4 \pm 1.6 \times 10^{-5} $ & 31 & 138 & $ 0.7 \pm 1.1 \times 10^{-4} $ & 26 & 115 \\
 & $ 10^{-5} $ & $ 1.9 \pm 1.2 \times 10^{-5} $ & 39 & 141 & $ 1.8 \pm 1.2 \times 10^{-5} $ & 46 & 181 & $ 6.2 \pm 7.5 \times 10^{-5} $ & 27 & 127 & $ 8.9 \pm 7.7 \times 10^{-5} $ & 31 & 135 \\
 & $ 10^{-4} $ & $ 2.2 \pm 1.1 \times 10^{-5} $ & 37 & 141 & $ 1.1 \pm 0.9 \times 10^{-5} $ & 33 & 142 & $ 1.3 \pm 1.0 \times 10^{-4} $ & 35 & 147 & $ 3.5 \pm 4.3 \times 10^{-5} $ & 37 & 160 \\
 & $ 10^{-3} $ & $ 7.0 \pm 7.2 \times 10^{-5} $ & 31 & 113 & $ 2.9 \pm 5.1 \times 10^{-4} $ & 23 & 98 & $ 1.0 \pm 2.1 \times 10^{-4} $ & 26 & 116 & $ 1.5 \pm 1.8 \times 10^{-4} $ & 21 & 111 \\
 & $ 10^{-2} $ & $ 2.3 \pm 3.7 \times 10^{-3} $ & 23 & 85 & $ 2.5 \pm 4.4 \times 10^{-3} $ & 19 & 89 & $ 7.7 \pm 6.2 \times 10^{-4} $ & 28 & 119 & $ 3.9 \pm 4.8 \times 10^{-4} $ & 24 & 120 \\
 & $ 10^{-1} $ & $ 1.3 \pm 1.6 \times 10^{-3} $ & 23 & 85 & $ 1.7 \pm 3.2 \times 10^{-3} $ & 19 & 89 & $ 0.8 \pm 1.1 \times 10^{-2} $ & 18 & 83 & $ 3.4 \pm 6.6 \times 10^{-3} $ & 22 & 81 \\
\midrule
\multirow[c]{7}{*}{16} & $0.0$ & $ 1.4 \pm 1.2 \times 10^{-5} $ & 111 & 280 & $ 1.5 \pm 1.0 \times 10^{-5} $ & 144 & 434 & $ 9.6 \pm 7.8 \times 10^{-5} $ & 69 & 186 & $ 0.7 \pm 1.0 \times 10^{-4} $ & 82 & 238 \\
 & $ 10^{-6} $ & $ 7.4 \pm 4.2 \times 10^{-6} $ & 175 & 441 & $ 6.8 \pm 4.0 \times 10^{-6} $ & 142 & 460 & $ 2.0 \pm 1.5 \times 10^{-5} $ & 86 & 228 & $ 5.0 \pm 5.9 \times 10^{-5} $ & 88 & 326 \\
 & $ 10^{-5} $ & $ 7.4 \pm 5.6 \times 10^{-6} $ & 127 & 320 & $ 1.2 \pm 0.8 \times 10^{-5} $ & 157 & 479 & $ 3.2 \pm 2.8 \times 10^{-5} $ & 114 & 371 & $ 2.9 \pm 2.5 \times 10^{-5} $ & 91 & 271 \\
 & $ 10^{-4} $ & $ 1.3 \pm 1.0 \times 10^{-5} $ & 111 & 280 & $ 2.4 \pm 2.0 \times 10^{-5} $ & 83 & 273 & $ 1.3 \pm 1.0 \times 10^{-4} $ & 65 & 199 & $ 4.2 \pm 2.6 \times 10^{-5} $ & 92 & 315 \\
 & $ 10^{-3} $ & $ 2.0 \pm 2.1 \times 10^{-4} $ & 78 & 200 & $ 8.8 \pm 9.1 \times 10^{-5} $ & 62 & 193 & $ 1.5 \pm 1.5 \times 10^{-4} $ & 77 & 197 & $ 3.0 \pm 3.1 \times 10^{-4} $ & 60 & 209 \\
 & $ 10^{-2} $ & $ 4.5 \pm 5.4 \times 10^{-4} $ & 62 & 160 & $ 0.6 \pm 1.0 \times 10^{-3} $ & 50 & 175 & $ 1.7 \pm 2.2 \times 10^{-4} $ & 62 & 175 & $ 5.7 \pm 7.2 \times 10^{-4} $ & 56 & 153 \\
 & $ 10^{-1} $ & $ 5.0 \pm 5.9 \times 10^{-3} $ & 47 & 120 & $ 3.7 \pm 4.2 \times 10^{-3} $ & 42 & 129 & $ 1.1 \pm 1.1 \times 10^{-2} $ & 42 & 113 & $ 0.8 \pm 1.2 \times 10^{-3} $ & 53 & 187 \\
\midrule
\multirow[c]{7}{*}{32} & $0.0$ & $ 1.0 \pm 1.2 \times 10^{-5} $ & 757 & 1360 & $ 9.6 \pm 8.8 \times 10^{-6} $ & 674 & 1448 & $ 1.7 \pm 2.4 \times 10^{-5} $ & 289 & 544 & $ 7.6 \pm 8.2 \times 10^{-6} $ & 318 & 593 \\
 & $ 10^{-6} $ & $ 9.2 \pm 6.8 \times 10^{-6} $ & 658 & 1190 & $ 1.8 \pm 1.4 \times 10^{-5} $ & 496 & 1081 & $ 2.1 \pm 1.8 \times 10^{-5} $ & 261 & 549 & $ 7.7 \pm 4.8 \times 10^{-5} $ & 299 & 573 \\
 & $ 10^{-5} $ & $ 1.1 \pm 1.2 \times 10^{-5} $ & 472 & 850 & $ 4.4 \pm 4.0 \times 10^{-5} $ & 206 & 475 & $ 3.9 \pm 3.8 \times 10^{-5} $ & 405 & 1076 & $ 1.5 \pm 0.8 \times 10^{-5} $ & 496 & 933 \\
 & $ 10^{-4} $ & $ 2.9 \pm 2.6 \times 10^{-5} $ & 223 & 396 & $ 3.9 \pm 2.7 \times 10^{-5} $ & 196 & 522 & $ 6.4 \pm 7.4 \times 10^{-5} $ & 171 & 321 & $ 2.2 \pm 2.1 \times 10^{-5} $ & 193 & 367 \\
 & $ 10^{-3} $ & $ 1.0 \pm 1.0 \times 10^{-4} $ & 191 & 340 & $ 1.5 \pm 2.0 \times 10^{-4} $ & 136 & 330 & $ 1.5 \pm 1.2 \times 10^{-4} $ & 194 & 368 & $ 1.1 \pm 1.2 \times 10^{-4} $ & 149 & 279 \\
 & $ 10^{-2} $ & $ 1.9 \pm 1.7 \times 10^{-4} $ & 159 & 283 & $ 0.9 \pm 1.8 \times 10^{-3} $ & 111 & 235 & $ 3.5 \pm 4.1 \times 10^{-4} $ & 125 & 260 & $ 1.1 \pm 1.2 \times 10^{-3} $ & 132 & 257 \\
 & $ 10^{-1} $ & $ 1.0 \pm 1.3 \times 10^{-2} $ & 95 & 170 & $ 5.5 \pm 6.0 \times 10^{-3} $ & 75 & 193 & $ 3.4 \pm 4.1 \times 10^{-3} $ & 102 & 200 & $ 1.3 \pm 2.1 \times 10^{-3} $ & 111 & 210 \\
\midrule
\multirow[c]{7}{*}{64} & $0.0$ & $ 2.1 \pm 1.3 \times 10^{-5} $ & 3242 & 4085 & $ 9.2 \pm 4.4 \times 10^{-6} $ & 3485 & 4738 & $ 1.2 \pm 1.0 \times 10^{-4} $ & 810 & 1038 & $ 8.1 \pm 5.3 \times 10^{-5} $ & 2916 & 4244 \\
 & $ 10^{-6} $ & $ 5.4 \pm 5.9 \times 10^{-5} $ & 1423 & 2002 & $ 3.2 \pm 1.0 \times 10^{-5} $ & 2506 & 3675 & $ 2.8 \pm 1.3 \times 10^{-5} $ & 1953 & 3550 & $ 1.5 \pm 0.6 \times 10^{-4} $ & 788 & 2380 \\
 & $ 10^{-5} $ & $ 1.1 \pm 2.1 \times 10^{-4} $ & 952 & 1201 & $ 7.6 \pm 4.1 \times 10^{-5} $ & 1745 & 3059 & $ 8.7 \pm 4.6 \times 10^{-5} $ & 1544 & 2982 & $ 7.1 \pm 3.2 \times 10^{-5} $ & 1280 & 2770 \\
 & $ 10^{-4} $ & $ 6.6 \pm 6.8 \times 10^{-5} $ & 988 & 1281 & $ 2.0 \pm 1.4 \times 10^{-4} $ & 584 & 953 & $ 8.1 \pm 5.0 \times 10^{-4} $ & 628 & 1347 & $ 2.3 \pm 2.1 \times 10^{-4} $ & 370 & 472 \\
 & $ 10^{-3} $ & $ 1.9 \pm 1.8 \times 10^{-4} $ & 917 & 1361 & $ 1.7 \pm 1.3 \times 10^{-4} $ & 732 & 1309 & $ 1.0 \pm 1.0 \times 10^{-3} $ & 1218 & 2552 & $ 3.2 \pm 3.2 \times 10^{-4} $ & 777 & 1989 \\
 & $ 10^{-2} $ & $ 1.2 \pm 2.1 \times 10^{-3} $ & 431 & 560 & $ 4.4 \pm 3.5 \times 10^{-4} $ & 267 & 480 & $ 3.0 \pm 2.9 \times 10^{-4} $ & 470 & 632 & $ 3.1 \pm 7.3 \times 10^{-3} $ & 467 & 635 \\
 & $ 10^{-1} $ & $ 2.6 \pm 2.6 \times 10^{-4} $ & 317 & 400 & $ 4.2 \pm 3.1 \times 10^{-3} $ & 199 & 440 & $ 7.7 \pm 7.3 \times 10^{-3} $ & 238 & 339 & $ 3.2 \pm 3.4 \times 10^{-3} $ & 269 & 409 \\
\bottomrule
\end{tabular}

%% file: tables_phases/mean_75/fourier_5/initial_0/min_phases_table_phases.tex
\begin{tabular}{cc|cc|cc|cc|cc|cc|cc}
\toprule
 &  & \multicolumn{2}{c}{\texttt{Full}} & \multicolumn{2}{c}{\texttt{Diagonal (adaptive)}} & \multicolumn{2}{c}{\texttt{Diagonal (multiple)}} & \multicolumn{2}{c}{\texttt{Grid}} \vline & \multicolumn{2}{c}{\citet{PhysRevA.92.040303}} & \multicolumn{2}{c}{\citet{job2023efficient}} \\
 \multicolumn{1}{c}{d} & \multicolumn{1}{c}{Infidelity $\leq$} \vline & Phases & T $[\mu s]$ & Phases & T $[\mu s]$ & Phases & T $[\mu s]$ & Phases & T $[\mu s]$ & Phases & T $[\mu s]$ & Phases & T $[\mu s]$ \\
\midrule
\multirow[c]{5}{*}{8} & $10^{-1}$ & 8 & 28 & \textbf{2} & 20 & \textbf{4} & 40 & \underline{\textbf{1}} & 10 & 63 & 298 & 63 & 298 \\
 & $10^{-2}$ & 16 & 56 & \underline{\textbf{10}} & 77 & \underline{\textbf{10}} & 60 & \underline{\textbf{10}} & 76 & 119 & 569 & 119 & 569 \\
 & $10^{-3}$ & 24 & 85 & \textbf{16} & 88 & \underline{\textbf{15}} & 92 & \textbf{16} & 87 & 119 & 569 & 175 & 840 \\
 & $10^{-4}$ & 32 & 113 & \textbf{22} & 114 & \underline{\textbf{19}} & 106 & \textbf{20} & 113 & 175 & 840 & 231 & 1110 \\
 & $10^{-5}$ & 40 & 141 & \textbf{28} & 128 & \underline{\textbf{22}} & 115 & \textbf{28} & 139 & 287 & 1381 & 399 & 1923 \\
\midrule
\multirow[c]{5}{*}{16} & $10^{-1}$ & 16 & 40 & \textbf{4} & 40 & \underline{\textbf{2}} & 20 & \textbf{3} & 24 & 255 & 852 & 255 & 852 \\
 & $10^{-2}$ & 48 & 120 & \underline{\textbf{16}} & 143 & \textbf{21} & 149 & \textbf{26} & 155 & 495 & 1664 & 495 & 1664 \\
 & $10^{-3}$ & 80 & 200 & \underline{\textbf{54}} & 220 & \textbf{76} & 214 & \textbf{55} & 195 & 735 & 2477 & 735 & 2477 \\
 & $10^{-4}$ & 80 & 200 & \underline{\textbf{72}} & 253 & 87 & 228 & \textbf{77} & 255 & 975 & 3289 & 1455 & 4914 \\
 & $10^{-5}$ & 128 & 320 & \underline{\textbf{110}} & 346 & - & - & - & - & 1695 & 5726 & 2415 & 8164 \\
\midrule
\multirow[c]{5}{*}{32} & $10^{-1}$ & 64 & 113 & \underline{\textbf{10}} & 103 & \underline{\textbf{10}} & 77 & \textbf{12} & 68 & 2015 & 4777 & 2015 & 4777 \\
 & $10^{-2}$ & 96 & 170 & \underline{\textbf{34}} & 242 & \textbf{44} & 306 & \textbf{35} & 204 & 2015 & 4777 & 3007 & 7136 \\
 & $10^{-3}$ & 288 & 510 & \underline{\textbf{152}} & 591 & \textbf{213} & 814 & \textbf{195} & 752 & 3007 & 7136 & 3999 & 9495 \\
 & $10^{-4}$ & 544 & 963 & \underline{\textbf{348}} & 911 & 663 & 1435 & 555 & 1304 & 5983 & 14214 & 6975 & 16573 \\
 & $10^{-5}$ & - & - & - & - & - & - & - & - & \underline{8959} & 21291 & 11935 & 28369 \\
\midrule
\multirow[c]{5}{*}{64} & $10^{-1}$ & 128 & 160 & \textbf{42} & 111 & \textbf{42} & 261 & \underline{\textbf{24}} & 68 & 8127 & 13596 & 8127 & 13596 \\
 & $10^{-2}$ & 192 & 240 & \textbf{97} & 239 & \underline{\textbf{84}} & 218 & \textbf{106} & 184 & 12159 & 20351 & 12159 & 20351 \\
 & $10^{-3}$ & 768 & 961 & \underline{\textbf{248}} & 754 & \textbf{394} & 1466 & \textbf{402} & 1456 & 16191 & 27106 & 20223 & 33861 \\
 & $10^{-4}$ & 2880 & 3604 & \underline{\textbf{548}} & 1213 & \textbf{1067} & 2425 & \textbf{588} & 1895 & 28287 & 47372 & 36351 & 60882 \\
 & $10^{-5}$ & - & - & \underline{2910} & 4267 & - & - & - & - & 48447 & 81148 & 64575 & 108169 \\
\bottomrule
\end{tabular}

%% file: tables_phases/mean_75/random_gaussian/initial_0/min_phases_table_phases.tex
\begin{tabular}{cc|cc|cc|cc|cc|cc|cc}
\toprule
 &  & \multicolumn{2}{c}{\texttt{Full}} & \multicolumn{2}{c}{\texttt{Diagonal (adaptive)}} & \multicolumn{2}{c}{\texttt{Diagonal (multiple)}} & \multicolumn{2}{c}{\texttt{Grid}} \vline & \multicolumn{2}{c}{\citet{PhysRevA.92.040303}} & \multicolumn{2}{c}{\citet{job2023efficient}} \\
 \multicolumn{1}{c}{d} & \multicolumn{1}{c}{Infidelity $\leq$} \vline & Phases & T $[\mu s]$ & Phases & T $[\mu s]$ & Phases & T $[\mu s]$ & Phases & T $[\mu s]$ & Phases & T $[\mu s]$ & Phases & T $[\mu s]$ \\
\midrule
\multirow[c]{5}{*}{8} & $10^{-1}$ & 16 & 56 & \underline{\textbf{6}} & 60 & \textbf{7} & 58 & \textbf{8} & 56 & 64 & 299 & 64 & 299 \\
 & $10^{-2}$ & 24 & 85 & \textbf{16} & 98 & \underline{\textbf{15}} & 92 & \underline{\textbf{15}} & 100 & 120 & 570 & 120 & 570 \\
 & $10^{-3}$ & 32 & 113 & \underline{\textbf{19}} & 104 & \underline{\textbf{19}} & 106 & \textbf{22} & 120 & 176 & 841 & 176 & 841 \\
 & $10^{-4}$ & 32 & 113 & \underline{\textbf{24}} & 117 & \underline{\textbf{24}} & 128 & \textbf{26} & 132 & 232 & 1112 & 288 & 1383 \\
 & $10^{-5}$ & 40 & 141 & \underline{\textbf{34}} & 142 & \underline{\textbf{34}} & 154 & 45 & 164 & 400 & 1925 & 512 & 2467 \\
\midrule
\multirow[c]{5}{*}{16} & $10^{-1}$ & 32 & 80 & \underline{\textbf{16}} & 131 & \textbf{21} & 134 & \textbf{20} & 106 & 256 & 853 & 256 & 853 \\
 & $10^{-2}$ & 48 & 120 & \underline{\textbf{30}} & 155 & \textbf{37} & 148 & \textbf{35} & 169 & 496 & 1666 & 496 & 1666 \\
 & $10^{-3}$ & 64 & 160 & \underline{\textbf{44}} & 161 & \textbf{57} & 182 & \textbf{54} & 241 & 736 & 2478 & 976 & 3291 \\
 & $10^{-4}$ & 80 & 200 & \underline{\textbf{66}} & 243 & \textbf{70} & 186 & \textbf{69} & 185 & 1216 & 4103 & 1456 & 4915 \\
 & $10^{-5}$ & 112 & 280 & \underline{\textbf{98}} & 296 & - & - & \textbf{108} & 275 & 1936 & 6540 & 2416 & 8165 \\
\midrule
\multirow[c]{5}{*}{32} & $10^{-1}$ & 96 & 170 & \underline{\textbf{36}} & 274 & \textbf{49} & 240 & \textbf{38} & 212 & 2016 & 4778 & 2016 & 4778 \\
 & $10^{-2}$ & 96 & 170 & \underline{\textbf{64}} & 210 & 96 & 192 & \textbf{85} & 275 & 2016 & 4778 & 3008 & 7137 \\
 & $10^{-3}$ & 128 & 226 & \underline{\textbf{96}} & 238 & 129 & 251 & \textbf{127} & 225 & 3008 & 7137 & 4000 & 9496 \\
 & $10^{-4}$ & 160 & 283 & \underline{\textbf{142}} & 393 & 197 & 370 & 161 & 308 & 5984 & 14214 & 6976 & 16574 \\
 & $10^{-5}$ & 448 & 793 & 673 & 1428 & - & - & \underline{\textbf{320}} & 593 & 9952 & 23651 & 11936 & 28370 \\
\midrule
\multirow[c]{5}{*}{64} & $10^{-1}$ & 192 & 240 & \underline{\textbf{92}} & 429 & \textbf{112} & 473 & \textbf{106} & 440 & 8128 & 13596 & 8128 & 13596 \\
 & $10^{-2}$ & 320 & 400 & \underline{\textbf{186}} & 358 & \textbf{207} & 284 & \textbf{219} & 324 & 12160 & 20351 & 12160 & 20351 \\
 & $10^{-3}$ & 320 & 400 & \underline{\textbf{266}} & 538 & \textbf{292} & 412 & 351 & 494 & 16192 & 27107 & 20224 & 33862 \\
 & $10^{-4}$ & 896 & 1121 & 1686 & 2586 & \underline{\textbf{799}} & 1175 & 954 & 1236 & 28288 & 47372 & 36352 & 60883 \\
 & $10^{-5}$ & - & - & \underline{3570} & 4738 & - & - & - & - & 48448 & 81148 & 64576 & 108169 \\
\bottomrule
\end{tabular}

%% file: tables_time/mean_75/fourier_5/initial_0/min_phases_table_time.tex
\begin{tabular}{cc|cc|cc|cc|cc|cc|cc}
\toprule
 &  & \multicolumn{2}{c}{\texttt{Full}} & \multicolumn{2}{c}{\texttt{Diagonal (adaptive)}} & \multicolumn{2}{c}{\texttt{Diagonal (multiple)}} & \multicolumn{2}{c}{\texttt{Grid}} \vline & \multicolumn{2}{c}{\citet{PhysRevA.92.040303}} & \multicolumn{2}{c}{\citet{job2023efficient}} \\
 \multicolumn{1}{c}{d} & \multicolumn{1}{c}{Infidelity $\leq$} \vline & Phases & T $[\mu s]$ & Phases & T $[\mu s]$ & Phases & T $[\mu s]$ & Phases & T $[\mu s]$ & Phases & T $[\mu s]$ & Phases & T $[\mu s]$ \\
\midrule
\multirow[c]{5}{*}{8} & $10^{-1}$ & 8 & 28 & 2 & \textbf{20} & 1 & \underline{\textbf{10}} & 1 & \underline{\textbf{10}} & 63 & 298 & 63 & 298 \\
 & $10^{-2}$ & 16 & \underline{56} & 12 & 59 & 15 & 65 & 14 & 64 & 119 & 569 & 119 & 569 \\
 & $10^{-3}$ & 24 & 85 & 16 & \underline{\textbf{68}} & 23 & \textbf{83} & 20 & \textbf{76} & 119 & 569 & 175 & 840 \\
 & $10^{-4}$ & 32 & 113 & 22 & \underline{\textbf{93}} & 24 & \textbf{96} & 23 & \textbf{94} & 175 & 840 & 231 & 1110 \\
 & $10^{-5}$ & 40 & 141 & 30 & \underline{\textbf{122}} & 29 & \textbf{131} & 30 & \textbf{133} & 287 & 1381 & 399 & 1923 \\
\midrule
\multirow[c]{5}{*}{16} & $10^{-1}$ & 16 & 40 & 4 & \textbf{28} & 3 & \underline{\textbf{24}} & 3 & \underline{\textbf{24}} & 255 & 852 & 255 & 852 \\
 & $10^{-2}$ & 48 & 120 & 22 & \textbf{93} & 29 & \underline{\textbf{91}} & 34 & \textbf{98} & 495 & 1664 & 495 & 1664 \\
 & $10^{-3}$ & 80 & 200 & 60 & \textbf{188} & 62 & \underline{\textbf{174}} & 65 & \textbf{178} & 735 & 2477 & 735 & 2477 \\
 & $10^{-4}$ & 96 & 240 & 78 & \textbf{215} & 74 & \underline{\textbf{208}} & 77 & 255 & 975 & 3289 & 1455 & 4914 \\
 & $10^{-5}$ & 144 & \underline{360} & 158 & 452 & 143 & 395 & - & - & 1695 & 5726 & 2415 & 8164 \\
\midrule
\multirow[c]{5}{*}{32} & $10^{-1}$ & 64 & 113 & 26 & \textbf{71} & 12 & \textbf{60} & 15 & \underline{\textbf{52}} & 2015 & 4777 & 2015 & 4777 \\
 & $10^{-2}$ & 96 & 170 & 44 & \underline{\textbf{110}} & 51 & \textbf{112} & 53 & \textbf{116} & 2015 & 4777 & 3007 & 7136 \\
 & $10^{-3}$ & 256 & \underline{453} & 226 & 496 & 273 & 495 & 260 & 509 & 3007 & 7136 & 3999 & 9495 \\
 & $10^{-4}$ & 640 & 1133 & 398 & \textbf{1109} & - & - & 260 & \underline{\textbf{509}} & 5983 & 14214 & 6975 & 16573 \\
 & $10^{-5}$ & - & - & - & - & - & - & - & - & 8959 & \underline{21291} & 11935 & 28369 \\
\midrule
\multirow[c]{5}{*}{64} & $10^{-1}$ & 128 & 160 & 42 & \textbf{111} & 67 & \textbf{97} & 36 & \underline{\textbf{82}} & 8127 & 13596 & 8127 & 13596 \\
 & $10^{-2}$ & 192 & 240 & 92 & \textbf{187} & 114 & \underline{\textbf{174}} & 116 & \textbf{176} & 12159 & 20351 & 12159 & 20351 \\
 & $10^{-3}$ & 640 & 801 & 250 & \underline{\textbf{724}} & 565 & \textbf{749} & 600 & 808 & 16191 & 27106 & 20223 & 33861 \\
 & $10^{-4}$ & 2112 & 2643 & 1776 & 2815 & 566 & \underline{\textbf{1893}} & 1297 & \textbf{2561} & 28287 & 47372 & 36351 & 60882 \\
 & $10^{-5}$ & - & - & - & - & - & - & - & - & 48447 & \underline{81148} & 64575 & 108169 \\
\bottomrule
\end{tabular}

%% file: tables_time/mean_75/random_gaussian/initial_0/min_phases_table_time.tex
\begin{tabular}{cc|cc|cc|cc|cc|cc|cc}
\toprule
 &  & \multicolumn{2}{c}{\texttt{Full}} & \multicolumn{2}{c}{\texttt{Diagonal (adaptive)}} & \multicolumn{2}{c}{\texttt{Diagonal (multiple)}} & \multicolumn{2}{c}{\texttt{Grid}} \vline & \multicolumn{2}{c}{\citet{PhysRevA.92.040303}} & \multicolumn{2}{c}{\citet{job2023efficient}} \\
 \multicolumn{1}{c}{d} & \multicolumn{1}{c}{Infidelity $\leq$} \vline & Phases & T $[\mu s]$ & Phases & T $[\mu s]$ & Phases & T $[\mu s]$ & Phases & T $[\mu s]$ & Phases & T $[\mu s]$ & Phases & T $[\mu s]$ \\
\midrule
\multirow[c]{5}{*}{8} & $10^{-1}$ & 16 & 56 & 8 & \underline{\textbf{47}} & 13 & \textbf{50} & 9 & \textbf{48} & 64 & 299 & 64 & 299 \\
 & $10^{-2}$ & 24 & 85 & 16 & \textbf{79} & 20 & \textbf{76} & 19 & \underline{\textbf{75}} & 120 & 570 & 120 & 570 \\
 & $10^{-3}$ & 24 & \underline{85} & 24 & 98 & 22 & 93 & 23 & 96 & 176 & 841 & 176 & 841 \\
 & $10^{-4}$ & 32 & 113 & 24 & \underline{\textbf{109}} & 25 & \textbf{111} & 31 & \textbf{111} & 232 & 1112 & 288 & 1383 \\
 & $10^{-5}$ & 40 & \underline{141} & 42 & 171 & - & - & 45 & 164 & 400 & 1925 & 512 & 2467 \\
\midrule
\multirow[c]{5}{*}{16} & $10^{-1}$ & 32 & 80 & 26 & 87 & 25 & \textbf{70} & 24 & \underline{\textbf{68}} & 256 & 853 & 256 & 853 \\
 & $10^{-2}$ & 48 & 120 & 38 & 122 & 43 & \textbf{113} & 41 & \underline{\textbf{110}} & 496 & 1666 & 496 & 1666 \\
 & $10^{-3}$ & 64 & 160 & 50 & \textbf{157} & 53 & \underline{\textbf{144}} & 55 & \textbf{147} & 736 & 2478 & 976 & 3291 \\
 & $10^{-4}$ & 80 & 200 & 63 & \textbf{193} & 70 & \underline{\textbf{186}} & 76 & \textbf{195} & 1216 & 4103 & 1456 & 4915 \\
 & $10^{-5}$ & 128 & \underline{320} & 129 & 423 & - & - & - & - & 1936 & 6540 & 2416 & 8165 \\
\midrule
\multirow[c]{5}{*}{32} & $10^{-1}$ & 96 & 170 & 65 & \textbf{138} & 68 & \textbf{139} & 64 & \underline{\textbf{134}} & 2016 & 4778 & 2016 & 4778 \\
 & $10^{-2}$ & 128 & 226 & 86 & \textbf{184} & 88 & \underline{\textbf{162}} & 87 & \textbf{181} & 2016 & 4778 & 3008 & 7137 \\
 & $10^{-3}$ & 128 & 226 & 112 & 235 & 116 & 236 & 126 & \underline{\textbf{224}} & 3008 & 7137 & 4000 & 9496 \\
 & $10^{-4}$ & 160 & \underline{283} & 148 & 320 & 173 & 321 & 161 & 308 & 5984 & 14214 & 6976 & 16574 \\
 & $10^{-5}$ & 672 & 1190 & 673 & 1428 & - & - & 320 & \underline{\textbf{593}} & 9952 & 23651 & 11936 & 28370 \\
\midrule
\multirow[c]{5}{*}{64} & $10^{-1}$ & 192 & 240 & 144 & \textbf{206} & 115 & \underline{\textbf{175}} & 119 & \textbf{180} & 8128 & 13596 & 8128 & 13596 \\
 & $10^{-2}$ & 320 & 400 & 256 & 419 & 239 & \underline{\textbf{308}} & 269 & \textbf{364} & 12160 & 20351 & 12160 & 20351 \\
 & $10^{-3}$ & 320 & 400 & 292 & 480 & 296 & \underline{\textbf{384}} & 371 & 472 & 16192 & 27107 & 20224 & 33862 \\
 & $10^{-4}$ & 896 & 1121 & 1128 & 1801 & 771 & \underline{\textbf{1000}} & 954 & 1236 & 28288 & 47372 & 36352 & 60883 \\
 & $10^{-5}$ & - & - & 3570 & \underline{4738} & - & - & - & - & 48448 & 81148 & 64576 & 108169 \\
\bottomrule
\end{tabular}

%% file: photon_loss/tables_phases/mean_75/fourier_5/initial_0/photon_loss_table.tex
\begin{tabular}{lcl|c|c|c|c}
\toprule
d & $\beta$ & $T_1$ [$ms$] & \texttt{Full} & \texttt{Diagonal (adaptive)} & \texttt{Diagonal (multiple)} & \texttt{Grid} \\
\midrule
\multirow[c]{16}{*}{8} & \multirow[c]{4}{*}{$0.0$} & $ 10^{-1} $ & $ 8.8 \pm 1.0 \times 10^{-2} $ & $ 1.5 \pm 0.5 \times 10^{-1} $ & $ 9.2 \pm 1.0 \times 10^{-2} $ & $ 8.3 \pm 0.4 \times 10^{-2} $ \\
 &  & $ 10^{0} $ & $ 2.7 \pm 0.4 \times 10^{-2} $ & $ 4.4 \pm 1.3 \times 10^{-2} $ & $ 2.9 \pm 0.3 \times 10^{-2} $ & $ 2.7 \pm 0.4 \times 10^{-2} $ \\
 &  & $ 10^{1} $ & $ 3.5 \pm 0.5 \times 10^{-3} $ & $ 6.2 \pm 1.3 \times 10^{-3} $ & $ 4.1 \pm 0.7 \times 10^{-3} $ & $ 3.8 \pm 0.6 \times 10^{-3} $ \\
 &  & $ 10^{2} $ & $ 5.0 \pm 0.9 \times 10^{-4} $ & $ 7.0 \pm 1.6 \times 10^{-4} $ & $ 7.7 \pm 2.5 \times 10^{-4} $ & $ 5.9 \pm 1.1 \times 10^{-4} $ \\
\cline{2-7}
 & \multirow[c]{4}{*}{$ 10^{-5} $} & $ 10^{-1} $ & $ 9.3 \pm 2.1 \times 10^{-2} $ & $ 1.5 \pm 0.5 \times 10^{-1} $ & $ 1.4 \pm 0.7 \times 10^{-1} $ & $ 9.1 \pm 1.8 \times 10^{-2} $ \\
 &  & $ 10^{0} $ & $ 2.8 \pm 0.5 \times 10^{-2} $ & $ 4.4 \pm 1.3 \times 10^{-2} $ & $ 4.6 \pm 1.3 \times 10^{-2} $ & $ 2.8 \pm 0.5 \times 10^{-2} $ \\
 &  & $ 10^{1} $ & $ 3.6 \pm 0.6 \times 10^{-3} $ & $ 6.2 \pm 1.4 \times 10^{-3} $ & $ 6.3 \pm 1.5 \times 10^{-3} $ & $ 3.5 \pm 0.5 \times 10^{-3} $ \\
 &  & $ 10^{2} $ & $ 4.4 \pm 0.7 \times 10^{-4} $ & $ 8.6 \pm 1.3 \times 10^{-4} $ & $ 7.9 \pm 2.1 \times 10^{-4} $ & $ 4.9 \pm 0.9 \times 10^{-4} $ \\
\cline{2-7}
 & \multirow[c]{4}{*}{$ 10^{-3} $} & $ 10^{-1} $ & $ 9.1 \pm 2.1 \times 10^{-2} $ & $ 1.6 \pm 0.5 \times 10^{-1} $ & $ 1.8 \pm 0.9 \times 10^{-1} $ & $ 1.4 \pm 0.6 \times 10^{-1} $ \\
 &  & $ 10^{0} $ & $ 2.9 \pm 0.4 \times 10^{-2} $ & $ 5.1 \pm 1.4 \times 10^{-2} $ & $ 5.5 \pm 1.6 \times 10^{-2} $ & $ 4.4 \pm 1.1 \times 10^{-2} $ \\
 &  & $ 10^{1} $ & $ 3.5 \pm 0.6 \times 10^{-3} $ & $ 8.7 \pm 2.1 \times 10^{-3} $ & $ 7.8 \pm 1.8 \times 10^{-3} $ & $ 6.2 \pm 1.5 \times 10^{-3} $ \\
 &  & $ 10^{2} $ & $ 5.0 \pm 1.3 \times 10^{-4} $ & $ 1.5 \pm 0.3 \times 10^{-3} $ & $ 9.5 \pm 2.0 \times 10^{-4} $ & $ 8.1 \pm 1.9 \times 10^{-4} $ \\
\cline{2-7}
 & \multirow[c]{4}{*}{$ 10^{-1} $} & $ 10^{-1} $ & $ 1.0 \pm 0.3 \times 10^{-1} $ & $ 1.6 \pm 0.8 \times 10^{-1} $ & $ 9.2 \pm 2.1 \times 10^{-2} $ & $ 9.4 \pm 3.0 \times 10^{-2} $ \\
 &  & $ 10^{0} $ & $ 6.3 \pm 2.5 \times 10^{-2} $ & $ 6.5 \pm 1.9 \times 10^{-2} $ & $ 3.5 \pm 0.8 \times 10^{-2} $ & $ 3.8 \pm 1.1 \times 10^{-2} $ \\
 &  & $ 10^{1} $ & $ 6.3 \pm 2.8 \times 10^{-2} $ & $ 1.4 \pm 0.5 \times 10^{-2} $ & $ 2.3 \pm 1.2 \times 10^{-2} $ & $ 1.5 \pm 0.9 \times 10^{-2} $ \\
 &  & $ 10^{2} $ & $ 7.0 \pm 3.2 \times 10^{-2} $ & $ 6.1 \pm 6.8 \times 10^{-3} $ & $ 2.1 \pm 1.2 \times 10^{-2} $ & $ 1.3 \pm 1.0 \times 10^{-2} $ \\
\midrule
\multirow[c]{16}{*}{16} & \multirow[c]{4}{*}{$0.0$} & $ 10^{-1} $ & $ 1.5 \pm 0.0 \times 10^{-1} $ & $ 1.6 \pm 0.0 \times 10^{-1} $ & $ 1.4 \pm 0.0 \times 10^{-1} $ & $ 1.5 \pm 0.0 \times 10^{-1} $ \\
 &  & $ 10^{0} $ & $ 4.9 \pm 0.3 \times 10^{-2} $ & $ 6.9 \pm 1.0 \times 10^{-2} $ & $ 4.9 \pm 0.3 \times 10^{-2} $ & $ 4.9 \pm 0.3 \times 10^{-2} $ \\
 &  & $ 10^{1} $ & $ 9.6 \pm 1.1 \times 10^{-3} $ & $ 1.3 \pm 0.2 \times 10^{-2} $ & $ 1.3 \pm 0.2 \times 10^{-2} $ & $ 1.0 \pm 0.1 \times 10^{-2} $ \\
 &  & $ 10^{2} $ & $ 1.3 \pm 0.1 \times 10^{-3} $ & $ 1.7 \pm 0.2 \times 10^{-3} $ & $ 2.6 \pm 0.8 \times 10^{-3} $ & $ 1.5 \pm 0.2 \times 10^{-3} $ \\
\cline{2-7}
 & \multirow[c]{4}{*}{$ 10^{-5} $} & $ 10^{-1} $ & $ 1.5 \pm 0.0 \times 10^{-1} $ & $ 1.5 \pm 0.0 \times 10^{-1} $ & $ 1.5 \pm 0.0 \times 10^{-1} $ & $ 1.5 \pm 0.0 \times 10^{-1} $ \\
 &  & $ 10^{0} $ & $ 4.9 \pm 0.3 \times 10^{-2} $ & $ 5.2 \pm 0.4 \times 10^{-2} $ & $ 4.8 \pm 0.2 \times 10^{-2} $ & $ 5.0 \pm 0.2 \times 10^{-2} $ \\
 &  & $ 10^{1} $ & $ 1.1 \pm 0.1 \times 10^{-2} $ & $ 1.2 \pm 0.2 \times 10^{-2} $ & $ 1.3 \pm 0.2 \times 10^{-2} $ & $ 1.1 \pm 0.2 \times 10^{-2} $ \\
 &  & $ 10^{2} $ & $ 1.5 \pm 0.2 \times 10^{-3} $ & $ 1.6 \pm 0.3 \times 10^{-3} $ & $ 1.8 \pm 0.4 \times 10^{-3} $ & $ 1.6 \pm 0.3 \times 10^{-3} $ \\
\cline{2-7}
 & \multirow[c]{4}{*}{$ 10^{-3} $} & $ 10^{-1} $ & $ 1.5 \pm 0.0 \times 10^{-1} $ & $ 1.8 \pm 0.1 \times 10^{-1} $ & $ 1.5 \pm 0.0 \times 10^{-1} $ & $ 1.5 \pm 0.0 \times 10^{-1} $ \\
 &  & $ 10^{0} $ & $ 4.9 \pm 0.3 \times 10^{-2} $ & $ 8.0 \pm 0.8 \times 10^{-2} $ & $ 4.9 \pm 0.3 \times 10^{-2} $ & $ 4.8 \pm 0.3 \times 10^{-2} $ \\
 &  & $ 10^{1} $ & $ 1.1 \pm 0.2 \times 10^{-2} $ & $ 1.4 \pm 0.2 \times 10^{-2} $ & $ 9.9 \pm 1.4 \times 10^{-3} $ & $ 1.0 \pm 0.2 \times 10^{-2} $ \\
 &  & $ 10^{2} $ & $ 1.4 \pm 0.3 \times 10^{-3} $ & $ 2.1 \pm 0.6 \times 10^{-3} $ & $ 1.3 \pm 0.2 \times 10^{-3} $ & $ 1.3 \pm 0.2 \times 10^{-3} $ \\
\cline{2-7}
 & \multirow[c]{4}{*}{$ 10^{-1} $} & $ 10^{-1} $ & $ 1.6 \pm 0.0 \times 10^{-1} $ & $ 1.8 \pm 0.0 \times 10^{-1} $ & $ 1.2 \pm 0.1 \times 10^{-1} $ & $ 1.4 \pm 0.0 \times 10^{-1} $ \\
 &  & $ 10^{0} $ & $ 6.7 \pm 0.1 \times 10^{-2} $ & $ 6.0 \pm 0.5 \times 10^{-2} $ & $ 4.3 \pm 0.4 \times 10^{-2} $ & $ 5.9 \pm 0.8 \times 10^{-2} $ \\
 &  & $ 10^{1} $ & $ 5.1 \pm 0.5 \times 10^{-2} $ & $ 1.7 \pm 0.4 \times 10^{-2} $ & $ 2.0 \pm 0.3 \times 10^{-2} $ & $ 4.0 \pm 0.9 \times 10^{-2} $ \\
 &  & $ 10^{2} $ & $ 4.7 \pm 0.3 \times 10^{-2} $ & $ 1.2 \pm 0.4 \times 10^{-2} $ & $ 1.7 \pm 0.3 \times 10^{-2} $ & $ 3.8 \pm 0.8 \times 10^{-2} $ \\
\midrule
\multirow[c]{16}{*}{32} & \multirow[c]{4}{*}{$0.0$} & $ 10^{-1} $ & $ 2.9 \pm 0.0 \times 10^{-1} $ & $ 2.8 \pm 0.0 \times 10^{-1} $ & $ 2.8 \pm 0.0 \times 10^{-1} $ & $ 2.7 \pm 0.0 \times 10^{-1} $ \\
 &  & $ 10^{0} $ & $ 9.9 \pm 0.1 \times 10^{-2} $ & $ 1.1 \pm 0.1 \times 10^{-1} $ & $ 1.0 \pm 0.0 \times 10^{-1} $ & $ 1.0 \pm 0.1 \times 10^{-1} $ \\
 &  & $ 10^{1} $ & $ 2.4 \pm 0.1 \times 10^{-2} $ & $ 3.7 \pm 0.6 \times 10^{-2} $ & $ 2.5 \pm 0.2 \times 10^{-2} $ & $ 2.5 \pm 0.2 \times 10^{-2} $ \\
 &  & $ 10^{2} $ & $ 4.8 \pm 0.7 \times 10^{-3} $ & $ 7.2 \pm 1.1 \times 10^{-3} $ & $ 7.3 \pm 1.1 \times 10^{-3} $ & $ 5.8 \pm 1.0 \times 10^{-3} $ \\
\cline{2-7}
 & \multirow[c]{4}{*}{$ 10^{-5} $} & $ 10^{-1} $ & $ 2.9 \pm 0.0 \times 10^{-1} $ & $ 2.7 \pm 0.0 \times 10^{-1} $ & $ 2.9 \pm 0.0 \times 10^{-1} $ & $ 2.8 \pm 0.0 \times 10^{-1} $ \\
 &  & $ 10^{0} $ & $ 9.9 \pm 0.1 \times 10^{-2} $ & $ 1.1 \pm 0.1 \times 10^{-1} $ & $ 1.1 \pm 0.0 \times 10^{-1} $ & $ 1.0 \pm 0.1 \times 10^{-1} $ \\
 &  & $ 10^{1} $ & $ 2.5 \pm 0.1 \times 10^{-2} $ & $ 3.7 \pm 0.7 \times 10^{-2} $ & $ 2.5 \pm 0.2 \times 10^{-2} $ & $ 2.5 \pm 0.2 \times 10^{-2} $ \\
 &  & $ 10^{2} $ & $ 5.3 \pm 1.0 \times 10^{-3} $ & $ 7.1 \pm 1.3 \times 10^{-3} $ & $ 7.3 \pm 0.9 \times 10^{-3} $ & $ 6.9 \pm 1.1 \times 10^{-3} $ \\
\cline{2-7}
 & \multirow[c]{4}{*}{$ 10^{-3} $} & $ 10^{-1} $ & $ 2.9 \pm 0.0 \times 10^{-1} $ & $ 3.3 \pm 0.1 \times 10^{-1} $ & $ 3.0 \pm 0.0 \times 10^{-1} $ & $ 2.7 \pm 0.0 \times 10^{-1} $ \\
 &  & $ 10^{0} $ & $ 9.9 \pm 0.2 \times 10^{-2} $ & $ 2.3 \pm 0.2 \times 10^{-1} $ & $ 1.5 \pm 0.1 \times 10^{-1} $ & $ 1.2 \pm 0.1 \times 10^{-1} $ \\
 &  & $ 10^{1} $ & $ 2.4 \pm 0.1 \times 10^{-2} $ & $ 4.7 \pm 0.3 \times 10^{-2} $ & $ 3.2 \pm 0.4 \times 10^{-2} $ & $ 2.7 \pm 0.2 \times 10^{-2} $ \\
 &  & $ 10^{2} $ & $ 5.5 \pm 1.1 \times 10^{-3} $ & $ 7.8 \pm 1.0 \times 10^{-3} $ & $ 9.4 \pm 1.3 \times 10^{-3} $ & $ 8.4 \pm 0.9 \times 10^{-3} $ \\
\cline{2-7}
 & \multirow[c]{4}{*}{$ 10^{-1} $} & $ 10^{-1} $ & $ 2.9 \pm 0.0 \times 10^{-1} $ & $ 3.2 \pm 0.1 \times 10^{-1} $ & $ 3.0 \pm 0.1 \times 10^{-1} $ & $ 2.6 \pm 0.1 \times 10^{-1} $ \\
 &  & $ 10^{0} $ & $ 1.1 \pm 0.1 \times 10^{-1} $ & $ 1.4 \pm 0.1 \times 10^{-1} $ & $ 1.3 \pm 0.1 \times 10^{-1} $ & $ 8.5 \pm 0.6 \times 10^{-2} $ \\
 &  & $ 10^{1} $ & $ 3.2 \pm 0.1 \times 10^{-2} $ & $ 3.2 \pm 0.4 \times 10^{-2} $ & $ 5.5 \pm 2.5 \times 10^{-2} $ & $ 2.5 \pm 0.3 \times 10^{-2} $ \\
 &  & $ 10^{2} $ & $ 1.4 \pm 0.1 \times 10^{-2} $ & $ 1.5 \pm 0.5 \times 10^{-2} $ & $ 2.5 \pm 1.1 \times 10^{-2} $ & $ 1.4 \pm 0.3 \times 10^{-2} $ \\
\midrule
\multirow[c]{16}{*}{64} & \multirow[c]{4}{*}{$0.0$} & $ 10^{-1} $ & $ 4.5 \pm 0.0 \times 10^{-1} $ & $ 4.7 \pm 0.0 \times 10^{-1} $ & $ 4.8 \pm 0.5 \times 10^{-1} $ & $ 4.9 \pm 0.0 \times 10^{-1} $ \\
 &  & $ 10^{0} $ & $ 2.0 \pm 0.0 \times 10^{-1} $ & $ 4.1 \pm 0.2 \times 10^{-1} $ & $ 2.2 \pm 0.2 \times 10^{-1} $ & $ 2.6 \pm 0.3 \times 10^{-1} $ \\
 &  & $ 10^{1} $ & $ 4.9 \pm 0.4 \times 10^{-2} $ & $ 1.8 \pm 0.3 \times 10^{-1} $ & $ 5.2 \pm 0.1 \times 10^{-2} $ & $ 9.0 \pm 1.6 \times 10^{-2} $ \\
 &  & $ 10^{2} $ & $ 1.3 \pm 0.1 \times 10^{-2} $ & $ 3.4 \pm 0.3 \times 10^{-2} $ & $ 2.0 \pm 0.4 \times 10^{-2} $ & $ 2.9 \pm 0.3 \times 10^{-2} $ \\
\cline{2-7}
 & \multirow[c]{4}{*}{$ 10^{-5} $} & $ 10^{-1} $ & $ 4.5 \pm 0.0 \times 10^{-1} $ & $ 4.2 \pm 0.0 \times 10^{-1} $ & $ 4.6 \pm 0.1 \times 10^{-1} $ & $ 4.5 \pm 0.0 \times 10^{-1} $ \\
 &  & $ 10^{0} $ & $ 2.0 \pm 0.0 \times 10^{-1} $ & $ 3.7 \pm 0.8 \times 10^{-1} $ & $ 2.3 \pm 0.1 \times 10^{-1} $ & $ 2.0 \pm 0.0 \times 10^{-1} $ \\
 &  & $ 10^{1} $ & $ 4.8 \pm 0.4 \times 10^{-2} $ & $ 1.7 \pm 0.2 \times 10^{-1} $ & $ 5.1 \pm 0.4 \times 10^{-2} $ & $ 4.4 \pm 0.3 \times 10^{-2} $ \\
 &  & $ 10^{2} $ & $ 1.1 \pm 0.1 \times 10^{-2} $ & $ 3.0 \pm 0.2 \times 10^{-2} $ & $ 2.2 \pm 0.7 \times 10^{-2} $ & $ 1.2 \pm 0.1 \times 10^{-2} $ \\
\cline{2-7}
 & \multirow[c]{4}{*}{$ 10^{-3} $} & $ 10^{-1} $ & $ 4.5 \pm 0.0 \times 10^{-1} $ & $ 4.8 \pm 0.1 \times 10^{-1} $ & $ 4.5 \pm 0.0 \times 10^{-1} $ & $ 4.6 \pm 0.0 \times 10^{-1} $ \\
 &  & $ 10^{0} $ & $ 2.2 \pm 0.1 \times 10^{-1} $ & $ 4.3 \pm 0.3 \times 10^{-1} $ & $ 2.2 \pm 0.1 \times 10^{-1} $ & $ 2.2 \pm 0.1 \times 10^{-1} $ \\
 &  & $ 10^{1} $ & $ 5.3 \pm 0.6 \times 10^{-2} $ & $ 1.3 \pm 0.1 \times 10^{-1} $ & $ 5.1 \pm 0.5 \times 10^{-2} $ & $ 5.4 \pm 0.6 \times 10^{-2} $ \\
 &  & $ 10^{2} $ & $ 1.1 \pm 0.1 \times 10^{-2} $ & $ 2.1 \pm 0.2 \times 10^{-2} $ & $ 1.0 \pm 0.1 \times 10^{-2} $ & $ 1.9 \pm 0.2 \times 10^{-2} $ \\
\cline{2-7}
 & \multirow[c]{4}{*}{$ 10^{-1} $} & $ 10^{-1} $ & $ 4.5 \pm 0.0 \times 10^{-1} $ & $ 4.5 \pm 0.0 \times 10^{-1} $ & $ 4.6 \pm 0.1 \times 10^{-1} $ & $ 4.1 \pm 0.0 \times 10^{-1} $ \\
 &  & $ 10^{0} $ & $ 2.3 \pm 0.1 \times 10^{-1} $ & $ 2.6 \pm 0.1 \times 10^{-1} $ & $ 3.1 \pm 0.6 \times 10^{-1} $ & $ 1.8 \pm 0.1 \times 10^{-1} $ \\
 &  & $ 10^{1} $ & $ 9.2 \pm 7.5 \times 10^{-2} $ & $ 5.9 \pm 0.4 \times 10^{-2} $ & $ 8.7 \pm 3.1 \times 10^{-2} $ & $ 5.5 \pm 1.0 \times 10^{-2} $ \\
 &  & $ 10^{2} $ & $ 1.9 \pm 0.1 \times 10^{-2} $ & $ 1.4 \pm 0.2 \times 10^{-2} $ & $ 2.5 \pm 0.7 \times 10^{-2} $ & $ 4.1 \pm 1.2 \times 10^{-2} $ \\
\bottomrule
\end{tabular}

%% file: photon_loss/tables_phases/mean_75/random_gaussian/initial_0/photon_loss_table.tex
\begin{tabular}{lcl|c|c|c|c}
\toprule
d & $\beta$ & $T_1$ [$ms$] & \texttt{Full} & \texttt{Diagonal (adaptive)} & \texttt{Diagonal (multiple)} & \texttt{Grid} \\
\midrule
\multirow[c]{16}{*}{8} & \multirow[c]{4}{*}{$0.0$} & $ 10^{-1} $ & $ 5.2 \pm 0.0 \times 10^{-1} $ & $ 5.4 \pm 0.2 \times 10^{-1} $ & $ 5.1 \pm 0.1 \times 10^{-1} $ & $ 5.3 \pm 0.0 \times 10^{-1} $ \\
 &  & $ 10^{0} $ & $ 1.6 \pm 0.2 \times 10^{-1} $ & $ 1.6 \pm 0.2 \times 10^{-1} $ & $ 1.6 \pm 0.2 \times 10^{-1} $ & $ 1.6 \pm 0.2 \times 10^{-1} $ \\
 &  & $ 10^{1} $ & $ 1.8 \pm 0.1 \times 10^{-2} $ & $ 1.9 \pm 0.2 \times 10^{-2} $ & $ 1.9 \pm 0.3 \times 10^{-2} $ & $ 1.8 \pm 0.2 \times 10^{-2} $ \\
 &  & $ 10^{2} $ & $ 2.0 \pm 0.2 \times 10^{-3} $ & $ 2.3 \pm 0.5 \times 10^{-3} $ & $ 2.2 \pm 0.3 \times 10^{-3} $ & $ 2.0 \pm 0.2 \times 10^{-3} $ \\
\cline{2-7}
 & \multirow[c]{4}{*}{$ 10^{-5} $} & $ 10^{-1} $ & $ 5.2 \pm 0.0 \times 10^{-1} $ & $ 5.3 \pm 0.1 \times 10^{-1} $ & $ 5.2 \pm 0.0 \times 10^{-1} $ & $ 5.2 \pm 0.0 \times 10^{-1} $ \\
 &  & $ 10^{0} $ & $ 1.6 \pm 0.2 \times 10^{-1} $ & $ 1.6 \pm 0.2 \times 10^{-1} $ & $ 1.6 \pm 0.3 \times 10^{-1} $ & $ 1.6 \pm 0.1 \times 10^{-1} $ \\
 &  & $ 10^{1} $ & $ 1.8 \pm 0.2 \times 10^{-2} $ & $ 1.9 \pm 0.2 \times 10^{-2} $ & $ 1.8 \pm 0.3 \times 10^{-2} $ & $ 1.8 \pm 0.2 \times 10^{-2} $ \\
 &  & $ 10^{2} $ & $ 2.2 \pm 0.3 \times 10^{-3} $ & $ 2.0 \pm 0.3 \times 10^{-3} $ & $ 2.1 \pm 0.4 \times 10^{-3} $ & $ 2.1 \pm 0.3 \times 10^{-3} $ \\
\cline{2-7}
 & \multirow[c]{4}{*}{$ 10^{-3} $} & $ 10^{-1} $ & $ 5.2 \pm 0.0 \times 10^{-1} $ & $ 5.2 \pm 0.1 \times 10^{-1} $ & $ 5.1 \pm 0.0 \times 10^{-1} $ & $ 5.2 \pm 0.0 \times 10^{-1} $ \\
 &  & $ 10^{0} $ & $ 1.6 \pm 0.1 \times 10^{-1} $ & $ 1.6 \pm 0.2 \times 10^{-1} $ & $ 1.6 \pm 0.2 \times 10^{-1} $ & $ 1.6 \pm 0.2 \times 10^{-1} $ \\
 &  & $ 10^{1} $ & $ 1.8 \pm 0.2 \times 10^{-2} $ & $ 1.9 \pm 0.3 \times 10^{-2} $ & $ 2.1 \pm 0.3 \times 10^{-2} $ & $ 1.9 \pm 0.3 \times 10^{-2} $ \\
 &  & $ 10^{2} $ & $ 2.1 \pm 0.3 \times 10^{-3} $ & $ 2.7 \pm 0.5 \times 10^{-3} $ & $ 2.8 \pm 0.5 \times 10^{-3} $ & $ 2.2 \pm 0.3 \times 10^{-3} $ \\
\cline{2-7}
 & \multirow[c]{4}{*}{$ 10^{-1} $} & $ 10^{-1} $ & $ 5.3 \pm 0.0 \times 10^{-1} $ & $ 5.3 \pm 0.1 \times 10^{-1} $ & $ 5.2 \pm 0.1 \times 10^{-1} $ & $ 5.1 \pm 0.1 \times 10^{-1} $ \\
 &  & $ 10^{0} $ & $ 1.7 \pm 0.2 \times 10^{-1} $ & $ 1.6 \pm 0.3 \times 10^{-1} $ & $ 1.6 \pm 0.2 \times 10^{-1} $ & $ 1.7 \pm 0.3 \times 10^{-1} $ \\
 &  & $ 10^{1} $ & $ 2.5 \pm 0.6 \times 10^{-2} $ & $ 4.0 \pm 2.5 \times 10^{-2} $ & $ 2.6 \pm 0.9 \times 10^{-2} $ & $ 3.1 \pm 1.6 \times 10^{-2} $ \\
 &  & $ 10^{2} $ & $ 6.1 \pm 4.7 \times 10^{-3} $ & $ 3.2 \pm 3.5 \times 10^{-2} $ & $ 5.8 \pm 3.8 \times 10^{-3} $ & $ 7.6 \pm 7.6 \times 10^{-3} $ \\
\midrule
\multirow[c]{16}{*}{16} & \multirow[c]{4}{*}{$0.0$} & $ 10^{-1} $ & $ 7.6 \pm 0.0 \times 10^{-1} $ & $ 7.7 \pm 0.1 \times 10^{-1} $ & $ 7.6 \pm 0.0 \times 10^{-1} $ & $ 7.5 \pm 0.1 \times 10^{-1} $ \\
 &  & $ 10^{0} $ & $ 3.2 \pm 0.2 \times 10^{-1} $ & $ 3.9 \pm 0.2 \times 10^{-1} $ & $ 3.3 \pm 0.2 \times 10^{-1} $ & $ 3.2 \pm 0.2 \times 10^{-1} $ \\
 &  & $ 10^{1} $ & $ 4.7 \pm 0.4 \times 10^{-2} $ & $ 5.6 \pm 0.4 \times 10^{-2} $ & $ 4.8 \pm 0.6 \times 10^{-2} $ & $ 5.0 \pm 0.6 \times 10^{-2} $ \\
 &  & $ 10^{2} $ & $ 5.3 \pm 0.4 \times 10^{-3} $ & $ 6.4 \pm 0.5 \times 10^{-3} $ & $ 5.7 \pm 0.7 \times 10^{-3} $ & $ 6.3 \pm 0.7 \times 10^{-3} $ \\
\cline{2-7}
 & \multirow[c]{4}{*}{$ 10^{-5} $} & $ 10^{-1} $ & $ 7.6 \pm 0.0 \times 10^{-1} $ & $ 7.7 \pm 0.1 \times 10^{-1} $ & $ 7.5 \pm 0.0 \times 10^{-1} $ & $ 7.4 \pm 0.1 \times 10^{-1} $ \\
 &  & $ 10^{0} $ & $ 3.3 \pm 0.2 \times 10^{-1} $ & $ 3.6 \pm 0.2 \times 10^{-1} $ & $ 3.3 \pm 0.2 \times 10^{-1} $ & $ 3.3 \pm 0.3 \times 10^{-1} $ \\
 &  & $ 10^{1} $ & $ 4.6 \pm 0.4 \times 10^{-2} $ & $ 5.3 \pm 0.5 \times 10^{-2} $ & $ 4.7 \pm 0.5 \times 10^{-2} $ & $ 5.5 \pm 0.6 \times 10^{-2} $ \\
 &  & $ 10^{2} $ & $ 5.1 \pm 0.4 \times 10^{-3} $ & $ 6.3 \pm 0.6 \times 10^{-3} $ & $ 5.7 \pm 0.5 \times 10^{-3} $ & $ 6.5 \pm 0.7 \times 10^{-3} $ \\
\cline{2-7}
 & \multirow[c]{4}{*}{$ 10^{-3} $} & $ 10^{-1} $ & $ 7.6 \pm 0.0 \times 10^{-1} $ & $ 7.8 \pm 0.1 \times 10^{-1} $ & $ 7.5 \pm 0.0 \times 10^{-1} $ & $ 7.5 \pm 0.0 \times 10^{-1} $ \\
 &  & $ 10^{0} $ & $ 3.3 \pm 0.2 \times 10^{-1} $ & $ 4.2 \pm 0.2 \times 10^{-1} $ & $ 3.2 \pm 0.2 \times 10^{-1} $ & $ 3.3 \pm 0.3 \times 10^{-1} $ \\
 &  & $ 10^{1} $ & $ 4.8 \pm 0.5 \times 10^{-2} $ & $ 6.4 \pm 0.5 \times 10^{-2} $ & $ 4.8 \pm 0.6 \times 10^{-2} $ & $ 5.2 \pm 0.7 \times 10^{-2} $ \\
 &  & $ 10^{2} $ & $ 5.7 \pm 0.6 \times 10^{-3} $ & $ 7.5 \pm 0.5 \times 10^{-3} $ & $ 5.4 \pm 0.5 \times 10^{-3} $ & $ 5.8 \pm 0.6 \times 10^{-3} $ \\
\cline{2-7}
 & \multirow[c]{4}{*}{$ 10^{-1} $} & $ 10^{-1} $ & $ 7.6 \pm 0.0 \times 10^{-1} $ & $ 7.9 \pm 0.1 \times 10^{-1} $ & $ 7.4 \pm 0.1 \times 10^{-1} $ & $ 7.3 \pm 0.1 \times 10^{-1} $ \\
 &  & $ 10^{0} $ & $ 3.4 \pm 0.2 \times 10^{-1} $ & $ 4.1 \pm 0.1 \times 10^{-1} $ & $ 3.6 \pm 0.2 \times 10^{-1} $ & $ 3.6 \pm 0.2 \times 10^{-1} $ \\
 &  & $ 10^{1} $ & $ 6.0 \pm 1.2 \times 10^{-2} $ & $ 7.1 \pm 1.3 \times 10^{-2} $ & $ 8.3 \pm 1.8 \times 10^{-2} $ & $ 8.6 \pm 2.0 \times 10^{-2} $ \\
 &  & $ 10^{2} $ & $ 1.2 \pm 0.6 \times 10^{-2} $ & $ 1.7 \pm 1.0 \times 10^{-2} $ & $ 3.6 \pm 1.9 \times 10^{-2} $ & $ 4.6 \pm 2.2 \times 10^{-2} $ \\
\midrule
\multirow[c]{16}{*}{32} & \multirow[c]{4}{*}{$0.0$} & $ 10^{-1} $ & $ 8.5 \pm 0.0 \times 10^{-1} $ & $ 8.7 \pm 0.0 \times 10^{-1} $ & $ 8.5 \pm 0.0 \times 10^{-1} $ & $ 8.5 \pm 0.0 \times 10^{-1} $ \\
 &  & $ 10^{0} $ & $ 5.7 \pm 0.1 \times 10^{-1} $ & $ 7.2 \pm 0.2 \times 10^{-1} $ & $ 5.7 \pm 0.1 \times 10^{-1} $ & $ 5.8 \pm 0.1 \times 10^{-1} $ \\
 &  & $ 10^{1} $ & $ 1.3 \pm 0.1 \times 10^{-1} $ & $ 1.7 \pm 0.1 \times 10^{-1} $ & $ 1.3 \pm 0.1 \times 10^{-1} $ & $ 1.4 \pm 0.1 \times 10^{-1} $ \\
 &  & $ 10^{2} $ & $ 1.7 \pm 0.1 \times 10^{-2} $ & $ 2.0 \pm 0.1 \times 10^{-2} $ & $ 1.6 \pm 0.1 \times 10^{-2} $ & $ 1.7 \pm 0.2 \times 10^{-2} $ \\
\cline{2-7}
 & \multirow[c]{4}{*}{$ 10^{-5} $} & $ 10^{-1} $ & $ 8.5 \pm 0.0 \times 10^{-1} $ & $ 8.8 \pm 0.0 \times 10^{-1} $ & $ 8.5 \pm 0.0 \times 10^{-1} $ & $ 8.5 \pm 0.0 \times 10^{-1} $ \\
 &  & $ 10^{0} $ & $ 5.8 \pm 0.1 \times 10^{-1} $ & $ 7.2 \pm 0.1 \times 10^{-1} $ & $ 5.8 \pm 0.1 \times 10^{-1} $ & $ 6.2 \pm 0.1 \times 10^{-1} $ \\
 &  & $ 10^{1} $ & $ 1.3 \pm 0.1 \times 10^{-1} $ & $ 1.7 \pm 0.1 \times 10^{-1} $ & $ 1.5 \pm 0.1 \times 10^{-1} $ & $ 1.5 \pm 0.1 \times 10^{-1} $ \\
 &  & $ 10^{2} $ & $ 1.6 \pm 0.1 \times 10^{-2} $ & $ 2.4 \pm 0.3 \times 10^{-2} $ & $ 1.9 \pm 0.2 \times 10^{-2} $ & $ 1.8 \pm 0.1 \times 10^{-2} $ \\
\cline{2-7}
 & \multirow[c]{4}{*}{$ 10^{-3} $} & $ 10^{-1} $ & $ 8.5 \pm 0.0 \times 10^{-1} $ & $ 8.7 \pm 0.0 \times 10^{-1} $ & $ 8.5 \pm 0.0 \times 10^{-1} $ & $ 8.5 \pm 0.0 \times 10^{-1} $ \\
 &  & $ 10^{0} $ & $ 5.8 \pm 0.1 \times 10^{-1} $ & $ 7.5 \pm 0.1 \times 10^{-1} $ & $ 5.8 \pm 0.1 \times 10^{-1} $ & $ 5.8 \pm 0.1 \times 10^{-1} $ \\
 &  & $ 10^{1} $ & $ 1.3 \pm 0.1 \times 10^{-1} $ & $ 1.7 \pm 0.1 \times 10^{-1} $ & $ 1.3 \pm 0.1 \times 10^{-1} $ & $ 1.3 \pm 0.1 \times 10^{-1} $ \\
 &  & $ 10^{2} $ & $ 1.6 \pm 0.1 \times 10^{-2} $ & $ 2.1 \pm 0.1 \times 10^{-2} $ & $ 1.6 \pm 0.2 \times 10^{-2} $ & $ 1.6 \pm 0.1 \times 10^{-2} $ \\
\cline{2-7}
 & \multirow[c]{4}{*}{$ 10^{-1} $} & $ 10^{-1} $ & $ 8.5 \pm 0.0 \times 10^{-1} $ & $ 8.8 \pm 0.0 \times 10^{-1} $ & $ 8.5 \pm 0.0 \times 10^{-1} $ & $ 8.5 \pm 0.1 \times 10^{-1} $ \\
 &  & $ 10^{0} $ & $ 5.8 \pm 0.1 \times 10^{-1} $ & $ 7.4 \pm 0.1 \times 10^{-1} $ & $ 5.7 \pm 0.1 \times 10^{-1} $ & $ 6.2 \pm 0.2 \times 10^{-1} $ \\
 &  & $ 10^{1} $ & $ 1.5 \pm 0.3 \times 10^{-1} $ & $ 1.8 \pm 0.1 \times 10^{-1} $ & $ 1.4 \pm 0.2 \times 10^{-1} $ & $ 1.8 \pm 0.2 \times 10^{-1} $ \\
 &  & $ 10^{2} $ & $ 3.9 \pm 2.6 \times 10^{-2} $ & $ 2.6 \pm 0.4 \times 10^{-2} $ & $ 1.9 \pm 0.5 \times 10^{-2} $ & $ 3.8 \pm 1.6 \times 10^{-2} $ \\
\midrule
\multirow[c]{16}{*}{64} & \multirow[c]{4}{*}{$0.0$} & $ 10^{-1} $ & $ 8.6 \pm 0.0 \times 10^{-1} $ & $ 8.9 \pm 0.5 \times 10^{-1} $ & $ 9.1 \pm 0.6 \times 10^{-1} $ & $ 8.7 \pm 0.4 \times 10^{-1} $ \\
 &  & $ 10^{0} $ & $ 7.6 \pm 0.1 \times 10^{-1} $ & $ 8.1 \pm 0.2 \times 10^{-1} $ & $ 7.6 \pm 0.0 \times 10^{-1} $ & $ 7.7 \pm 0.0 \times 10^{-1} $ \\
 &  & $ 10^{1} $ & $ 3.2 \pm 0.1 \times 10^{-1} $ & $ 4.4 \pm 0.1 \times 10^{-1} $ & $ 3.2 \pm 0.1 \times 10^{-1} $ & $ 3.5 \pm 0.1 \times 10^{-1} $ \\
 &  & $ 10^{2} $ & $ 5.1 \pm 0.5 \times 10^{-2} $ & $ 9.0 \pm 1.8 \times 10^{-2} $ & $ 5.4 \pm 1.4 \times 10^{-2} $ & $ 6.4 \pm 0.9 \times 10^{-2} $ \\
\cline{2-7}
 & \multirow[c]{4}{*}{$ 10^{-5} $} & $ 10^{-1} $ & $ 8.6 \pm 0.0 \times 10^{-1} $ & $ 8.6 \pm 0.0 \times 10^{-1} $ & $ 8.6 \pm 0.0 \times 10^{-1} $ & $ 8.8 \pm 0.4 \times 10^{-1} $ \\
 &  & $ 10^{0} $ & $ 7.7 \pm 0.0 \times 10^{-1} $ & $ 8.4 \pm 0.2 \times 10^{-1} $ & $ 7.7 \pm 0.1 \times 10^{-1} $ & $ 7.9 \pm 0.1 \times 10^{-1} $ \\
 &  & $ 10^{1} $ & $ 3.7 \pm 0.1 \times 10^{-1} $ & $ 5.6 \pm 0.4 \times 10^{-1} $ & $ 4.4 \pm 0.1 \times 10^{-1} $ & $ 3.9 \pm 0.1 \times 10^{-1} $ \\
 &  & $ 10^{2} $ & $ 7.8 \pm 1.2 \times 10^{-2} $ & $ 1.4 \pm 0.2 \times 10^{-1} $ & $ 1.1 \pm 0.1 \times 10^{-1} $ & $ 7.9 \pm 1.6 \times 10^{-2} $ \\
\cline{2-7}
 & \multirow[c]{4}{*}{$ 10^{-3} $} & $ 10^{-1} $ & $ 8.6 \pm 0.0 \times 10^{-1} $ & $ 9.2 \pm 0.4 \times 10^{-1} $ & $ 8.6 \pm 0.0 \times 10^{-1} $ & $ 8.5 \pm 0.1 \times 10^{-1} $ \\
 &  & $ 10^{0} $ & $ 7.6 \pm 0.1 \times 10^{-1} $ & $ 8.8 \pm 0.2 \times 10^{-1} $ & $ 7.6 \pm 0.1 \times 10^{-1} $ & $ 8.4 \pm 0.2 \times 10^{-1} $ \\
 &  & $ 10^{1} $ & $ 3.1 \pm 0.1 \times 10^{-1} $ & $ 6.7 \pm 0.2 \times 10^{-1} $ & $ 3.2 \pm 0.1 \times 10^{-1} $ & $ 4.9 \pm 0.2 \times 10^{-1} $ \\
 &  & $ 10^{2} $ & $ 6.1 \pm 1.1 \times 10^{-2} $ & $ 1.7 \pm 0.1 \times 10^{-1} $ & $ 5.0 \pm 1.0 \times 10^{-2} $ & $ 1.3 \pm 0.1 \times 10^{-1} $ \\
\cline{2-7}
 & \multirow[c]{4}{*}{$ 10^{-1} $} & $ 10^{-1} $ & $ 8.7 \pm 0.4 \times 10^{-1} $ & $ 8.6 \pm 0.0 \times 10^{-1} $ & $ 8.5 \pm 0.0 \times 10^{-1} $ & $ 8.5 \pm 0.1 \times 10^{-1} $ \\
 &  & $ 10^{0} $ & $ 7.7 \pm 0.1 \times 10^{-1} $ & $ 8.4 \pm 0.0 \times 10^{-1} $ & $ 7.7 \pm 0.1 \times 10^{-1} $ & $ 8.0 \pm 0.2 \times 10^{-1} $ \\
 &  & $ 10^{1} $ & $ 3.5 \pm 0.2 \times 10^{-1} $ & $ 4.5 \pm 0.1 \times 10^{-1} $ & $ 4.1 \pm 0.2 \times 10^{-1} $ & $ 4.5 \pm 0.1 \times 10^{-1} $ \\
 &  & $ 10^{2} $ & $ 6.9 \pm 1.7 \times 10^{-2} $ & $ 8.7 \pm 1.7 \times 10^{-2} $ & $ 9.4 \pm 1.8 \times 10^{-2} $ & $ 1.5 \pm 0.2 \times 10^{-1} $ \\
\bottomrule
\end{tabular}

%% file: photon_loss/tables_time/mean_75/fourier_5/initial_0/photon_loss_table.tex
\begin{tabular}{lcl|c|c|c|c}
\toprule
d & $\beta$ & $T_1$ [$ms$] & \texttt{Full} & \texttt{Diagonal (adaptive)} & \texttt{Diagonal (multiple)} & \texttt{Grid} \\
\midrule
\multirow[c]{12}{*}{8} & \multirow[c]{4}{*}{$ 10^{-5} $} & $ 10^{-1} $ & $ 9.7 \pm 2.3 \times 10^{-2} $ & $ 1.3 \pm 0.3 \times 10^{-1} $ & $ 8.8 \pm 0.7 \times 10^{-2} $ & $ 1.0 \pm 0.3 \times 10^{-1} $ \\
 &  & $ 10^{0} $ & $ 2.9 \pm 0.5 \times 10^{-2} $ & $ 4.1 \pm 0.8 \times 10^{-2} $ & $ 2.8 \pm 0.4 \times 10^{-2} $ & $ 3.5 \pm 0.9 \times 10^{-2} $ \\
 &  & $ 10^{1} $ & $ 3.6 \pm 0.6 \times 10^{-3} $ & $ 5.4 \pm 1.1 \times 10^{-3} $ & $ 4.4 \pm 1.1 \times 10^{-3} $ & $ 4.9 \pm 1.6 \times 10^{-3} $ \\
 &  & $ 10^{2} $ & $ 4.4 \pm 0.8 \times 10^{-4} $ & $ 8.4 \pm 1.7 \times 10^{-4} $ & $ 6.1 \pm 1.6 \times 10^{-4} $ & $ 8.5 \pm 3.2 \times 10^{-4} $ \\
\cline{2-7}
 & \multirow[c]{4}{*}{$ 10^{-3} $} & $ 10^{-1} $ & $ 9.8 \pm 2.6 \times 10^{-2} $ & $ 1.2 \pm 0.3 \times 10^{-1} $ & $ 1.0 \pm 0.3 \times 10^{-1} $ & $ 1.1 \pm 0.3 \times 10^{-1} $ \\
 &  & $ 10^{0} $ & $ 2.7 \pm 0.4 \times 10^{-2} $ & $ 4.6 \pm 1.3 \times 10^{-2} $ & $ 3.0 \pm 0.5 \times 10^{-2} $ & $ 3.0 \pm 0.6 \times 10^{-2} $ \\
 &  & $ 10^{1} $ & $ 3.6 \pm 0.6 \times 10^{-3} $ & $ 6.5 \pm 1.6 \times 10^{-3} $ & $ 3.9 \pm 0.7 \times 10^{-3} $ & $ 4.4 \pm 1.4 \times 10^{-3} $ \\
 &  & $ 10^{2} $ & $ 5.0 \pm 1.2 \times 10^{-4} $ & $ 1.0 \pm 0.2 \times 10^{-3} $ & $ 7.9 \pm 2.1 \times 10^{-4} $ & $ 6.2 \pm 1.9 \times 10^{-4} $ \\
\cline{2-7}
 & \multirow[c]{4}{*}{$ 10^{-1} $} & $ 10^{-1} $ & $ 9.4 \pm 2.2 \times 10^{-2} $ & $ 9.8 \pm 2.7 \times 10^{-2} $ & $ 9.0 \pm 1.6 \times 10^{-2} $ & $ 9.2 \pm 1.9 \times 10^{-2} $ \\
 &  & $ 10^{0} $ & $ 2.9 \pm 0.5 \times 10^{-2} $ & $ 4.2 \pm 1.2 \times 10^{-2} $ & $ 3.0 \pm 0.6 \times 10^{-2} $ & $ 3.8 \pm 0.9 \times 10^{-2} $ \\
 &  & $ 10^{1} $ & $ 6.7 \pm 4.3 \times 10^{-3} $ & $ 1.2 \pm 0.6 \times 10^{-2} $ & $ 8.1 \pm 5.0 \times 10^{-3} $ & $ 1.5 \pm 1.0 \times 10^{-2} $ \\
 &  & $ 10^{2} $ & $ 1.6 \pm 0.8 \times 10^{-3} $ & $ 4.6 \pm 4.2 \times 10^{-3} $ & $ 3.0 \pm 2.9 \times 10^{-3} $ & $ 1.4 \pm 1.0 \times 10^{-2} $ \\
\midrule
\multirow[c]{12}{*}{16} & \multirow[c]{4}{*}{$ 10^{-5} $} & $ 10^{-1} $ & $ 1.5 \pm 0.0 \times 10^{-1} $ & $ 1.6 \pm 0.0 \times 10^{-1} $ & $ 1.5 \pm 0.0 \times 10^{-1} $ & $ 1.5 \pm 0.0 \times 10^{-1} $ \\
 &  & $ 10^{0} $ & $ 4.8 \pm 0.3 \times 10^{-2} $ & $ 6.2 \pm 0.6 \times 10^{-2} $ & $ 4.8 \pm 0.3 \times 10^{-2} $ & $ 4.9 \pm 0.2 \times 10^{-2} $ \\
 &  & $ 10^{1} $ & $ 9.7 \pm 1.3 \times 10^{-3} $ & $ 1.4 \pm 0.2 \times 10^{-2} $ & $ 1.1 \pm 0.2 \times 10^{-2} $ & $ 1.1 \pm 0.2 \times 10^{-2} $ \\
 &  & $ 10^{2} $ & $ 1.3 \pm 0.1 \times 10^{-3} $ & $ 1.7 \pm 0.3 \times 10^{-3} $ & $ 1.4 \pm 0.2 \times 10^{-3} $ & $ 1.5 \pm 0.3 \times 10^{-3} $ \\
\cline{2-7}
 & \multirow[c]{4}{*}{$ 10^{-3} $} & $ 10^{-1} $ & $ 1.5 \pm 0.0 \times 10^{-1} $ & $ 1.6 \pm 0.0 \times 10^{-1} $ & $ 1.5 \pm 0.0 \times 10^{-1} $ & $ 1.5 \pm 0.0 \times 10^{-1} $ \\
 &  & $ 10^{0} $ & $ 5.0 \pm 0.2 \times 10^{-2} $ & $ 6.3 \pm 0.8 \times 10^{-2} $ & $ 4.9 \pm 0.4 \times 10^{-2} $ & $ 4.9 \pm 0.4 \times 10^{-2} $ \\
 &  & $ 10^{1} $ & $ 1.1 \pm 0.2 \times 10^{-2} $ & $ 1.3 \pm 0.2 \times 10^{-2} $ & $ 1.1 \pm 0.2 \times 10^{-2} $ & $ 1.1 \pm 0.1 \times 10^{-2} $ \\
 &  & $ 10^{2} $ & $ 1.4 \pm 0.2 \times 10^{-3} $ & $ 1.7 \pm 0.3 \times 10^{-3} $ & $ 1.6 \pm 0.3 \times 10^{-3} $ & $ 1.5 \pm 0.2 \times 10^{-3} $ \\
\cline{2-7}
 & \multirow[c]{4}{*}{$ 10^{-1} $} & $ 10^{-1} $ & $ 1.5 \pm 0.0 \times 10^{-1} $ & $ 1.7 \pm 0.0 \times 10^{-1} $ & $ 1.5 \pm 0.0 \times 10^{-1} $ & $ 1.5 \pm 0.0 \times 10^{-1} $ \\
 &  & $ 10^{0} $ & $ 5.4 \pm 0.4 \times 10^{-2} $ & $ 5.4 \pm 0.2 \times 10^{-2} $ & $ 4.9 \pm 0.2 \times 10^{-2} $ & $ 5.3 \pm 0.2 \times 10^{-2} $ \\
 &  & $ 10^{1} $ & $ 2.2 \pm 0.2 \times 10^{-2} $ & $ 2.0 \pm 0.2 \times 10^{-2} $ & $ 1.6 \pm 0.1 \times 10^{-2} $ & $ 2.1 \pm 0.1 \times 10^{-2} $ \\
 &  & $ 10^{2} $ & $ 1.4 \pm 0.1 \times 10^{-2} $ & $ 1.4 \pm 0.2 \times 10^{-2} $ & $ 8.9 \pm 1.9 \times 10^{-3} $ & $ 1.4 \pm 0.1 \times 10^{-2} $ \\
\midrule
\multirow[c]{12}{*}{32} & \multirow[c]{4}{*}{$ 10^{-5} $} & $ 10^{-1} $ & $ 2.9 \pm 0.0 \times 10^{-1} $ & $ 3.1 \pm 0.0 \times 10^{-1} $ & $ 2.9 \pm 0.0 \times 10^{-1} $ & $ 2.8 \pm 0.2 \times 10^{-1} $ \\
 &  & $ 10^{0} $ & $ 9.8 \pm 0.2 \times 10^{-2} $ & $ 1.9 \pm 0.5 \times 10^{-1} $ & $ 1.0 \pm 0.0 \times 10^{-1} $ & $ 9.4 \pm 0.4 \times 10^{-2} $ \\
 &  & $ 10^{1} $ & $ 2.4 \pm 0.1 \times 10^{-2} $ & $ 6.3 \pm 1.1 \times 10^{-2} $ & $ 2.6 \pm 0.2 \times 10^{-2} $ & $ 2.5 \pm 0.2 \times 10^{-2} $ \\
 &  & $ 10^{2} $ & $ 4.3 \pm 0.6 \times 10^{-3} $ & $ 9.9 \pm 1.2 \times 10^{-3} $ & $ 5.7 \pm 0.9 \times 10^{-3} $ & $ 7.1 \pm 0.9 \times 10^{-3} $ \\
\cline{2-7}
 & \multirow[c]{4}{*}{$ 10^{-3} $} & $ 10^{-1} $ & $ 2.9 \pm 0.0 \times 10^{-1} $ & $ 2.7 \pm 0.0 \times 10^{-1} $ & $ 2.9 \pm 0.0 \times 10^{-1} $ & $ 2.9 \pm 0.0 \times 10^{-1} $ \\
 &  & $ 10^{0} $ & $ 1.0 \pm 0.0 \times 10^{-1} $ & $ 1.3 \pm 0.2 \times 10^{-1} $ & $ 1.1 \pm 0.0 \times 10^{-1} $ & $ 9.9 \pm 0.5 \times 10^{-2} $ \\
 &  & $ 10^{1} $ & $ 2.5 \pm 0.2 \times 10^{-2} $ & $ 3.7 \pm 0.4 \times 10^{-2} $ & $ 2.8 \pm 0.1 \times 10^{-2} $ & $ 2.6 \pm 0.4 \times 10^{-2} $ \\
 &  & $ 10^{2} $ & $ 6.3 \pm 1.2 \times 10^{-3} $ & $ 6.8 \pm 1.2 \times 10^{-3} $ & $ 6.7 \pm 1.1 \times 10^{-3} $ & $ 6.1 \pm 1.5 \times 10^{-3} $ \\
\cline{2-7}
 & \multirow[c]{4}{*}{$ 10^{-1} $} & $ 10^{-1} $ & $ 2.9 \pm 0.0 \times 10^{-1} $ & $ 2.8 \pm 0.0 \times 10^{-1} $ & $ 2.9 \pm 0.0 \times 10^{-1} $ & $ 2.9 \pm 0.0 \times 10^{-1} $ \\
 &  & $ 10^{0} $ & $ 1.0 \pm 0.0 \times 10^{-1} $ & $ 1.1 \pm 0.1 \times 10^{-1} $ & $ 9.6 \pm 0.1 \times 10^{-2} $ & $ 9.6 \pm 0.2 \times 10^{-2} $ \\
 &  & $ 10^{1} $ & $ 2.8 \pm 0.2 \times 10^{-2} $ & $ 2.5 \pm 0.1 \times 10^{-2} $ & $ 2.4 \pm 0.1 \times 10^{-2} $ & $ 2.3 \pm 0.1 \times 10^{-2} $ \\
 &  & $ 10^{2} $ & $ 1.1 \pm 0.1 \times 10^{-2} $ & $ 1.1 \pm 0.1 \times 10^{-2} $ & $ 9.8 \pm 0.8 \times 10^{-3} $ & $ 9.2 \pm 0.7 \times 10^{-3} $ \\
\midrule
\multirow[c]{12}{*}{64} & \multirow[c]{4}{*}{$ 10^{-5} $} & $ 10^{-1} $ & $ 4.5 \pm 0.0 \times 10^{-1} $ & $ 4.2 \pm 0.0 \times 10^{-1} $ & $ 4.5 \pm 0.1 \times 10^{-1} $ & $ 4.6 \pm 0.1 \times 10^{-1} $ \\
 &  & $ 10^{0} $ & $ 2.1 \pm 0.0 \times 10^{-1} $ & $ 2.1 \pm 0.4 \times 10^{-1} $ & $ 2.4 \pm 0.2 \times 10^{-1} $ & $ 2.2 \pm 0.1 \times 10^{-1} $ \\
 &  & $ 10^{1} $ & $ 4.8 \pm 0.4 \times 10^{-2} $ & $ 1.2 \pm 0.3 \times 10^{-1} $ & $ 6.0 \pm 0.9 \times 10^{-2} $ & $ 6.1 \pm 0.8 \times 10^{-2} $ \\
 &  & $ 10^{2} $ & $ 1.3 \pm 0.1 \times 10^{-2} $ & $ 2.8 \pm 0.4 \times 10^{-2} $ & $ 2.7 \pm 0.5 \times 10^{-2} $ & $ 2.6 \pm 0.5 \times 10^{-2} $ \\
\cline{2-7}
 & \multirow[c]{4}{*}{$ 10^{-3} $} & $ 10^{-1} $ & $ 4.5 \pm 0.0 \times 10^{-1} $ & $ 4.5 \pm 0.2 \times 10^{-1} $ & $ 4.7 \pm 0.0 \times 10^{-1} $ & $ 4.5 \pm 0.0 \times 10^{-1} $ \\
 &  & $ 10^{0} $ & $ 2.1 \pm 0.1 \times 10^{-1} $ & $ 2.5 \pm 0.5 \times 10^{-1} $ & $ 3.1 \pm 0.2 \times 10^{-1} $ & $ 2.1 \pm 0.1 \times 10^{-1} $ \\
 &  & $ 10^{1} $ & $ 4.9 \pm 0.3 \times 10^{-2} $ & $ 1.1 \pm 0.2 \times 10^{-1} $ & $ 1.1 \pm 0.7 \times 10^{-1} $ & $ 4.8 \pm 0.4 \times 10^{-2} $ \\
 &  & $ 10^{2} $ & $ 1.2 \pm 0.1 \times 10^{-2} $ & $ 2.1 \pm 0.2 \times 10^{-2} $ & $ 4.4 \pm 0.8 \times 10^{-2} $ & $ 1.1 \pm 0.1 \times 10^{-2} $ \\
\cline{2-7}
 & \multirow[c]{4}{*}{$ 10^{-1} $} & $ 10^{-1} $ & $ 4.5 \pm 0.0 \times 10^{-1} $ & $ 4.2 \pm 0.0 \times 10^{-1} $ & $ 4.5 \pm 0.0 \times 10^{-1} $ & $ 4.5 \pm 0.0 \times 10^{-1} $ \\
 &  & $ 10^{0} $ & $ 2.3 \pm 0.1 \times 10^{-1} $ & $ 2.4 \pm 0.2 \times 10^{-1} $ & $ 2.2 \pm 0.0 \times 10^{-1} $ & $ 2.0 \pm 0.1 \times 10^{-1} $ \\
 &  & $ 10^{1} $ & $ 5.2 \pm 0.1 \times 10^{-2} $ & $ 8.2 \pm 0.8 \times 10^{-2} $ & $ 5.4 \pm 0.2 \times 10^{-2} $ & $ 5.3 \pm 0.4 \times 10^{-2} $ \\
 &  & $ 10^{2} $ & $ 1.2 \pm 0.3 \times 10^{-2} $ & $ 1.4 \pm 0.3 \times 10^{-2} $ & $ 1.6 \pm 0.2 \times 10^{-2} $ & $ 1.4 \pm 0.4 \times 10^{-2} $ \\
\bottomrule
\end{tabular}

%% file: photon_loss/tables_time/mean_75/random_gaussian/initial_0/photon_loss_table.tex
\begin{tabular}{lcl|c|c|c|c}
\toprule
d & $\beta$ & $T_1$ [$ms$] & \texttt{Full} & \texttt{Diagonal (adaptive)} & \texttt{Diagonal (multiple)} & \texttt{Grid} \\
\midrule
\multirow[c]{12}{*}{8} & \multirow[c]{4}{*}{$ 10^{-5} $} & $ 10^{-1} $ & $ 5.3 \pm 0.0 \times 10^{-1} $ & $ 5.2 \pm 0.1 \times 10^{-1} $ & $ 5.2 \pm 0.1 \times 10^{-1} $ & $ 5.3 \pm 0.0 \times 10^{-1} $ \\
 &  & $ 10^{0} $ & $ 1.6 \pm 0.2 \times 10^{-1} $ & $ 1.5 \pm 0.2 \times 10^{-1} $ & $ 1.6 \pm 0.2 \times 10^{-1} $ & $ 1.6 \pm 0.2 \times 10^{-1} $ \\
 &  & $ 10^{1} $ & $ 1.7 \pm 0.2 \times 10^{-2} $ & $ 1.8 \pm 0.2 \times 10^{-2} $ & $ 2.0 \pm 0.3 \times 10^{-2} $ & $ 1.9 \pm 0.2 \times 10^{-2} $ \\
 &  & $ 10^{2} $ & $ 2.2 \pm 0.3 \times 10^{-3} $ & $ 2.2 \pm 0.3 \times 10^{-3} $ & $ 2.4 \pm 0.4 \times 10^{-3} $ & $ 2.4 \pm 0.3 \times 10^{-3} $ \\
\cline{2-7}
 & \multirow[c]{4}{*}{$ 10^{-3} $} & $ 10^{-1} $ & $ 5.2 \pm 0.0 \times 10^{-1} $ & $ 5.2 \pm 0.0 \times 10^{-1} $ & $ 5.2 \pm 0.0 \times 10^{-1} $ & $ 5.2 \pm 0.0 \times 10^{-1} $ \\
 &  & $ 10^{0} $ & $ 1.6 \pm 0.2 \times 10^{-1} $ & $ 1.5 \pm 0.2 \times 10^{-1} $ & $ 1.5 \pm 0.2 \times 10^{-1} $ & $ 1.6 \pm 0.3 \times 10^{-1} $ \\
 &  & $ 10^{1} $ & $ 1.9 \pm 0.2 \times 10^{-2} $ & $ 1.8 \pm 0.2 \times 10^{-2} $ & $ 1.7 \pm 0.2 \times 10^{-2} $ & $ 1.8 \pm 0.3 \times 10^{-2} $ \\
 &  & $ 10^{2} $ & $ 2.3 \pm 0.3 \times 10^{-3} $ & $ 2.2 \pm 0.2 \times 10^{-3} $ & $ 1.9 \pm 0.3 \times 10^{-3} $ & $ 2.4 \pm 0.5 \times 10^{-3} $ \\
\cline{2-7}
 & \multirow[c]{4}{*}{$ 10^{-1} $} & $ 10^{-1} $ & $ 5.2 \pm 0.0 \times 10^{-1} $ & $ 5.3 \pm 0.1 \times 10^{-1} $ & $ 5.2 \pm 0.0 \times 10^{-1} $ & $ 5.2 \pm 0.0 \times 10^{-1} $ \\
 &  & $ 10^{0} $ & $ 1.7 \pm 0.3 \times 10^{-1} $ & $ 1.7 \pm 0.3 \times 10^{-1} $ & $ 1.7 \pm 0.2 \times 10^{-1} $ & $ 1.6 \pm 0.2 \times 10^{-1} $ \\
 &  & $ 10^{1} $ & $ 2.4 \pm 0.7 \times 10^{-2} $ & $ 2.6 \pm 1.1 \times 10^{-2} $ & $ 3.0 \pm 1.4 \times 10^{-2} $ & $ 2.3 \pm 0.6 \times 10^{-2} $ \\
 &  & $ 10^{2} $ & $ 6.4 \pm 6.7 \times 10^{-3} $ & $ 6.8 \pm 9.8 \times 10^{-3} $ & $ 7.9 \pm 8.1 \times 10^{-3} $ & $ 6.5 \pm 6.2 \times 10^{-3} $ \\
\midrule
\multirow[c]{12}{*}{16} & \multirow[c]{4}{*}{$ 10^{-5} $} & $ 10^{-1} $ & $ 7.6 \pm 0.0 \times 10^{-1} $ & $ 7.7 \pm 0.1 \times 10^{-1} $ & $ 7.5 \pm 0.0 \times 10^{-1} $ & $ 7.6 \pm 0.0 \times 10^{-1} $ \\
 &  & $ 10^{0} $ & $ 3.2 \pm 0.2 \times 10^{-1} $ & $ 3.9 \pm 0.2 \times 10^{-1} $ & $ 3.4 \pm 0.2 \times 10^{-1} $ & $ 3.3 \pm 0.2 \times 10^{-1} $ \\
 &  & $ 10^{1} $ & $ 4.9 \pm 0.5 \times 10^{-2} $ & $ 5.8 \pm 0.5 \times 10^{-2} $ & $ 5.2 \pm 0.5 \times 10^{-2} $ & $ 4.8 \pm 0.4 \times 10^{-2} $ \\
 &  & $ 10^{2} $ & $ 5.1 \pm 0.4 \times 10^{-3} $ & $ 6.6 \pm 0.5 \times 10^{-3} $ & $ 5.9 \pm 0.5 \times 10^{-3} $ & $ 5.5 \pm 0.6 \times 10^{-3} $ \\
\cline{2-7}
 & \multirow[c]{4}{*}{$ 10^{-3} $} & $ 10^{-1} $ & $ 7.6 \pm 0.0 \times 10^{-1} $ & $ 7.8 \pm 0.1 \times 10^{-1} $ & $ 7.6 \pm 0.0 \times 10^{-1} $ & $ 7.6 \pm 0.0 \times 10^{-1} $ \\
 &  & $ 10^{0} $ & $ 3.3 \pm 0.2 \times 10^{-1} $ & $ 4.1 \pm 0.2 \times 10^{-1} $ & $ 3.3 \pm 0.2 \times 10^{-1} $ & $ 3.2 \pm 0.2 \times 10^{-1} $ \\
 &  & $ 10^{1} $ & $ 4.7 \pm 0.4 \times 10^{-2} $ & $ 6.0 \pm 0.5 \times 10^{-2} $ & $ 4.7 \pm 0.3 \times 10^{-2} $ & $ 4.8 \pm 0.5 \times 10^{-2} $ \\
 &  & $ 10^{2} $ & $ 5.4 \pm 0.5 \times 10^{-3} $ & $ 6.3 \pm 0.4 \times 10^{-3} $ & $ 5.6 \pm 0.6 \times 10^{-3} $ & $ 5.9 \pm 0.6 \times 10^{-3} $ \\
\cline{2-7}
 & \multirow[c]{4}{*}{$ 10^{-1} $} & $ 10^{-1} $ & $ 7.6 \pm 0.0 \times 10^{-1} $ & $ 7.7 \pm 0.1 \times 10^{-1} $ & $ 7.5 \pm 0.0 \times 10^{-1} $ & $ 7.6 \pm 0.0 \times 10^{-1} $ \\
 &  & $ 10^{0} $ & $ 3.3 \pm 0.2 \times 10^{-1} $ & $ 3.5 \pm 0.2 \times 10^{-1} $ & $ 3.3 \pm 0.2 \times 10^{-1} $ & $ 3.2 \pm 0.2 \times 10^{-1} $ \\
 &  & $ 10^{1} $ & $ 5.0 \pm 0.7 \times 10^{-2} $ & $ 5.8 \pm 0.9 \times 10^{-2} $ & $ 4.9 \pm 0.6 \times 10^{-2} $ & $ 5.3 \pm 0.7 \times 10^{-2} $ \\
 &  & $ 10^{2} $ & $ 7.6 \pm 2.1 \times 10^{-3} $ & $ 1.2 \pm 0.6 \times 10^{-2} $ & $ 6.5 \pm 1.3 \times 10^{-3} $ & $ 7.1 \pm 1.4 \times 10^{-3} $ \\
\midrule
\multirow[c]{12}{*}{32} & \multirow[c]{4}{*}{$ 10^{-5} $} & $ 10^{-1} $ & $ 8.5 \pm 0.0 \times 10^{-1} $ & $ 8.8 \pm 0.0 \times 10^{-1} $ & $ 8.5 \pm 0.0 \times 10^{-1} $ & $ 8.5 \pm 0.0 \times 10^{-1} $ \\
 &  & $ 10^{0} $ & $ 5.7 \pm 0.1 \times 10^{-1} $ & $ 7.5 \pm 0.2 \times 10^{-1} $ & $ 5.9 \pm 0.1 \times 10^{-1} $ & $ 5.7 \pm 0.1 \times 10^{-1} $ \\
 &  & $ 10^{1} $ & $ 1.3 \pm 0.1 \times 10^{-1} $ & $ 2.0 \pm 0.1 \times 10^{-1} $ & $ 1.5 \pm 0.1 \times 10^{-1} $ & $ 1.3 \pm 0.1 \times 10^{-1} $ \\
 &  & $ 10^{2} $ & $ 1.6 \pm 0.1 \times 10^{-2} $ & $ 2.7 \pm 0.2 \times 10^{-2} $ & $ 1.8 \pm 0.1 \times 10^{-2} $ & $ 1.6 \pm 0.1 \times 10^{-2} $ \\
\cline{2-7}
 & \multirow[c]{4}{*}{$ 10^{-3} $} & $ 10^{-1} $ & $ 8.5 \pm 0.0 \times 10^{-1} $ & $ 8.8 \pm 0.0 \times 10^{-1} $ & $ 8.5 \pm 0.0 \times 10^{-1} $ & $ 8.5 \pm 0.0 \times 10^{-1} $ \\
 &  & $ 10^{0} $ & $ 5.8 \pm 0.1 \times 10^{-1} $ & $ 7.8 \pm 0.1 \times 10^{-1} $ & $ 5.7 \pm 0.1 \times 10^{-1} $ & $ 6.0 \pm 0.2 \times 10^{-1} $ \\
 &  & $ 10^{1} $ & $ 1.3 \pm 0.1 \times 10^{-1} $ & $ 2.1 \pm 0.1 \times 10^{-1} $ & $ 1.3 \pm 0.1 \times 10^{-1} $ & $ 1.4 \pm 0.1 \times 10^{-1} $ \\
 &  & $ 10^{2} $ & $ 1.5 \pm 0.1 \times 10^{-2} $ & $ 2.5 \pm 0.1 \times 10^{-2} $ & $ 1.7 \pm 0.1 \times 10^{-2} $ & $ 1.6 \pm 0.1 \times 10^{-2} $ \\
\cline{2-7}
 & \multirow[c]{4}{*}{$ 10^{-1} $} & $ 10^{-1} $ & $ 8.5 \pm 0.0 \times 10^{-1} $ & $ 8.8 \pm 0.0 \times 10^{-1} $ & $ 8.5 \pm 0.0 \times 10^{-1} $ & $ 8.5 \pm 0.0 \times 10^{-1} $ \\
 &  & $ 10^{0} $ & $ 5.7 \pm 0.1 \times 10^{-1} $ & $ 7.2 \pm 0.1 \times 10^{-1} $ & $ 5.7 \pm 0.1 \times 10^{-1} $ & $ 5.8 \pm 0.1 \times 10^{-1} $ \\
 &  & $ 10^{1} $ & $ 1.4 \pm 0.1 \times 10^{-1} $ & $ 1.6 \pm 0.1 \times 10^{-1} $ & $ 1.3 \pm 0.1 \times 10^{-1} $ & $ 1.4 \pm 0.2 \times 10^{-1} $ \\
 &  & $ 10^{2} $ & $ 1.8 \pm 0.4 \times 10^{-2} $ & $ 2.3 \pm 0.5 \times 10^{-2} $ & $ 1.9 \pm 0.5 \times 10^{-2} $ & $ 2.0 \pm 0.5 \times 10^{-2} $ \\
\midrule
\multirow[c]{12}{*}{64} & \multirow[c]{4}{*}{$ 10^{-5} $} & $ 10^{-1} $ & $ 8.8 \pm 0.4 \times 10^{-1} $ & $ 8.7 \pm 0.4 \times 10^{-1} $ & $ 8.7 \pm 0.2 \times 10^{-1} $ & $ 8.7 \pm 0.4 \times 10^{-1} $ \\
 &  & $ 10^{0} $ & $ 7.6 \pm 0.0 \times 10^{-1} $ & $ 8.4 \pm 0.1 \times 10^{-1} $ & $ 7.8 \pm 0.1 \times 10^{-1} $ & $ 7.8 \pm 0.1 \times 10^{-1} $ \\
 &  & $ 10^{1} $ & $ 3.2 \pm 0.1 \times 10^{-1} $ & $ 5.6 \pm 0.3 \times 10^{-1} $ & $ 3.8 \pm 0.2 \times 10^{-1} $ & $ 3.8 \pm 0.3 \times 10^{-1} $ \\
 &  & $ 10^{2} $ & $ 5.0 \pm 0.6 \times 10^{-2} $ & $ 1.2 \pm 0.1 \times 10^{-1} $ & $ 7.9 \pm 1.6 \times 10^{-2} $ & $ 8.3 \pm 1.2 \times 10^{-2} $ \\
\cline{2-7}
 & \multirow[c]{4}{*}{$ 10^{-3} $} & $ 10^{-1} $ & $ 8.6 \pm 0.0 \times 10^{-1} $ & $ 8.6 \pm 0.0 \times 10^{-1} $ & $ 8.6 \pm 0.0 \times 10^{-1} $ & $ 8.8 \pm 0.4 \times 10^{-1} $ \\
 &  & $ 10^{0} $ & $ 7.6 \pm 0.0 \times 10^{-1} $ & $ 8.4 \pm 0.2 \times 10^{-1} $ & $ 7.8 \pm 0.1 \times 10^{-1} $ & $ 7.9 \pm 0.2 \times 10^{-1} $ \\
 &  & $ 10^{1} $ & $ 3.1 \pm 0.1 \times 10^{-1} $ & $ 5.8 \pm 0.2 \times 10^{-1} $ & $ 3.8 \pm 0.2 \times 10^{-1} $ & $ 3.9 \pm 0.1 \times 10^{-1} $ \\
 &  & $ 10^{2} $ & $ 6.0 \pm 0.8 \times 10^{-2} $ & $ 1.2 \pm 0.2 \times 10^{-1} $ & $ 6.8 \pm 0.8 \times 10^{-2} $ & $ 8.7 \pm 1.0 \times 10^{-2} $ \\
\cline{2-7}
 & \multirow[c]{4}{*}{$ 10^{-1} $} & $ 10^{-1} $ & $ 8.7 \pm 0.4 \times 10^{-1} $ & $ 8.6 \pm 0.0 \times 10^{-1} $ & $ 9.0 \pm 0.5 \times 10^{-1} $ & $ 9.1 \pm 0.6 \times 10^{-1} $ \\
 &  & $ 10^{0} $ & $ 7.6 \pm 0.0 \times 10^{-1} $ & $ 8.4 \pm 0.0 \times 10^{-1} $ & $ 7.6 \pm 0.0 \times 10^{-1} $ & $ 7.6 \pm 0.1 \times 10^{-1} $ \\
 &  & $ 10^{1} $ & $ 3.1 \pm 0.1 \times 10^{-1} $ & $ 4.1 \pm 0.1 \times 10^{-1} $ & $ 3.1 \pm 0.2 \times 10^{-1} $ & $ 3.1 \pm 0.1 \times 10^{-1} $ \\
 &  & $ 10^{2} $ & $ 5.8 \pm 1.3 \times 10^{-2} $ & $ 6.3 \pm 0.8 \times 10^{-2} $ & $ 5.7 \pm 1.4 \times 10^{-2} $ & $ 6.0 \pm 1.0 \times 10^{-2} $ \\
\bottomrule
\end{tabular}